\pdfoutput=1
\documentclass[12pt]{article}

\font\grande=cmr9.5 scaled \magstep4
\font\medio=cmr9.5 scaled \magstep2
\outer\def\beginsection#1\par{\medbreak\bigskip
      \message{#1}\leftline{\bf#1}\nobreak\medskip
\vskip-\parskip
      \noindent}
\usepackage{graphicx} 
\begin{document}
\bibliographystyle {unsrt}

\titlepage

\begin{flushright}
CERN-PH-TH/2015-137
\end{flushright}

\vspace{10mm}
\begin{center}
{\grande Inflationary magnetogenesis, derivative couplings }\\
\vspace{10mm}
{\grande and relativistic Van der Waals interactions}\\
\vspace{1.5cm}
 Massimo Giovannini
 \footnote{Electronic address: massimo.giovannini@cern.ch}\\
\vspace{1cm}
{{\sl Department of Physics, 
Theory Division, CERN, 1211 Geneva 23, Switzerland }}\\
\vspace{0.5cm}
{{\sl INFN, Section of Milan-Bicocca, 20126 Milan, Italy}}
\vspace*{0.5cm}
\end{center}

\vskip 0.5cm
\centerline{\medio  Abstract}
When the gauge fields have derivative couplings to scalars, like in the case 
of the relativistic theory of Van der Waals (or Casimir-Polder) interactions, conformal invariance is broken but the magnetic and electric susceptibilities 
are not bound to coincide. We analyze the formation of large-scale magnetic fields in slow-roll inflation and find that they are generated at the level of a few hundredths 
of a nG and over typical length scales between few Mpc and $100$ Mpc. Using a new time parametrization that reduces to conformal time but only for coincident susceptibilities, 
the gauge action is quantized while the evolution equations of the corresponding mode functions are more easily solvable.
The power spectra depend on the normalized rates of variation of the two susceptibilities (or of the corresponding gauge couplings) and on the absolute value of their ratio at the beginning of inflation.
We pin down explicit regions in the parameter space where all the physical requirements (i.e. the  backreaction constraints, the magnetogenesis bounds  and the 
naturalness of the initial conditions of the scenario) are jointly satisfied. Weakly coupled initial data are favoured if the gauge couplings are of the same order at the 
end of inflation. Duality is systematically used to simplify the analysis of the wide parameter space of the model.

\vskip 0.5cm

\noindent

\vspace{5mm}

\vfill
\newpage
\renewcommand{\theequation}{1.\arabic{equation}}
\setcounter{equation}{0}
\section{Introduction}
\label{sec1}
Magnetic fields with typical correlation scales exceeding the astronomical unit (i.e. roughly $10^{13}$ cm) 
permeate the interstellar and intergalactic plasmas which are, in many respects, very similar to the one we can produce in 
terrestrial experiments \cite{b1,b2}.  In spite of this,  the origin of large-scale magnetism is still under intense debate both theoretically and observationally \cite{rev}.
As  proposed in the last few years  \cite{magnetized1}  the temperature and the polarization anisotropies 
of the Cosmic Microwave Background (CMB in what follows) may be magnetized. This 
observation offers the unique opportunity of direct limits on the large-scale magnetism prior to matter-radiation equality since the large-scale magnetic 
fields affect directly the initial conditions of the Einstein-Boltzmann hierarchy. 
The current Planck explorer data can be used to set bounds on large-scale magnetic fields \cite{magnetized2} as previoulsy done with the WMAP 
3-yr and 9-yr releases (see, respectively, first and second papers of Ref.  \cite{magnetized2}). The results can be summarized by saying that
the WMAP9 \cite{data1,data2} and Planck data \cite{data3} are compatible, they are both sensitive to magnetic fields in the nG range ($1\,\mathrm{nG} = 10^{-9} \,\mathrm{G}$) 
for magnetic spectral indices $n_{B} = {\mathcal O}(1.3)$ using the available temperature and polarization power spectra\footnote{We use here 
the same conventions employed to assign the curvature power spectra: the scale invariant magnetic power spectra are realized for $n_{B} \to 1$
\cite{magnetized1,magnetized2}. Within these conventions the magnetic power spectrum (i.e. the Fourier transform 
of the two-point function) has the same dimensions of the magnetic energy density. While other conventions stipulate that the scale-invariant 
limit is realized for $n_{B} \to -3$, in the present paper, the Fourier transform will be consistently assigned for 
the scalar modes of the geometry and for the magnetic fields implying that  $n_{B} \to 1$ is the scale-invariant limit.}.

It has been repeatedly argued that large-scale magnetic fields might well be generated during a stage of inflationary expansion.
The rationale for this requirement has to do with the correlation scale of the produced field that must be sufficiently large 
and the onset of the rotation of the protogalaxy. A very promising framework for generating magnetic fields with comoving correlation 
scales exceeding the Mpc is represented by models where the gauge fields couple directly to one or more scalar fields (see \cite{DT1,DT2,DT3,DT4} for an incomplete list of references). The scalar fields may coincide with one or (more inflatons) or even with multiple spectator fields. The conventional class of models is based on the following action\footnote{In the notations employed on this paper $Y^{\mu\nu}$ and $\tilde{Y}^{\mu\nu}$ are, respectively, the gauge field strength and its dual; $g = \mathrm{det}g_{\mu\nu}$ is the determinant of the four-dimensional metric with signature mostly minus.}: 
\begin{equation}
S = - \frac{1}{16 \pi} \int \, d^{4} x\, \sqrt{-g} \biggl[ \lambda(\varphi,\psi) Y_{\alpha\beta} \, Y^{\alpha\beta} + \lambda_{pseudo}(\varphi,\psi) 
Y_{\alpha\beta}\, \tilde{Y}^{\alpha\beta}\biggr], 
\label{first}
\end{equation}
where $\varphi$ and $\psi$ may denote, for instance, a generic inflaton field and a generic spectator field. Various situations 
can be envisaged and most of them have been investigated in the literature. The presence of $\lambda(\varphi,\psi)$  in Eq. (\ref{first}) 
 is more relevant than the pseudoscalar  (axion-like \cite{c0,c0a}) coupling which will be ignored even if it has been studied by many authors \cite{c1,c1a,c2}
 in the context of the magnetic field generation. For the amplification of the magnetic field itself the pseudoscalar vertex is not so efficient
  but it is relevant when one wants to generate magnetic fields whose flux lines are linked or twisted as originally discussed in \cite{mm1}. The 
  produced Chern-Simons condensate leads to a viable mechanism for baryogenesis via hypermagnetic knots \cite{mm1,mm2}. These  helical fields 
 play also a role in anomalous magnetohydrodynamics where the evolution of the magnetic fields 
 at finite conductivity is analyzed in the presence of anomalous charges \cite{cme1}. In the collisions of heavy ions this phenomenon 
 is often dubbed chiral magnetic effect \cite{cme2}.
 
It has been recently argued  \cite{SUSC1} that the class of models pinned down by Eq. (\ref{first}) can be complemented by further terms: 
\begin{eqnarray}
S = - \frac{1}{16 \pi} \int \, d^{4} x\, \sqrt{-g} \biggl[ {\mathcal M}_{\sigma}^{\rho}(\varphi,\psi) 
Y_{\rho\alpha}\, Y^{\sigma\alpha} - {\mathcal N}_{\sigma}^{\rho}(\varphi,\psi) 
\tilde{Y}_{\rho\alpha}\, \tilde{Y}^{\sigma\alpha} \biggr].
\label{second}
\end{eqnarray} 
Equation (\ref{second}) leads to unequal electric and magnetic susceptibilities \cite{SUSC1}. As a special case Eq. (\ref{second}) includes the typical derivative coupling arising in the relativistic theory of Casimir-Polder and Van der Waals interactions \cite{such}:
\begin{equation}
S = - \int d^{4} x \, \sqrt{-g}\,\biggl[ g_{1} \partial_{\alpha} \varphi \partial_{\beta} \varphi^{*} \, Y^{\alpha\rho} \, Y^{\beta}_{\, \rho} + g_{2} |\varphi|^2\, Y_{\alpha\beta} \, Y^{\alpha\beta} \biggr].
\label{third}
\end{equation}
Other terms potentially present in Eq. (\ref{second}) (such as ${\mathcal P}_{\rho}^{\sigma} Y_{\sigma\alpha}\tilde{Y}^{\rho\alpha}$) will be neglected 
even if, as explained above, they might be relevant for the evolution of the magnetic helicity.
 
The simplest parametrization of the coupling functions ${\mathcal M}_{\rho\sigma}$ and ${\mathcal N}_{\rho\sigma}$ is \cite{SUSC1}:
\begin{equation}
{\mathcal M}_{\rho\sigma}(\varphi) = \lambda_{E}(\varphi) u_{\rho}(\varphi)\,u_{\sigma}(\varphi), \qquad {\mathcal N}_{\rho\sigma}(\psi) = \lambda_{B}(\psi) \overline{u}_{\rho}(\psi) \,\overline{u}_{\sigma}(\psi).
\label{fourth}
\end{equation}
Note that  $u_{\rho}(\varphi)$ and $\overline{u}_{\rho}(\psi)$ appearing in Eq. (\ref{fourth}) are the normalized gradients of the corresponding scalar fields but, in the context of a purely 
hydrodynamical model, they can also play the role of the four-velocities of a relativistic fluid.
The full action has been taken to be the sum of the first term of Eq. (\ref{first}) and of the remaining two terms in Eq. (\ref{third}) with 
the parametrization of Eq. (\ref{fourth}). In this case the electric and the magnetic susceptibilities are given by:
\begin{equation}
\chi_{E} = \sqrt{\lambda + \frac{\lambda_{E}}{2}}, \qquad \chi_{B} = \sqrt{\lambda + \frac{\lambda_{B}}{2}}.
\label{fifth}
\end{equation}

In this framework viable magnetogeneis models can be formulated in different dynamical situations and 
some of these possibilities have been already swiftly examined \cite{SUSC1}.
The purpose of the present paper is to undertake a comprehensive analysis of the parameter space of this scenario
in terms of the initial conditions and of the evolution of the gauge couplings. To achieve this goal 
the expressions of ${\mathcal M}_{\rho\sigma}$ and ${\mathcal N}_{\rho\sigma}$ shall be first generalized. 
The three essential parameters of this class of models are the normalized rate of variation 
 of the electric and magnetic gauge couplings (denoted, respectively, by $F_{E}$ and $F_{B}$) and also the 
 initial value of their ratio.

From a pragmatic viewpoint, the whole parameter space can be reduced 
to the first quadrant of the $(F_{B},\, F_{E})$ plane where $F_{B}$ and $F_{E}$ are both positive: the other three quadrants can be charted through the systematic use of the duality transformations\footnote{The duality 
transformations used here are a simple generalization of the strandard electromagnetic duality \cite{duality1}. While in the standard case the electric and the 
magnetic gauge couplings coincide, in the present situation they differ. See section \ref{sec2} for a more detailed discussion.}.
To save time we first impose the backreaction constraints over different scales, then discuss the naturalness of the initial conditions and finally 
verify if the obtained region of the parameter space satisfies the magnetogensis requirements. 

A technical aspect of the present analysis concerns the quantization of the action and the evolution of the 
related mode functions. This problem can be discussed in the conventional manner by using the conformal time 
coordinate $\tau$. However the analysis becomes much more transparent in terms of a newly time variable, denoted hereunder by $\eta$, which reduces to the conformal time but only in the case of coincident  electric and the magnetic gauge couplings. 

This paper is organized as follows. In  section \ref{sec2} we present the generally covariant decomposition 
of the coupling functions ${\mathcal M}_{\rho\sigma}$ and ${\mathcal N}_{\rho\sigma}$ and study their symmetry properties 
in the light of the equations of motion. We also show how the obtained parametrization fits 
within large field and small field inflationary models. In section \ref{sec3} we show how the power spectra can be
derived by avoiding the standard conformal time parametrization and by expressing the full action in terms of a new time variable directly 
related to the mismatch between the magnetic and the electric susceptibilities. The magnetic and electric power spectra 
are derived in section \ref{sec4}; we shall also analyze how the different regions of the parameter space are transformed under 
duality and conclude that the most relevant region (from the practical viewpoint) is represented by the first quadrant of the $(F_{B},\, F_{E})$ plane.
In section \ref{sec5} the different portions of the $(F_{B},\, F_{E})$ plane are scrutinized in the light of the backreaction constraints (and for different initial 
conditions of the gauge couplings). The allowed region of the parameter space is explicitly obtained and discussed in detail.
Section \ref{sec6} contains our concluding remarks and a summary of the main findings.

\renewcommand{\theequation}{2.\arabic{equation}}
\setcounter{equation}{0}
\section{General parametrization of the coupling functions}
\label{sec2}
The general form of the coupling functions shall now be discussed 
first in the context of relativistic hydrodynamics and then in the case of conventional inflationary models. This 
analysis complements the parametrization already outlined in Eq. (\ref{fourth}). 

\subsection{Covariant decompositions} 

To begin with let us recall the very well known covariant decomposition of $\nabla_{\beta} u_{\alpha}$ namely:
\begin{equation}
\nabla_{\beta} u_{\alpha}  = u^{\gamma}u_{\beta}\,\nabla_{\gamma} u_{\alpha} + \sigma_{\alpha\beta} + \omega_{\alpha\beta} + \frac{\Theta}{3} {\mathcal P}_{\alpha\beta},
\label{M1}
\end{equation}
where $\Theta = \nabla_{\alpha} u^{\alpha}$ and the projector is defined as ${\mathcal P}_{\alpha\beta} =( g_{\alpha\beta} - u_{\alpha} u_{\beta})$. The remaining terms in Eq. (\ref{M1}) are given by:
\begin{eqnarray}
\sigma_{\alpha\beta} &=& \frac{1}{2} (\nabla_{\beta} u_{\alpha} + \nabla_{\alpha} u_{\beta}) - \frac{1}{2} 
u^{\gamma} [(\nabla_{\gamma} u_{\alpha}) u_{\beta} + (\nabla_{\gamma} u_{\beta}) u_{\alpha}]   - \frac{\Theta}{3} {\mathcal P}_{\alpha\beta},
\label{M2}\\
\omega_{\alpha\beta} &=&  \frac{1}{2} (\nabla_{\beta} u_{\alpha} - \nabla_{\alpha} u_{\beta}) - \frac{1}{2} 
u^{\gamma} [(\nabla_{\gamma} u_{\alpha}) u_{\beta} - (\nabla_{\gamma} u_{\beta}) u_{\alpha}].
\label{M3}
\end{eqnarray}
In what follows, we shall assume 
that $\omega_{\alpha\beta} \to 0$. Thus the symmetric coupling tensors 
${\mathcal M}_{\rho\sigma}$ and ${\mathcal N}_{\rho\sigma}$  can be written as:
\begin{eqnarray}
&& {\mathcal M}_{\rho\sigma} = \lambda_{E} u_{\rho} u_{\sigma} + \frac{{\mathcal D}_{E}}{2 M} \biggl( \nabla_{\rho} u_{\sigma} + \nabla_{\sigma} u_{\rho} \biggr) + \frac{\overline{{\mathcal D}}_{E}}{2 M} u^{\gamma} \biggl[ (\nabla_{\gamma} u_{\rho}) u_{\sigma} + (\nabla_{\gamma} u_{\sigma}) u_{\rho}\biggr],
\label{M4}\\
&& {\mathcal N}_{\rho\sigma} = \lambda_{B} \overline{u}_{\rho} \overline{u}_{\sigma} + \frac{{\mathcal D}_{B}}{2 M} \biggl( \nabla_{\rho} \overline{u}_{\sigma} + \nabla_{\sigma} \overline{u}_{\rho}\biggl) + \frac{\overline{{\mathcal D}}_{B}}{2 M} \overline{u}^{\gamma} \biggl[ (\nabla_{\gamma} \overline{u}_{\rho}) \overline{u}_{\sigma} + (\nabla_{\gamma} \overline{u}_{\sigma}) \overline{u}_{\rho}\biggr],
\label{M5}
\end{eqnarray}
where $M$ denotes a generic mass scale; $u_{\rho}$ (and $\overline{u}_{\rho}$) obey $g^{\rho\sigma} u_{\rho} u_{\sigma} =1$ and $g^{\rho\sigma} \overline{u}_{\rho} \overline{u}_{\sigma} =1$.  Notice that we dropped the terms containing $g_{\rho\sigma}$ since it can be reabsorbed by a redefinition 
of $\lambda$ in Eq. (\ref{first}).

In the analysis of \cite{SUSC1} the ${\mathcal D}$-terms have been neglected, i.e. ${\mathcal D}_{E} = \overline{{\mathcal D}}_{E}=0$ and 
${\mathcal D}_{B} = \overline{{\mathcal D}}_{B}=0$. All the terms appearing in Eqs. (\ref{M4}) and (\ref{M5}) will now be considered on equal 
footing. From Eqs. (\ref{first}) and (\ref{second}) the explicit form of the evolution equations of the gauge fields is:
\begin{eqnarray}
&& \nabla_{\alpha} \biggl( \lambda \, Y^{\alpha\beta}\biggr) + \frac{1}{2} \nabla_{\alpha} {\mathcal Z}^{\alpha\beta}  - \frac{1}{2} \nabla_{\alpha} {\mathcal W}^{\alpha\beta} = 4 \pi j^{\beta},
\label{FF1}\\
&& \nabla_{\alpha} \tilde{Y}^{\alpha\beta} =0.
\label{FF2}
\end{eqnarray}
Defining $E^{\alpha\beta\rho\zeta}= \epsilon^{\alpha\beta\rho\zeta}/\sqrt{-g}$ (where $\epsilon^{\alpha\beta\rho\zeta}$ is the total antisymmetric pseudotensor of fourth rank) the two antisymmetric tensors ${\mathcal Z}^{\alpha\beta}$ and ${\mathcal W}^{\alpha\beta}$ are given by:
\begin{equation}
{\mathcal Z}^{\alpha\beta} = {\mathcal M}^{\alpha}_{\sigma} \,\,Y^{\sigma\beta} - {\mathcal M}^{\beta}_{\sigma} \,\, Y^{\sigma\alpha},\qquad 
{\mathcal W}^{\alpha\beta} = E^{\alpha\beta\rho\zeta} \,\, \tilde{Y}_{\sigma\zeta}\, \, {\mathcal N}^{\sigma}_{\rho}.
\label{WZ}
\end{equation}
We shall now focus on the case where the background metric is conformally flat:
\begin{equation}
g_{\mu\nu}(\tau) = a^{2}(\tau) \eta_{\mu\nu}, \qquad {\mathcal H} = \frac{a^{\prime}}{a},
\label{MM}
\end{equation}
where $\eta_{\mu\nu}$ is the Minkowski metric and the prime denotes a derivation with respect to the conformal time coordinate $\tau$.
Recalling that the standard Hubble rate is $H = {\mathcal H}/a$,
the explicit components of ${\mathcal Z}^{\alpha\beta}$ and ${\mathcal W}^{\alpha\beta}$ are\footnote{We recall that  
in the conformally flat metric of Eq. (\ref{MM})  the gauge field strength can be expressed in terms of the electric and magnetic fields as 
$Y^{i0} = e^{i}/a^2$ and $Y^{ij} = - \epsilon^{ijk} b_{k}/a^2$. The explicit components of ${\mathcal M}_{\rho}^{\sigma}$ and  ${\mathcal N}_{\rho}^{\sigma}$ 
can be explicitly obtained from Eqs. (\ref{M4}) and (\ref{M5}) and they are:
${\mathcal M}_{0}^{0} = \lambda_{E}$,  ${\mathcal N}_{0}^{0} = \lambda_{B}$, ${\mathcal M}_{i}^{j} = {\mathcal D}_{E} H \delta_{i}^{j}/M$ and ${\mathcal N}_{i}^{j} 
=  {\mathcal D}_{B}H\delta_{i}^{j}/M$.}:
\begin{eqnarray}
&&  {\mathcal Z}^{0i} = - {\mathcal Z}^{i0} = - \biggl( \lambda_{E} + \frac{{\mathcal D}_{E}}{M} H \biggr) \frac{e^{i}}{a^2},\qquad {\mathcal Z}^{ij} = - 2 H\frac{{\mathcal D}_{E}}{M} \frac{\epsilon^{i j k}}{a^2} b_{k},
\label{ZZ1}\\
&& {\mathcal W}^{0i} = -  {\mathcal W}^{i0}=  2 H \frac{{\mathcal D}_{B}}{M} \frac{e^{i}}{a^2},
\qquad {\mathcal W}^{ij} = \biggl( \lambda_{B} + \frac{{\mathcal D}_{B}}{M} H \biggr) \frac{\epsilon^{ijk} }{a^2}  b_{k}.
\label{WW1}
\end{eqnarray}

\subsection{Equations of motion and duality}
Thanks to Eqs. (\ref{ZZ1}) and (\ref{WW1}), Eqs. (\ref{FF1}) and (\ref{FF2})  imply the following form of the equations of motion:
\begin{eqnarray}
&& \vec{\nabla} \times \biggl( \sqrt{\Lambda_{B}} \vec{B} \biggr) = \partial_{\tau} \biggl( \sqrt{\Lambda_{E}} \vec{E} \biggr) + 4 \pi \vec{J},
\label{one}\\
&& \vec{\nabla} \times \biggl(\frac{\vec{E}}{\sqrt{\Lambda_{E}}}\biggr) + \partial_{\tau} \biggl(\frac{\vec{B}}{\sqrt{\Lambda_{B}}}\biggr) =0,
\label{two}\\
&& \vec{\nabla} \cdot \biggl(\frac{\vec{B}}{\sqrt{\Lambda_{B}}}\biggr)=0,\qquad \vec{\nabla}\cdot ( \sqrt{\Lambda_{E}}\, \vec{E} ) = 4 \pi \rho,
\label{three}
\end{eqnarray}
where we have introduced the following rescaled variables:
\begin{eqnarray}
\vec{B} &=& a^2 \, \sqrt{\Lambda_{B}}\, \vec{b}, \qquad \vec{E} = a^2 \, \sqrt{\Lambda_{E}}\, \vec{e},
\label{defEB}\\
\Lambda_{B} &=& \lambda + \frac{\lambda_{B}}{2} + \frac{{\mathcal D}_{B}}{2M} H +  \frac{{\mathcal D}_{E}}{M} H,
\label{LB}\\
\Lambda_{E} &=& \lambda + \frac{\lambda_{E}}{2} +  \frac{{\mathcal D}_{E}}{2 M} H+ \frac{{\mathcal D}_{B}}{M} H.
\label{LE}
\end{eqnarray}
Concerning Eqs. (\ref{defEB}), (\ref{LB}) and (\ref{LE}) three comments are in order:
\begin{itemize}
\item{} in the homogenous case the terms containing $\overline{\mathcal D}_{E}$ and $\overline{\mathcal D}_{B}$ do not contribute to the 
evolution of the gauge fields;
\item{} in the simultaneous limit ${\mathcal D}_{B}\to 0$ and ${\mathcal D}_{E} \to 0$ we have that $\Lambda_{B} = \lambda + \lambda_{B}/2$ 
and that $\Lambda_{E} = \lambda + \lambda_{E}/2$, as previously established;
\item{} if we transform $\lambda_{E} \to \lambda_{B}$,  ${\mathcal D}_{B}\to {\mathcal D}_{E}$ and ${\mathcal D}_{E} \to {\mathcal D}_{B}$ then $\Lambda_{E} \to \Lambda_{B}$ (and vice-versa).
\end{itemize}
The susceptibilities $\chi_{E}$, and $\chi_{B}$ are 
defined as $\chi_{E} = \sqrt{\Lambda_{E}}$ and $\chi_{B} = \sqrt{\Lambda_{B}}$. The corresponding 
gauge couplings are instead the inverse of the susceptibilities\footnote{Note that in the limit 
$\lambda_{B} = \lambda_{E} \to 0$ and ${\mathcal D}_{E} = {\mathcal D}_{B} \to 0$ $\chi_{E}$ and $\chi_{B}$ coincide and 
the same happens for the gauge couplings, i.e. $g_{E} = g_{B} = \sqrt{4\pi/\lambda}$.}
 \begin{equation}
 g_{E} = \sqrt{4\pi/\Lambda_{E}}, \qquad g_{B} = \sqrt{4\pi/\Lambda_{B}}.
 \label{DD1}
 \end{equation}
The mismatch between the two susceptibilities (i.e. $f = \chi_{E}^2/\chi_{B}^2 = \Lambda_{E}/\Lambda_{B}$) determines an effective refractive index affecting the evolution of the canonical fields. 

In Eqs. (\ref{one}) and (\ref{three}) $\vec{J}$ and $\rho$ denote, respectively, the rescaled current density and the 
rescaled charge density. In what follows we shall analyze the amplification of the gauge fields from vacuum fluctuations so we shall set the 
sources to zero. There are however different possible initial conditions like the ones stipulating that initially a 
globally neutral plasma is present, namely $\rho = 0$ with $\vec{J}$ given by the Ohmic current of the plasma. This possibility has
been investigated in the conventional case \cite{mgg} bust shall not be specifically analyzed here.

An important aspect that will be systematically used to simplify the discussion of the parameter space concerns the symmetries 
of the model. Neglecting the sources, Eqs. (\ref{one}), (\ref{two}) and (\ref{three}) are invariant under the generalized duality transformation \cite{SUSC1} that coincides with the conventional duality \cite{duality1} in the case of coincident electric and magnetic gauge couplings. 
When the gauge couplings are exchanged and inverted (i.e.  $g_{E} \to 1/g_{B}$ and $g_{B} \to 1/g_{E}$)
the generalized duality transformation stipulates that
Eqs. (\ref{one}), (\ref{two}) and (\ref{three}) keep exactly the same form provided  $\vec{E} \to - \vec{B}$ and  $\vec{B} \to \vec{E}$.

\subsection{Coupling functions and scalar fields}
We shall now examine form of the coupling functions  when the four-velocity is related to the covariant gradients of a scalar field (i.e. either the inflaton or some 
other spectator field); in this case from the general considerations discussed above we shall have 
\begin{eqnarray}
{\mathcal M}_{\rho\sigma}(\varphi) &=& {\mathcal B}_{1}(\varphi) \nabla_{\rho} \varphi \nabla_{\sigma} \varphi+ \frac{{\mathcal B}_{2}(\varphi)}{2} \biggl( \nabla_{\rho} \nabla_{\sigma} \varphi +  
 \nabla_{\sigma} \nabla_{\rho} \varphi\biggr)
 \nonumber\\
 &+& \frac{{\mathcal B}_{3}(\varphi)}{2} \nabla^{\gamma} \varphi\biggl( \nabla_{\rho} \varphi\,\, \nabla_{\gamma}\nabla_{\sigma}\varphi +  \nabla_{\sigma} \varphi\,\, \nabla_{\gamma}\nabla_{\rho} \varphi\biggr),
 \label{MN1}\\
{\mathcal N}_{\rho\sigma}(\psi) &=& {\mathcal C}_{1}(\psi) \nabla_{\rho} \psi \nabla_{\sigma} \psi+ \frac{{\mathcal C}_{2}(\psi)}{2} \biggl( \nabla_{\rho} \nabla_{\sigma} \psi +  
 \nabla_{\sigma} \nabla_{\rho} \psi\biggr)
 \nonumber\\
 &+& \frac{{\mathcal C}_{3}(\psi)}{2} \nabla^{\gamma} \psi \biggl( \nabla_{\rho} \psi\,\, \nabla_{\gamma}\nabla_{\sigma}\psi +  \nabla_{\sigma} \psi\,\, \nabla_{\gamma}\nabla_{\rho} \psi\biggr).
 \label{MN2}
 \end{eqnarray}
As discussed before ${\mathcal M}_{\rho\sigma}$ and ${\mathcal N}_{\rho\sigma}$ do not have to depend on the same field. For illustration  two different scalar fields (i.e $\varphi$ and $\psi$) have been used in Eqs. (\ref{MN1}) and (\ref{MN2}). The general form of Eqs. (\ref{MN1}) and (\ref{MN2}) suggests that it is always possible 
to select the arbitrary coupling functions in such a way that the electric and magnetic gauge couplings depend on the 
scale factor. During the inflationary phase their analytic dependence can always be written as 
\begin{eqnarray}
g_{E}(a) &=& \overline{g}_{E} \, \biggl(\frac{a}{a_{i}}\biggr)^{F_{E}}, \qquad g_{B}(a) = \overline{g}_{B} \biggl(\frac{a}{a_{i}}\biggr)^{F_{B}},
\label{par1}\\
f(a) &=&  f_{i}\biggl(\frac{a}{a_{i}}\biggr)^{F},\qquad f_{i} = \frac{\overline{g}_{B}^2}{\overline{g}_{E}^2} = \frac{\overline{\Lambda}_{B}}{\overline{\Lambda}_{E}},
\label{par2}
\end{eqnarray}
where $F_{E}$, $F_{B}$ and $F$ denote, respectively, the normalized rates of variation, i.e.\footnote{We recall that in this paper $\ln{}$ denotes the natural 
logarithm while $\log{}$ denotes the common logarithm.} 
\begin{equation}
F_{E} = \frac{\partial \ln{g_{E}}}{\partial \ln{a}}, \qquad F_{B} = \frac{\partial \ln{g_{B}}}{\partial \ln{a}}, \qquad F = \frac{\partial \ln{f}}{\partial \ln{a}},
\label{FAFB}
\end{equation}
and, by definition, $F= 2 (F_{B} - F_{E})$. In Eqs. (\ref{par1})--(\ref{par2})  $a_{i}$ denotes the value of the scale factor at the onset of the evolution of the gauge couplings; this moment will be taken to coincide, for simplicity, with the onset of the inflationary phase\footnote{If this identification is not made there will be a supplementary parameter fixing the origin of the evolution of the gauge couplings in comparison with the onset of the inflationary phase.}. 
 Recalling Eq. (\ref{DD1}),  the parametrization of Eqs. (\ref{par1}) and (\ref{par2}) implies that $\chi_{E}(a) \propto a^{-F_{E}}$ and $ \chi_{B}  \propto a^{-F_{B}}$. 

The parameter counting of  Eqs. (\ref{par1}) and (\ref{par2}) goes as follows. There are nominally four parameters in Eqs. (\ref{par1}) and (\ref{par2}), namely the rates $F_{E}$ and $F_{B}$ and the corresponding amplitudes, i.e. $\overline{g}_{E}$ and $\overline{g}_{B}$. These four parameters 
 can be identified with $F_{E}$, $F_{B}$, $f_{i}$ and with the initial value of one of the gauge couplings (be it either $\overline{g}_{B}$ or $\overline{g}_{E}$).
Neither $\overline{g}_{B}$ or $\overline{g}_{E}$ appear alone in the final electric and magnetic power spectra but only combined in $f_{i}$ which can be different from one and, in principle, also very large or very small depending on the specific model under consideration. On 
top of these three parameters there is also the total duration of the inflationary phase and, possibly, the overall duration of the pumping action of the gauge couplings. 
 
\subsection{Power-law backgrounds}
Before proceeding further, the parametrization of Eqs. (\ref{par1}) and (\ref{par2}) shall now be justified in the context 
of specific scenarios. This step is, in a sense, superfluous given the arbitrariness of the coupling functions. However it might 
be interesting to have this point treated in more depth. In what follows we shall assume the validity of the standard 
inflationary evolution and examine different specific models. 
The power law backgrounds are interesting insofar as they are exactly solvable. In this case it is immediately evident that 
the coupling functions must have an exponential or quasi-exponential form if we want to justify the validity of the parametrization of Eqs. 
(\ref{par1}) and (\ref{par2}). In  power-law backgrounds $\epsilon = - \dot{H}/H^2= 1/\alpha$ and 
\begin{equation}
\dot{\varphi}^2 = \frac{2 \alpha \overline{M}_{P}^2}{t^2}, \qquad \varphi = \varphi_{0} + \sqrt{\frac{2}{\epsilon}} \, \overline{M}_{P} \, \ln{(t/t_{i})},\qquad a(t) = (t/t_{i})^{\alpha}
 \label{PL1}
 \end{equation}
 where $t_{i}$ denotes the initial cosmic time of the inflationary expansion. 
 It can also happen that the susceptibilities do not depend on the inflaton but only on some spectator field 
that evolves during a quasi-de Sitter stage of expansion. In this case the evolution equation of $\psi$ can be solved even 
exactly like in the case of an exponential potential. If the field $\psi$ evolves in an exact power-law background its evolution 
will be given as 
\begin{equation}
\psi = \psi_{0} + {\mathcal K}_{1} \overline{M}_{P} \ln{(t/t_{i})} ,\qquad W(\psi) = W_{0} e^{- {\mathcal K}_{2} (\psi - \psi_{0})/\overline{M}_{P}},
\label{PL4}
\end{equation}
where ${\mathcal K}_{1} {\mathcal K}_{2} = 2$ and $W_{0} {\mathcal K}_{2} = H_{i}^2 \overline{M}_{P}^2 \epsilon (\epsilon -3)$. With 
this information the explicit couplings can be explicitly constructed and they will fall within the parametrization of Eqs. (\ref{par1}) and (\ref{par2}).
 
In the conventional case the coupling functions are exponentials \cite{DT1,DT2,DT3}. If we make the same choice here   
 $\lambda(\varphi) = Q_{0} \exp{[ q_{0} (\varphi - \varphi_{0})]}$ sets the overall normalization while  
the other couplings can be parametrized as:
\begin{equation}
{\mathcal B}_{1}(\varphi) =  Q_{1} \frac{e^{q_1 (\varphi - \varphi_{0})/\overline{M}_{P}}}{H_{i}^2 \overline{M}_{P}^2}, \qquad 
{\mathcal B}_{2}(\varphi) = Q_{2} \frac{e^{q_2 (\varphi  - \varphi_{0})/\overline{M}_{P}}}{ H_{i}^2 \overline{M}_{P}},\qquad 
{\mathcal B}_{3}(\varphi) = Q_{3} \frac{e^{q_3 (\varphi  - \varphi_{0})/\overline{M}_{P}}}{ H_{i}^4 \overline{M}_{P}^3}.
\label{AAA}
\end{equation}
Exactly the same parametrization can be envisaged for  ${\mathcal C}_{1}(\psi)$, ${\mathcal C}_{2}(\psi)$ and ${\mathcal C}_{3}(\psi)$ where 
we choose, for simplicity, $\psi=\varphi$:
\begin{equation}
{\mathcal C}_{1}(\varphi) =  P_{1} \frac{e^{p_1 (\varphi - \varphi_{0})/\overline{M}_{P}}}{H_{i}^2 \overline{M}_{P}^2}, \qquad 
{\mathcal C}_{2}(\varphi) = P_{2} \frac{e^{p_2 (\varphi  - \varphi_{0})/\overline{M}_{P}}}{ H_{i}^2 \overline{M}_{P}},\qquad 
{\mathcal C}_{3}(\varphi) = P_{3} \frac{e^{p_3 (\varphi  - \varphi_{0})/\overline{M}_{P}}}{ H_{i}^4 \overline{M}_{P}^3}.
\label{CCC}
\end{equation}
Equations (\ref{AAA}) and (\ref{CCC}) extend the class of models of Refs. \cite{DT1,DT2,DT3,DT4} to the case of derivative interactions obtained 
through the present construction.

The explicit form of the couplings of Eqs. (\ref{AAA}) and (\ref{CCC}) together with 
the background evolution of Eq. (\ref{PL1}) leads to the parametrization of  Eqs. (\ref{par1}) and (\ref{par2}).
While more complicated dynamical situations are certainly possible (and should be separately investigated) we shall now argue that the monotonic 
evolution of the gauge couplings is realized in various explicit scenarios. Let us assume, as above, $\psi=\varphi$ and also set ${\mathcal B}_{1}(\varphi) ={\mathcal B}_{2}(\varphi) = {\mathcal B}_{3}(\varphi) =0$ 
together with ${\mathcal C}_{3}(\varphi) =0$.  The explicit form of the susceptibilities, in the case of exponential couplings, will then be\footnote{The connection between cosmic and conformal time can always be written as $(t/t_{i}) = (- \tau/\tau_{*})^{1/(1-\alpha)}$ where $\tau_{*} = t_{i}/(\alpha -1)$.}:
 \begin{eqnarray}
 \Lambda_{E}(\tau) &=& Q_{0} \biggl(- \frac{\tau}{\tau_{*}} \biggr)^{ \sqrt{2 \epsilon} q_0/(\epsilon -1)} + P_{2} \sqrt{ 2 \epsilon}  \biggl(- \frac{\tau}{\tau_{*}} \biggr)^{\sqrt{2\epsilon} (p_{2} - \sqrt{2\epsilon})/(\epsilon -1)}
\label{PL2}\\
 \Lambda_{B}(\tau) &=&  Q_{0} \biggl(- \frac{\tau}{\tau_{*}} \biggr)^{ \sqrt{2 \epsilon} q_0/(\epsilon -1)}  + P_{1} \epsilon  \biggl(- \frac{\tau}{\tau_{*}} \biggr)^{\sqrt{2\epsilon} (p_{1} - \sqrt{2\epsilon})/(\epsilon -1)} 
 \nonumber\\
&+& \frac{P_{2}}{2} \sqrt{2\epsilon} (1 - \epsilon) \biggl(- \frac{\tau}{\tau_{*}} \biggr)^{\sqrt{2\epsilon} (p_{2} - \sqrt{2\epsilon})/(\epsilon -1)}.
\label{PL3}
\end{eqnarray}
Depending on the relative hierarchy of $q_{0}$, $p_{1}$ and $p_{2}$ the paranetrization of Eqs. (\ref{par1}) and (\ref{par2}) 
can be easily recovered  after some putative time related to $\tau_{*}$. This parametrization may also lead to more 
complicated situations where $f(\tau)$ may even have a non-monotonic behaviour. However it is sufficient to suppose 
that $p_{2} \gg q_{0}$ and $p_{1} \gg p_{2} \gg q_{0}$ to obtain 
\begin{equation}
\chi_{E}(a) = \overline{\chi}_{E} \biggl(\frac{a}{a_{*}}\biggr)^{- F_{E}}, \qquad \chi_{B}(a) = \overline{\chi}_{B} \biggl(\frac{a}{a_{*}}\biggr)^{- F_{B}},
\end{equation}
where $F_{E} = \sqrt{\epsilon/2} (p_{2} - \sqrt{2\epsilon})/(\epsilon -1)$ and  $F_{B} = \sqrt{\epsilon/2} (p_{1} - \sqrt{2\epsilon})/(\epsilon -1)$; moreover
$ \overline{\chi}^2_{E}  = \sqrt{2 \epsilon} P_{2}$ and  $\overline{\chi}^2_{B}  = \epsilon P_{1}$. The initial value of the susceptibilities and the 
relative strength of the couplings can be tuned by changing $P_{1}$ and $P_{2}$.

\subsection{Slow-roll models}

The previous discussion can be easily generalized to the case of generic slow-roll evolution. The simplest situation in this respect is given by the case 
where all the undetermined functions vanish except two of them. Let us therefore set ${\mathcal B}_{1}(\varphi) = {\mathcal B}_{2}(\varphi) = {\mathcal B}_{3}(\varphi) =0$ and also ${\mathcal C}_{3}(\varphi) = {\mathcal C}_{1}(\varphi) =0$. Given a certain inflationary potential we can always define 
\begin{equation}
{\mathcal I}(\varphi_{i}, \varphi) = \frac{1}{\overline{M}_{P}^2} \int_{\varphi_{i}}^{\varphi} \frac{W}{W_{\, , \varphi}} \, d\varphi.
\end{equation}
The coupling functions can then be fixed as:
\begin{eqnarray}
{\mathcal C}_{2}(\varphi) &=& \frac{6}{W_{\, , \varphi}} e^{ - 2 F_{E} {\mathcal I}(\varphi_{i}, \varphi)} \biggl[ e^{ -2 (F_{B} - F_{E}) {\mathcal I}(\varphi_{i}, \varphi)}-1\biggr],
\nonumber\\
\lambda(\varphi) &=& e^{ - 2 F_{E} {\mathcal I}(\varphi_{i}, \varphi)}\biggl[ 2 e^{- 2 (F_{B} -F_{E}) {\mathcal I}(\varphi_{i}, \varphi)}- 1\biggr].
\label{st1}
\end{eqnarray}
Another possibility is to set ${\mathcal C}_{2}(\varphi)= {\mathcal C}_{3}(\varphi) =0$ and, as before ${\mathcal B}_{1}(\varphi) = {\mathcal B}_{2}(\varphi) = {\mathcal B}_{3}(\varphi) =0$. The values of $\lambda(\varphi)$ and ${\mathcal C}_{1}(\varphi)$ are given, in this case, by:
\begin{equation}
\lambda(\varphi) = e^{- 2 F_{E} {\mathcal I}(\varphi_{i}, \varphi)}, \qquad {\mathcal C}_{1}(\varphi) = \frac{3}{W \, \epsilon} e^{ - 2 F_{E}{\mathcal I}(\varphi_{i}, \varphi)} \biggl[ e^{ -2 (F_{B} - F_{E}) {\mathcal I}(\varphi_{i}, \varphi)} -1 \biggr].
\label{st2}
\end{equation}
Equations (\ref{st1}) and (\ref{st2}) imply that the susceptibilities can be written, in a unified notation, as:
\begin{eqnarray}
\chi_{B}^2 &=& \lambda + \frac{{\mathcal C}_{1}(\varphi)}{2} \dot{\varphi}^2 - {\mathcal C}_{2}(\varphi) \biggl[ H \dot{\varphi} + \frac{W_{\,,\varphi}}{2} \biggr]= \overline{\chi}_{B}^2 \biggl(\frac{a}{a_{i}}\biggr)^{- 2 F_{B}},
\nonumber\\
\chi_{E}^2 &=& \lambda + {\mathcal C}_{2}(\varphi) H \dot{\varphi}= \overline{\chi}_{E}^2 \biggl(\frac{a}{a_{i}}\biggr)^{- 2 F_{E}},
\label{st3}
\end{eqnarray}
where we recall that in the slow-roll approximation $(a/a_{i}) = \exp{[- {\mathcal I}(\varphi_{i}, \varphi)]}$; furthermore the slow-roll relations have been used where appropriate. The same analysis leading to Eq. (\ref{st3}) can be repeated when all the ${\mathcal C}_{i}(\varphi)$ are equal to zero while, 
on top of $\lambda(\varphi)$, the only two non-vanishing coupling functions are either ${\mathcal B}_{1}(\varphi)$ or ${\mathcal B}_{2}(\varphi)$.

It is obvious that depending on the model the relation between $(a/a_{i})$ and $\varphi$ will be a bit different but still the general parametrization of Eqs. (\ref{par1}) and (\ref{par2}) will remain valid. 
In the case of simple monomial potentials we have $W = M^4 \phi^{p}$ and therefore we shall have that 
\begin{equation}
\biggl(\frac{a}{a_{i}}\biggr) =\exp{\biggl[ - \frac{\varphi^2 - \varphi_{i}^2}{2 p \overline{M}_{P}^2} \biggr]}.
\end{equation}
Other typical models can be analyzed  like  the case of small field 
and hybrid models where the potential can be written as $W(\varphi) = M^4 ( 1 \pm \kappa \varphi^{p})$ (where the plus corresponds 
to the hybrid models while the minus to the small field models). In the case of $R^2$ models the potential, in the Einstein frame, is given by 
\begin{equation}
W(\varphi) = \frac{3 M^2 \overline{M}_{P}^2}{4} \bigg( 1 - e^{- \sqrt{2/3} \varphi/\overline{M}_{P}}\biggr)^2,
\end{equation}
where we defined, for practical reasons, $\Phi= \sqrt{2/3} \varphi/\overline{M}_{P}$. Using the evolution equations 
we obtain:
\begin{equation}
\Phi = \ln{[ 2 M (t_{*} - t)/3]}, \qquad a = a_{*} \exp{3 [ \Phi -  e^{\Phi}]/4}.
\end{equation} 
In cosmic time the scale factor reads 
\begin{equation}
a(t) = [ M (t_{*} - t)]^{3/4} \, e^{ - M(t_{*} - t)/2}.
\end{equation}
where $t_{*}$ denotes in practice the final time of the inflationary expansion.  All the specific analyses 
discussed before can be easily applied also to this case with just one caveat. To obtain 
the specific equations relating the cosmic and the conformal time parametrization 
we must expand the above equations in the limit $t \ll t_{*}$. The result of this 
simple expansion can be written as
\begin{equation}
a \simeq a_{*} \, x_{*}^{3/4} \, e^{- x_{*}/2} e^{[2 x - 3 (x/x_{*})]/4},
\end{equation}
where $x= M t$ and $x_{*} = M t_{*}$. Thanks to the previous expression 
we can easily deduce the relation of $t$ to $\tau$ by recalling that 
$a(\tau) d\tau = dt$. In summary we have demonstrated that while the most general parametrization of the coupling functions 
may lead to non-monotonic susceptibilities, after some initial transient time the monotonic behaviour 
can be recovered. The parametrization of Eqs. (\ref{par1}) and (\ref{par2}) is therefore fully 
justified and it encompasses a variety of models. 

\renewcommand{\theequation}{3.\arabic{equation}}
\setcounter{equation}{0}
\section{Quantization and power spectra}
\label{sec3}

The problem is greatly simplified by abandoning the conformal time and by rescaling the canonical variables. This transformation leaves the action invariant and simplifies the evolution of the mode functions.

\subsection{Time reparametrization of the action and quantization}
The main idea is to use a new time parametrization in the action and a new set of rescaled vector potentials. 
In time-dependent (conformally flat) backgrounds and in the Coulomb gauge (i.e. $Y_{0} =0$  and $\vec{\nabla}\cdot \vec{Y} =0$)  that is preserved (unlike the Lorentz gauge condition)  under a conformal rescaling of the metric the action is 
\begin{equation}
S = \frac{1}{2}\int \,d\tau\, d^{3} x \, \biggl\{ \vec{A}^{\,\prime \,2} + \biggl(\frac{\chi_{E}^{\,\prime}}{\chi_{E}}\biggr)^2 
 \vec{A}^{\,2}  - 2 \frac{\chi_{E}'}{\chi_{E}} \vec{A} \cdot \vec{A}^{\,\prime} - \frac{\chi_{B}^2}{\chi_{E}^2}\partial_{i} \vec{A} \cdot \partial^{i} \vec{A}\biggr\},
\label{actold}
\end{equation}
where\footnote{The $1/\sqrt{4\pi}$ is purely conventional 
and its presence comes from the factor $16\pi$ included in the initial gauge action. } $\vec{A} = \sqrt{ \Lambda_{E}/(4\pi)} \vec{Y}$. We have assumed that $\chi_{E}$ and $\chi_{B}$ are only dependent on the conformal time coordinate $\tau$. In terms of the canonical momentum conjugate to $\vec{A}$ 
the canonical Hamiltonian is simply given by \cite{SUSC1}:
\begin{equation}
H_{A}(\tau) = \frac{1}{2} \int d^3 x \biggl[ \vec{\pi}^{2} + 2 \frac{\chi_{E}'}{\chi_{E}} \vec{\pi} \cdot \vec{A} + 
 \frac{\partial_{i} \vec{A} \cdot \partial^{i} \vec{A}}{f}\biggr], \qquad \vec{\pi} = \vec{A}^{\,\prime} - \frac{\chi_{E}'}{\chi_{E}} \vec{A}.
\label{hamold}
\end{equation}
We can now change the time parametrization in the action of Eq. (\ref{actold}) 
\begin{equation}
\tau\to \eta = \eta(\tau), \qquad  d\tau= \sqrt{f(\eta)} d\eta,
\label{transtime}
\end{equation}
and simultaneously redefine the vector potential:
\begin{equation}
\vec{A}(\tau, \vec{x})  \to {\mathcal A}(\eta,\vec{x})  = \frac{\vec{A}[\tau(\eta), \vec{x}]}{ f^{1/4}[\tau(\eta)]}. 
\label{transvector}
\end{equation}
Ultimately $\vec{{\mathcal A}}$ can be directly expressed in terms of $\vec{Y}$ as:
\begin{equation}
\vec{{\mathcal A}}(\eta, \vec{x})= \sqrt{\frac{\Lambda_{E}(\eta)}{4\pi \sqrt{f(\eta)} }} \vec{Y}(\eta, \vec{x}) = 
\frac{\sqrt[4]{\Lambda_{E}(\eta) \, \Lambda_{B}(\eta)}}{\sqrt{4 \pi} } \vec{Y}(\eta, \vec{x})= \frac{\vec{Y}(\eta, \vec{x})}{\sqrt{g_{E}(\eta) g_{B}(\eta)}},
\end{equation}
where the second and third equalities follow, respectively, from the mutual relations among the different susceptibilities 
and from the relation between each susceptibility and the corresponding gauge coupling. 
Using Eqs. (\ref{transtime}) and (\ref{transvector}) the action (\ref{actold}) becomes:
\begin{equation}
S =\frac{1}{2}\int  \frac{1}{2}\int \,d\eta \, d^{3} x \biggl\{ \partial_{\eta} \vec{{\mathcal A}}\cdot \partial_{\eta} \vec{{\mathcal A}}+ 
\biggl[\frac{(\sqrt{\chi_{E} \chi_{B}})^{\bullet}}{\sqrt{\chi_{E} \chi_{B}}}\biggr]^2 \vec{{\mathcal A}}^{\,2}  - 2 \frac{(\sqrt{\chi_{E} \chi_{B}})^{\bullet}}{\sqrt{\chi_{E} \chi_{B}}} \partial_{\eta} \vec{{\mathcal A}} \cdot \vec{{\mathcal A}} -\partial_{i} \vec{{\mathcal A}} \cdot \partial^{i} \vec{{\mathcal A}}\biggr\},
\label{actnew}
\end{equation}
where the thick overdot (or bullet) denotes a derivation with respect to $\eta$. From the action (\ref{actnew}) 
the new canonical Hamiltonian and the corresponding canonical momenta are now given by:
\begin{eqnarray}
H_{{\mathcal A}}(\eta) &=& \frac{1}{2} \int d^3 x \biggl[ \vec{\Pi}^{2} + 2   \frac{(\sqrt{\chi_{E} \chi_{B}})^{\bullet}}{\sqrt{\chi_{E} \chi_{B}}}\vec{\Pi} \cdot \vec{{\mathcal A}} + 
 \partial_{i} \vec{{\mathcal A}} \cdot \partial^{i} \vec{{\mathcal A}}\biggr],
 \label{hamnew}\\
 \vec{\Pi}(\eta,\vec{x}) &=& \partial_{\eta} \vec{{\mathcal A}} - \frac{(\sqrt{\chi_{E} \chi_{B}})^{\bullet}}{\sqrt{\chi_{E} \chi_{B}}}\vec{{\mathcal A}}.
\label{cannew}
\end{eqnarray}
The Fourier mode expansion for the canonical fields:
\begin{equation}
\vec{\Pi}(\vec{x},\eta) = \frac{1}{(2\pi)^{3/2}} \int d^{3} k\,\, \vec{\Pi}_{\vec{k}}(\eta) \,\,e^{-i \vec{k}\cdot\vec{x}}, \qquad 
 \vec{A}(\vec{x},\eta) = \frac{1}{(2\pi)^{3/2}} \int d^{3} k \,\, \vec{{\mathcal A}}_{\vec{k}}(\eta) \,\,e^{-i \vec{k}\cdot\vec{x}}
 \label{dual9b}
 \end{equation}
implies that the canonical Hamiltonian (\ref{hamnew}) can be rewritten as:
\begin{equation}
H_{A}(\eta) = \frac{1}{2} \int d^3 k \biggl[ \vec{\Pi}_{\vec{k}} \cdot \vec{\Pi}_{-\vec{k}} +   \frac{(\sqrt{\chi_{E} \chi_{B}})^{\bullet}}{\sqrt{\chi_{E} \chi_{B}}}
\biggl( \vec{\Pi}_{\vec{k}} \cdot \vec{{\mathcal A}}_{-\vec{k}}  +  \vec{\Pi}_{-\vec{k}} \cdot \vec{{\mathcal A}}_{\vec{k}}\biggr)
+ k^2 \,\vec{{\mathcal A}}_{\vec{k}} \cdot \vec{{\mathcal A}}_{-\vec{k}}\biggr].
\label{hamnew2}
\end{equation}
From Eq. (\ref{hamnew2}) the corresponding equations or motion are:
\begin{eqnarray}
&& \partial_{\eta} \vec{{\mathcal A}}_{\vec{k}}= \vec{\Pi}_{\vec{k}} + \sqrt{g_{B} g_{E}} \biggl(\frac{1}{\sqrt{g_{B} g_{E}}}\biggr)^{\bullet} \vec{{\mathcal A}}_{\vec{k}},
\label{neweq1}\\
&& \partial_{\eta} \vec{\Pi}_{\vec{k}} = - k^2\,\vec{{\mathcal A}}_{\vec{k}} - \sqrt{g_{B} g_{E}} \biggl(\frac{1}{\sqrt{g_{B} g_{E}}}\biggr)^{\bullet} \vec{\Pi}_{\vec{k}},
\label{neweq2}
\end{eqnarray}
where we have used the explicit relations connecting the susceptibilities to the gauge couplings. 
The duality transformation exchanges the canonical fields and the conjugate momenta so that Eqs. (\ref{neweq1}) and (\ref{neweq2}) 
go one into the other by virtue of the following transformation:
\begin{eqnarray}
 g_{E}(\eta) \to \frac{1}{g_{B}(\eta)}, \qquad g_{B}(\eta) \to \frac{1}{g_{E}(\eta)}, \qquad
 \vec{\Pi}_{\vec{k}}(\eta) \to - k \, \vec{{\mathcal A}}_{\vec{k}}(\eta), \qquad \vec{{\mathcal A}}_{\vec{k}}(\eta) \to  \frac{\vec{\Pi}_{\vec{k}}(\eta)}{k}.
\label{newtrans1}
\end{eqnarray}

Promoting the canonical fields to quantum operators (i.e. ${\mathcal A}_{i} \to \hat{{\mathcal A}}_{i}$ and $\pi_{i} \to \hat{\pi}_{i}$) the following
(equal time) commutation relations (in units $\hbar = c =1$) must hold:
\begin{equation}
[\hat{{\mathcal A}}_{i}(\vec{x}_{1},\eta),\hat{\Pi}_{j}(\vec{x}_{2},\eta)] = i \Delta_{ij}(\vec{x}_{1} - \vec{x}_{2}),\qquad 
\Delta_{ij}(\vec{x}_{1} - \vec{x}_{2}) = \int \frac{d^{3}k}{(2\pi)^3} e^{i \vec{k} \cdot (\vec{x}_{1} - \vec{x}_2)} P_{ij}(k), 
\label{dual17a}
\end{equation}
where $P_{ij}(k) = (\delta_{ij} - k_{i} k_{j}/k^2)$. The function $\Delta_{ij}(\vec{x}_{1} - \vec{x}_{2})$ is the transverse generalization of the Dirac delta function
ensuring that both $\vec{E}$ and $\vec{A}$ are divergenceless. The field operators can then be expanded in terms of the corresponding 
mode functions 
\begin{eqnarray}
\hat{{\mathcal A}}_{i}(\eta,\vec{x}) = \int\frac{d^{3} k}{(2\pi)^{3/2}} \sum_{\alpha} e^{(\alpha)}_{i}(k) \, 
\biggl[ \overline{F}_{k}(\eta) \, \hat{a}_{k,\alpha} e^{- i \vec{k} \cdot\vec{x}} +  \overline{F}_{k}^{*}(\eta) \, \hat{a}^{\dagger}_{k,\alpha} e^{ i \vec{k} \cdot\vec{x}}\biggr],
\label{expnew1}\\
\hat{\Pi}_{i}(\eta, \vec{x}) = \int\frac{d^{3} k}{(2\pi)^{3/2}} \sum_{\alpha} e^{(\alpha)}_{i}(k) \, 
\biggl[ \overline{G}_{k}(\eta) \, \hat{a}_{k,\alpha} e^{- i \vec{k} \cdot\vec{x}} +  \overline{G}_{k}^{*}(\eta) \, \hat{a}^{\dagger}_{k,\alpha} e^{ i \vec{k} \cdot\vec{x}}\biggr].
\label{expnew2}
\end{eqnarray}
where $\overline{F}_{k}(\eta)$ and $\overline{G}_{k}(\eta)$ are the mode functions in the $\eta$-parametrization. 

\subsection{General solutions for the mode functions}
As a consequence of the indetermination relations, Eqs. (\ref{expnew1}) and (\ref{expnew2}) imply that the mode functions must 
 obey the equations derived previously and the Wronskian normalization condition:
\begin{equation}
\overline{F}_{k}(\eta)\,\overline{G}_{k}^{*}(\eta) - \overline{F}_{k}^{*}(\eta)\,\overline{G}_{k}(\eta) =i.
\label{wron}
\end{equation}
From Eqs.(\ref{neweq1}) and (\ref{neweq2}), $\overline{F}_{k}(\eta)$ and $\overline{G}_{k}(\eta)$ can be shown to satisfy the 
following pair of equations:
\begin{eqnarray}
&& \frac{d \overline{F}_{k}}{d\eta} = \overline{G}_{k} +\sqrt{ g_{B} g_{E}} \biggl(\frac{1}{\sqrt{g_{B} g_{E}}}\biggr)^{\bullet} \overline{F}_{k}, 
\label{bar1}\\
&& \frac{d \overline{G}_{k}}{d\eta} = - k^2\overline{F}_{k} - \sqrt{g_{B} g_{E}} \biggl(\frac{1}{\sqrt{g_{B} g_{E}}}\biggr)^{\bullet}\overline{G}_{k}.
\label{bar2}
\end{eqnarray}
Combining Eqs. (\ref{bar1}) and (\ref{bar2}) we obtain two decoupled equations:
\begin{eqnarray}
&& \frac{d^{2} \overline{F}_{k}}{d\eta^2} + \biggl[k^2 -  \sqrt{g_{B} g_{E} }\biggl(\frac{1}{\sqrt{g_{B} g_{E}}}\biggl)^{\bullet\bullet} \biggr]  \overline{F}_{k}=0,
\label{bardec1}\\
&& \frac{d^{2} \overline{G}_{k}}{d\eta^2} + \biggl[k^2 -  \frac{(\sqrt{g_{B} g_{E}})^{\bullet\bullet}}{\sqrt{g_{B} g_{E}}}\biggr]\overline{G}_{k}=0.
\label{bardec2}
\end{eqnarray}
Equations (\ref{bardec1}) and (\ref{bardec2}) can also be written in terms of the susceptibilities 
$\chi_{E}$ and $\chi_{B}$ by simply taking into account that, as already mentioned, $g_{E} = \sqrt{4 \pi/\chi_{E}}$ and 
$g_{B} = \sqrt{4\pi/\chi_{B}}$. Equations (\ref{bar1})--(\ref{bar2}) and (\ref{bardec1})--(\ref{bardec2}) can be directly derived from 
the mode functions obtainable from the quantization in conformal time. Indeed, using the recipe 
\begin{equation}
\frac{d\tau}{\sqrt{f[\tau(\eta)]}} = d\eta, \qquad F_{k}(\tau)= f^{1/4}[\eta(\tau)] \overline{F}_{k}[\eta(\tau)], \qquad G_{k}[\eta(\tau)]= f^{-1/4}[\eta(\tau)] \overline{G}_{k}[\eta(\tau)],
\end{equation}
Eqs. (\ref{bar1})--(\ref{bar2}) become immediately 
\begin{equation}
F_{k}' = G_{k} + \frac{\chi_{E}'}{\chi_{E}} F_{k},\qquad  G_{k}' = -  \frac{k^2}{f} F_{k} - \frac{\chi_{E}'}{\chi_{E}} G_{k},
\label{oldeq}
\end{equation}
where the prime denotes a derivation with respect to $\tau$; Eqs. (\ref{oldeq}) coincide with the ones derived in \cite{SUSC1}. 

The general solutions of Eqs. (\ref{bardec1}) and (\ref{bardec2}) can be obtained either exactly (i.e. by solving 
the evolution in exact terms) or approximately (i.e. in different asymptotic limits). 
In a large class of physical situations the pump fields of Eqs. (\ref{bardec1}) and (\ref{bardec2}) are expected to decay 
for $|\eta| \to \infty$ faster than $1/\eta$ (see however at the end of this subsection for a different situation). 
The solution of Eq. (\ref{bardec1}) can then be obtained as:
\begin{eqnarray}
\overline{F}_{k}(\eta) &=& \frac{1}{\sqrt{2 k}} e^{\pm i k\eta}, \qquad k^2 \gg \bigg| \sqrt{g_{E} g_{B}} \biggl( \frac{1}{\sqrt{g_{B} g_{E}}}\biggr)^{\bullet\bullet}\bigg|,
\label{newsol1}\\
\overline{F}_{k}(\eta) &=&  \frac{D^{(1)}_{k}}{ \sqrt{g_{E} \, g_{B}}} + \frac{D^{(2)}_{k}}{ \sqrt{g_{E} \, g_{B}}} \int^{\eta} d\eta_{1} 
g_{E}(\eta_1) g_{B}(\eta_{1})
, \qquad k^2 \ll \bigg| \sqrt{g_{E} g_{B}} \biggl( \frac{1}{\sqrt{g_{B} g_{E}}}\biggr)^{\bullet\bullet}\bigg|.
\label{newsol2}
\end{eqnarray}
where $D^{(1)}_{k}$ and $D^{(2)}_{k}$ are the two arbitrary constants that must be fixed from the boundary conditions.
Recalling Eq. (\ref{bar1}), Eqs. (\ref{newsol1})--(\ref{newsol2}) also determine the explicit form of $\overline{G}_{k}(\eta)$. 
Alternatively Eq. (\ref{bardec2}) can be directly solved with the same strategy leading to Eqs. (\ref{newsol1}) and (\ref{newsol2}).  

\subsection{Exact solutions with asymmetric gauge couplings}
According to Eqs. (\ref{bardec1})--(\ref{bardec2}) and using the generalized parametrization of Eqs. (\ref{par1}) and (\ref{par2}), the evolution of the mode functions obeys:
\begin{equation}
 \frac{d^{2} \overline{F}_{k}}{d\eta^2} + \biggl[k^2 - \frac{\sigma^2-1/4}{\eta^2} \biggr]  \overline{F}_{k}=0,\qquad  \frac{d^{2} \overline{G}_{k}}{d\eta^2} + \biggl[k^2 -  \frac{ (\sigma-1)^2 -1/4}{\eta^2}\biggr]\overline{G}_{k}=0,
\end{equation}
whose solution is is given by:
\begin{eqnarray}
&& \overline{F}_{k}(\eta) = \frac{{\mathcal N}}{\sqrt{ 2 k}} \, \sqrt{- k \eta} \, H_{\sigma}^{(1)}(- k \eta),\qquad \sigma = \frac{1 - 2 F_{E}}{2( 1 + F_{B} - F_{E})}
\label{sol1}\\
&& \overline{G}_{k}(\eta) = - {\mathcal N}\, \sqrt{\frac{k}{2}}\, \sqrt{- k \eta} \, H_{\sigma-1}^{(1)}(- k \eta),
\label{sol2}
\end{eqnarray}
where $|{\mathcal N}|= \sqrt{\pi/2}$. The case $F_{E} = F_{B} + 1$ is a bit special since we have that $f = \overline{f} (\tau/ \tau_{i})^2$. But in this situation the susceptibilities can be parametrized, in general terms, as\footnote{In this specific case the relation between 
$\tau$ and $\eta$ is exponential since $\eta$ is the logarithm of $\tau$, i.e. $\ln{(-\tau/\tau_{i})} = - \sqrt{\overline{f}} (\eta/\tau_{i})$ and $\eta_{i}= \tau_{i}/\sqrt{\overline{f}}$.}:
\begin{equation}
\chi_{E}(\eta)= \overline{\chi}_{E} e^{ \alpha_{E} (\eta + \eta_{i})/\eta_{*}}, \qquad \chi_{B}(\eta)= \overline{\chi}_{B} e^{ \alpha_{B} (\eta + \eta_{i})/\eta_{*}},\qquad f(\eta) = \overline{f} e^{ 2 (\alpha_{E} - \alpha_{B})(\eta + \eta_{i})/\eta_{*}}.
\label{sing1}
\end{equation}
The evolution equation for $\overline{F}_{k}(\eta)$ can be written as:
\begin{equation}
 \frac{d^{2} \overline{F}_{k}}{d\eta^2} + \biggl[k^2 - \frac{(\alpha_{E} +\alpha_{B})^2}{4\eta_{*}^2} \biggr]  \overline{F}_{k}=0,
\label{sing2}
 \end{equation}
 and the solution of the previous equation is: 
 \begin{equation}
 \overline{F}_{k}(\eta) = \frac{1}{\sqrt{ 2 \Omega_{k}}} e^{ - i \Omega_{k} (\eta + \eta_{i})}, \qquad \Omega_{k} = \sqrt{ k^2 - \frac{(\alpha_{E} + \alpha_{B})}{4 \eta_{*}^2}}.
\label{sol1a}
 \end{equation}
 If $k \eta_{*} \gg 1$ then $\Omega_{k} \simeq k$ while in the opposite limit $\Omega_{k} \simeq i |\alpha_{E} + \alpha_{B}|/(2\eta_{*})$. Recalling the 
 expression of $\overline{G}_{k}$ we have that 
 \begin{equation}
 \overline{G}_{k} = - i \sqrt{\frac{\Omega_{k}}{2}} \biggl[ 1 - \frac{i (\alpha_{E} + \alpha_{B})}{2 \Omega\eta_{*}} \biggr] e^{ - i \Omega_{k} (\eta + \eta_{*})}. 
 \label{sol1b}
 \end{equation}
 Note that in the limit $ k \eta_{*} \ll 1$ we have that $\overline{G}_{k}(\eta) \to 0$ and $P_{E}(k,\eta) \to 0$ provided $\alpha_{E}$ and $\alpha_{B}$ are both positive semidefinite. Conversely, in the same limit, 
 \begin{equation}
 |\overline{F}_{k}(\eta)|^2 = \frac{4 \eta_{*}}{|\alpha_{E} + \alpha_{B}|} e^{ |\alpha_{E} +\alpha_{B}|(\eta + \eta_{i})/\eta_{*}}.
 \end{equation}
 
\subsection{Correlation functions}

In the $\eta$-parametrization the correlation functions of the electric and of the magnetic fields can be swiftly computed. It is 
sufficient to recall that the canonical fields can be directly expressed in terms 
of the comoving electric and magnetic fields, i.e.  $\vec{E}$ and $\vec{B}$:
\begin{equation}
\vec{E}(\eta,\vec{x}) = - \frac{\vec{\Pi}(\eta,\vec{x})}{\sqrt[4]{f(\eta)}}, \qquad \vec{B}(\eta,\vec{x}) = \vec{\nabla}\times \left[\frac{ \vec{{\mathcal A}}(\eta,\vec{x})}{\sqrt[4]{f(\eta)}} \right].
\label{rel1}
\end{equation}
In terms of the mode decomposition the comoving electric and magnetic fields are 
\begin{eqnarray}
&& \hat{B}_{i}(\eta, \vec{x}) = - \frac{i\, \epsilon_{m n i}}{(2\pi)^{3/2}\, \sqrt[4]{f(\eta)}}  \sum_{\alpha} \int d^{3} k k_{m} e^{(\alpha)}_{n} 
\biggl[ \overline{F}_{k}(\eta)\, \hat{a}_{\vec{k}, \alpha} e^{- i \vec{k} \cdot \vec{x}}  - \overline{F}_{k}^{*}(\eta) \hat{a}^{\dagger}_{\vec{k}, \alpha}
e^{ i \vec{k} \cdot \vec{x}}\biggr], 
\label{rel2}\\
&& \hat{E}_{i}(\eta,\vec{x}) = - \frac{1}{(2\pi)^{3/2}\, \sqrt[4]{f(\eta)}}  \sum_{\alpha} \int d^{3} k \,e^{(\alpha)}_{i} 
\biggl[ \overline{G}_{k}(\eta)  \hat{a}_{\vec{k}, \alpha} e^{- i \vec{k} \cdot \vec{x}}  + \overline{G}_{k}^{*}(\eta) \hat{a}^{\dagger}_{\vec{k}, \alpha}e^{ i \vec{k} \cdot \vec{x}} \biggr].
\label{rel3}
\end{eqnarray}
In terms of the mode functions, the Fourier components of the field operators $\hat{B}_{i}(\vec{x},\tau)$ and $\hat{E}_{i}(\vec{x},\tau)$ are respectively: 
\begin{eqnarray}
&& \hat{B}_{i}(\eta,\,\vec{q}) = - \frac{i}{\sqrt[4]{f(\eta)}}\, \epsilon_{m n i}\sum_{\alpha}  \, e^{(\alpha)}_{n} \, q_{m} \bigl[ \hat{a}_{\vec{q},\alpha} \,\,\overline{F}_{q}(\eta) 
+ \hat{a}^{\dagger}_{-\vec{q},\alpha} \,\,\overline{F}^{*}_{q}(\eta)\bigr], 
\label{rel4}\\
&& \hat{E}_{i}(\eta,\,\vec{q}) = \frac{1}{\sqrt[4]{f(\eta)}}\, \sum_{\alpha} e^{(\alpha)}_{i} \bigl[ \hat{a}_{\vec{q},\beta} \,\,\overline{G}_{q}(\eta) 
+ \hat{a}^{\dagger}_{-\vec{q},\beta} \,\,\overline{G}^{*}_{q}(\eta)\bigr].
\label{rel5}
\end{eqnarray}
The correlators in Fourier space are then given by:
\begin{eqnarray}
&& \langle B_{i}(\eta,\vec{k})\, B_{j}(\eta,\vec{p}) \rangle = \frac{2\pi^2}{k^3}\, P_{B}(k,\eta)\, P_{ij}(k)  \,\delta^{(3)}(\vec{k} + \vec{p}),
\label{cc1}\\
&& \langle E_{i}(\eta,\vec{k})\, E_{j}(\eta,\vec{p}) \rangle = \frac{2\pi^2}{k^3}\, P_{E}(k,\eta)\, P_{ij}(k)  \,\delta^{(3)}(\vec{k} + \vec{p}),
\label{cc2}
\end{eqnarray}
where $P_{ij}(k)=(\delta_{ij} - k_{i} k_{j}/k^2)$ and where the power spectra are: 
\begin{eqnarray}
P_{B}(k,\eta) &=& \frac{k^{5}}{2\, \pi^2\, a^4(\eta) \, \sqrt{f(\eta)}} \, |\overline{F}_{k}(\eta)|^2, 
\label{PBc}\\
P_{E}(k,\eta) &=& \frac{k^3}{2\, \pi^2\, a^4(\eta) \sqrt{f(\eta)}} \,  |\overline{G}_{k}(\eta)|^2.
\label{PEd}
\end{eqnarray}
Note that, as a consequence of the duality symmetry, the magnetic and the electric power spectra are 
interchanged when $g_{E} \to 1/g_{B}$ and $g_{B}\to 1/g_{E}$. In this case 
$\overline{G}_{k} \to - k \overline{F}_{k}$ and $\overline{F}_{k} \to \overline{G}_{k}/k$ so that 
$P_{B}(k,\eta) \to P_{E}(k,\eta)$ and vice-versa. It is very important to remark, for the following discussion, that 
the definition of power spectrum of Eqs. (\ref{cc1}) and (\ref{cc2}) is particularly suitable since $P_{B}(k,\eta) $ and 
$P_{E}(k,\eta)$ have both  dimensions of an energy density. Consequently we shall measure the amplitude 
of the power spectrum either in critical units of in $\mathrm{nG}^2$. We stress that this occurrence 
is not only mathematically useful but also physically sound: if the power spectrum is defined according to the 
present strategy it is evident, as it has to be, that the magnetic power spectrum is also (up to an irrelevant 
numerical factor) the magnetic energy density per logarithmic interval of wavenumbers (or logarithmic interval of frequency).

\renewcommand{\theequation}{4.\arabic{equation}}
\setcounter{equation}{0}
\section{Power spectra and initial data}
\label{sec4}
\subsection{General form of the power spectra}
The power spectra depend on the initial data 
of the electric and magnetic susceptibilities (or, which is the same, of the electric and magnetic gauge 
couplings).  Two complementary possibilities can be envisaged. The first possibility stipulates that $f(a_{i}) =f_{i} = 1$, i.e. $\chi_{E}(a_{i}) = \chi_{B}(a_{i})$: the susceptibilities are initially equal and then they diverge so that, 
depending on the number of efolds, they will still be different at the end of inflation.
The second possibility is the one where  $f(a_{f}) =f_{f} =1$; in this second case $\chi_{E}(a_{f}) = \chi_{B}(a_{f})$, i.e. 
the susceptibilities are initially very different but they become equal at the end of inflation. In this second case 
$f(a_{i})\gg 1$.  The various intermediate situations (stipulating that initially there is an arbitrary 
mismatch between the susceptibilities) will fall between the patterns of the two limiting cases $f_{i} =1$ and $f_{f} =1$.

With this distinction in mind, and recalling Eqs. (\ref{par1})--(\ref{par2}) and (\ref{FAFB}),
the solution of the mode function of Eqs. (\ref{sol1}) implies that the magnetic power spectrum becomes:
\begin{eqnarray}
P_{B}(k,\,a,\, \sigma) &=& H^4 \,\,{\mathcal Q}_{B}(\sigma,\mu) \, \, f_{i}^{|\sigma| -1} \,  \,\biggl(\frac{k}{a H}\biggr)^{5 - 2 |\sigma|} \, \, \biggl(\frac{a}{a_{i}}\biggr)^{2 \mu ( |\sigma|-1)},
\label{PSB}\\
{\mathcal Q}_{B}(\sigma,\,\mu) &=& \frac{\Gamma^2(|\sigma|)}{\pi^3} \, 2^{ 2 |\sigma| -3} \, |1+ \mu|^{ 2 |\sigma|-1},
\nonumber
\end{eqnarray}
where, for practical reasons, the following notation has been employed: 
\begin{equation}
\sigma = \frac{1 - 2 F_{E}}{1 + F_{B} - F_{E}}, \qquad \mu = \frac{F}{2} = F_{B} - F_{E}.
\label{defsigma}
\end{equation}
Similarly thanks to Eq. (\ref{sol2}) the electric power spectrum is: 
 \begin{eqnarray}
P_{E}(k,\,a,\,\sigma) &=& H^4 \,\,{\mathcal Q}_{E}(\sigma,\mu)  \, \, f_{i}^{|\sigma-1| -1} \,  \,\biggl(\frac{k}{a H}\biggr)^{5 - 2 |\sigma\, -\,1 |} \, \, \biggl(\frac{a}{a_{i}}\biggr)^{ 2 \mu ( |\sigma-1|-1)},
\label{PSE}\\
{\mathcal Q}_{E}(\sigma,\,\mu) &=&  \frac{\Gamma^2(|\sigma-1|)}{\pi^3} \, 2^{ 2 |\sigma-1| -3} \, |1+ \mu|^{ 2 |\sigma-1|-1}.
\nonumber
\end{eqnarray}
Note that in the plane $(F_{B},\, F_{E})$ there is a singular trajectory, namely $1 + F_{B} - F_{E} =0$ where 
 $\sigma$ diverges. This singularity is not physical and stems from the fact that for $F_{E} = F_{B} +1$ 
 the gauge couplings evolve exponentially in $\eta$.  
 
An essential feature of Eqs. (\ref{cc1}) and (\ref{cc2}) is that the electric and the magnetic power spectra both have dimensions 
of an energy density in Fourier space and this is related to the fact that the scale-invariant limit of the spectra occurs for $ n_{B}\to 1$ and 
$n_{E} \to 1$. As already mentioned we shall measure the magnetic power spectrum either in critical units of in $\mathrm{nG}^2$. 
Another way of measuring the amplitudes of the power spectra will be to use critical units and, in this case, the 
magnetic and electric power spectra can be expressed as:
\begin{eqnarray}
\Omega_{B}(k,\, a,\, \sigma,\,\mu,\, f_{i})  &=& \frac{8 \pi^2}{3} \,\, \epsilon \,\, {\mathcal A}_{{\mathcal R}} \, {\mathcal Q}_{B}(\sigma,\,\mu)\, \, f_{i}^{|\sigma| -1}
 \,\, \biggl(\frac{k}{a H}\biggr)^{5 - 2 |\sigma|} \,\, \biggl(\frac{a}{a_{i}}\biggr)^{2 \mu( |\sigma | -1)},
\label{om2}\\
\Omega_{E}(k,\, a,\, \sigma,\, \mu,\, f_{i})  &=& \frac{8 \pi^2}{3} \, \epsilon \, {\mathcal A}_{{\mathcal R}} \, {\mathcal Q}_{E}(\sigma,\,\mu) \, \, f_{i}^{|\sigma-1| -1}\,\, \biggl(\frac{k}{a H}\biggr)^{5 - 2 |\sigma-1|} \,\, \biggl(\frac{a}{a_{i}}\biggr)^{2 \mu( |\sigma -1 | -1)}.
\label{om3}
\end{eqnarray}
Notice that in Eqs. (\ref{om2}) and (\ref{om3}) we traded $H/M_{P}$ for ${\mathcal A}_{{\mathcal R}}$, i.e.  the amplitude of the power 
spectrum of curvature perturbations at the pivot scale $k_{p} =0.002\, \mathrm{Mpc}^{-1}$; note also that  the slow-roll parameter $\epsilon$ 
appears in the final expressions.
 
The magnetic and the electric spectral indices can be easily deduced from Eqs. (\ref{PSB})--(\ref{PSE}) and from Eqs. (\ref{om2})--(\ref{om3}) and they are:
\begin{equation}
n_{B} = 6 - 2 |\sigma|, \qquad n_{E} = 6 - 2 |\sigma-1|.
\label{om4}
\end{equation}
If a certain hierarchy between the gauge couplings is imposed ab initio for some theoretical consideration, then 
$f_{i}$ is fixed. Similarly $f_{i}$ can also be fixed by tuning $f(a_{f})$ at a certain specific value, be it $f_{f}$: in this case 
$f_{f} = f_{i} (a_{f}/a_{i})^{2 \mu}$. In the conformal time parametrization  the spatial gradient leads to a term $k/\sqrt{f}$ which can be either blueshifted or redshifted depending on $\sqrt{f}$ . Naively we would require that $\sqrt{f} >1$ to avoid a light speed that is larger than one. This observation, however, depends on the time parametrization. For instance in the cosmic time parametrization we have that the physical 
frequency is given by $k/a(t)$; if we would apply the same tenets we would exclude the interesting possibility of contracting Universes.
In the specific cases discussed below the cases $\sqrt{f} \ll 1$ are phenomenologically disfavoured and are anyway excluded for more
physical reasons. Let us finally note that to fix $f_{i}$ or $f_{f}$ is in general different. However in the framework of the monotonic 
parametrization of Eqs. (\ref{par1}) and (\ref{par2}) the value of $f_{f}$ is uniquely determined as a function to $f_{i}$ through 
the total number of inflationary efolds, i.e. $\ln{(f _{f}/f_{i})} = 2 (F_{B} - F_{E})\, N_{t}$. Therefore, for the purposes of the 
concrete discussion, the boundary conditions can be fixed either in terms of $f_{i}$ or in terms of $f_{f}$.

\subsection{Physical regions of the parameter space}
\begin{figure}[!ht]
\centering
\includegraphics[height=8cm]{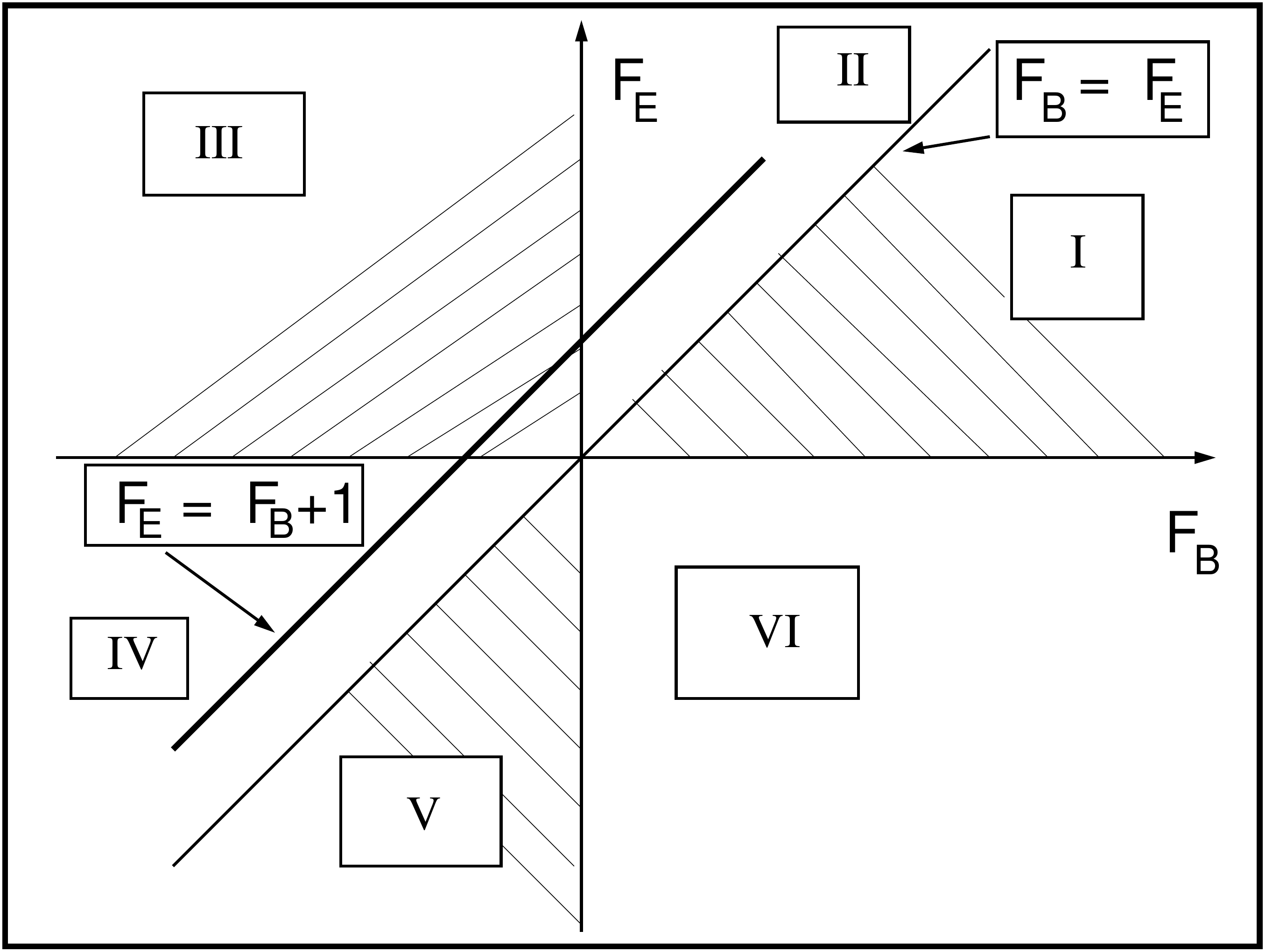}
\caption[a]{The parameter space of the model is schematically illustrated in terms of the normalized rates of variation of the gauge couplings.}
\label{Figure1}      
\end{figure}
The  $6$ different regions of the parameter space in the $(F_{E},\, F_{B})$ plane are illustrated in Fig. \ref{Figure1} where 
on the horizontal and vertical axes we report, respectively $F_{B}$ and $F_{E}$, i.e.  the rates of variation of $g_{B}$ and $g_{E}$.
The different regions appearing in Fig. (\ref{Figure1}) correspond to $6$ different dynamical situations that can be summarized 
for short as follows:
\begin{itemize}
\item[{\it (I)}] in the first region the gauge couplings are both increasing, i.e. $F_{B}\geq 0$ and $F_{E}\geq0$; furthermore also $f$ increases since $F_{B}\geq F_{E}$;
\item[{\it (II)}] in the second region the gauge couplings are also increasing but since $F_{E}\geq F_{B}$  we have that $F<0$ and $f$ is bound to decrease;
\item[{\it (III)}] in the third region  $F_{B} <0$ and $F_{E}>0$ (leading to $F < 0$);
\item[{\it (IV)}] in the fourth region both gauge couplings are decreasing, (i.e. $F_{B}<0$ and $F_{E}< 0$) while $F_{B}< F_{E}$ 
and consequently $F<0$;
\item[{\it (V)}] the gauge couplings are also decreasing simultaneously (i.e. $F_{E}< 0$ and $F_{B} <0$) 
but in this region $F >0$ (i.e. $f$ increases);
\item[{\it (VI)}] in the sixth and last region $F_{E} <0$ 
and $F_{B}>0$ implying $F >0$ and the corresponding growth of $f$.
\end{itemize}

In regions ${\it (I)}$ and ${\it (II)}$ the gauge couplings are both increasing; in regions 
${\it (IV)}$ and ${\it (V)}$ the gauge couplings are both decreasing (but at a different rate); finally in regions 
${\it (III)}$ and ${\it (VI)}$ one of the two gauge couplings is increasing while the other 
is decreasing.  Recalling Eq. (\ref{defsigma}) the spectral indices can be directly expressed in terms of $F_{E}$ and $F_{B}$:
\begin{equation}
n_{B}(F_{B},F_{E}) = 6 -  \biggl| \frac{1 - 2 F_{E}}{( 1 + F_{B} - F_{E})}\biggr|, \qquad 
n_{E}(F_{B},F_{E})  = 6 -  \biggl|\frac{1+ 2 F_{B}}{ (F_{E} - F_{B} -1)}\biggr|.
\label{SPINDEX1}
\end{equation}
Thanks to Eq. (\ref{SPINDEX1}) it is immediate to show that, under duality: 
\begin{equation}
F_{E} \to - F_{B}, \qquad F_{B} \to - F_{E},\qquad n_{B} \to n_{E}, \qquad n_{E} \to n_{B}.
\label{dualdual}
\end{equation}
Furthermore, recalling Eq. (\ref{defsigma}), under duality $\sigma \to \overline{\sigma} = 1 - \sigma$.

The magnetic power spectra solely expressed in terms of $F_{B}$ and $F_{E}$ become then:
\begin{eqnarray}
P_{B}(k,\, F_{B},\, F_{E},\, f_{i}) &=& H^4 \,\,{\mathcal Q}_{B}(F_{B},\, F_{E})  f_{i}^{[4 - n_{B}(F_{B},F_{E})]/2} \biggl(\frac{k}{a H}\biggr)^{n_{B}(F_{B},F_{E})-1} {\mathcal N}_{B}(N_{t},\, F_{B},\, F_{E}), 
\nonumber\\
{\mathcal Q}_{B}(F_{B},\, F_{E},\, f_{i}) &=& \frac{\Gamma^2[( 6 - n_{B}(F_{B},F_{E}))/2]}{\pi^3} \, 2^{ 3 - n_{B}(F_{B},F_{E})} \, |1+ F_{B} - F_{E}|^{ 5 - n_{B}(F_{B},F_{E})}.
\nonumber\\
 {\mathcal N}_{B}( F_{B},\, F_{E}) &=& e^{ N_{t} (F_{B} - F_{E}) [4 - n_{B}(F_{B},F_{E})]}.
 \label{PSB2}
 \end{eqnarray}
As a function of  $F_{B}$ and $F_{E}$ the electric power spectra become instead:
\begin{eqnarray}
P_{E}(k,\,N_{t},\, F_{B},\, F_{E}) &=& H^4 \,\,{\mathcal Q}_{E}(F_{B},\, F_{E}) f_{i}^{[4 - n_{E}(F_{B},F_{E})]/2} \biggl(\frac{k}{a H}\biggr)^{n_{E}(F_{B},F_{E})-1} {\mathcal N}_{E}(N_{t},\, F_{B},\, F_{E})
\nonumber\\
{\mathcal Q}_{E}(F_{B},\, F_{E}) &=& \frac{\Gamma^2[( 6- n_{E}(F_{B},F_{E}))/2]}{\pi^3} \, 2^{ 3 - n_{E}(F_{B},F_{E})} \, |1+ F_{B} - F_{E}|^{ 5 - n_{E}(F_{B},F_{E})},
\nonumber\\
{\mathcal N}_{E}(N_{t},\, F_{B},\, F_{E}) &=& e^{ N_{t} (F_{B} - F_{E}) [4 - n_{E}(F_{B},F_{E})]},
\label{PSE2}
\end{eqnarray}
where $n_{B}(F_{B},F_{E})$ and $n_{E}(F_{B},F_{E})$ are now given in Eq. (\ref{SPINDEX1}). With the same 
strategy the corresponding critical fractions of Eqs. (\ref{om2}) and (\ref{om3}) can be easily expressed in the $(F_{B},\, F_{E})$ plane. 
Let us finally mention that Eqs. (\ref{PSB2}) and (\ref{PSE2}) go
one into the other thanks to a duality transformation. This important property, already mentioned in the previous section, 
is now explicitly verifiable at the level of the final results and thanks to Eq. (\ref{dualdual}).

\subsection{Dualizing the regions of the parameter space}
\begin{figure}[!ht]
\centering
\includegraphics[height=8.5cm]{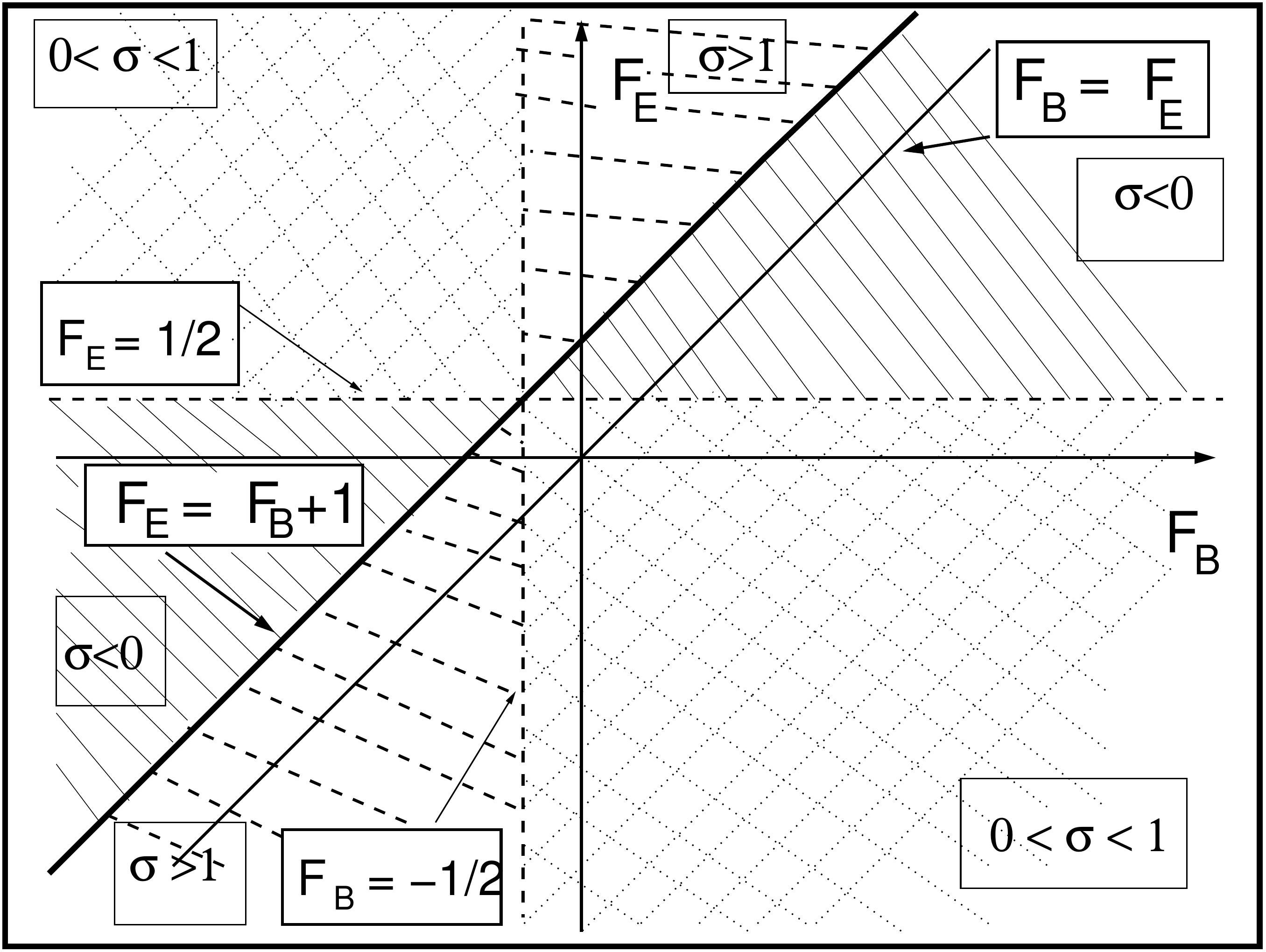}
\caption[a]{Dependence of the power spectra on the rates of the gauge couplings.}
\label{Figure2}      
\end{figure}

In each region of Fig. \ref{Figure1}, depending on the values of $F_{E}$, $F_{B}$ and $F$, the power 
spectra have different slopes and different amplitude.This aspect is illustrated in  Fig. \ref{Figure2} 
where we analyze the parameter space from the viewpoint of the slope of the power spectra.
The shaded regions of  Fig. \ref{Figure2} correspond to different spectral indices and different values 
of $\sigma$. Under duality the various regions are interchanged.  Consider, for instance, the first quadrant of Fig. \ref{Figure2} where the gauge couplings are both 
increasing, i.e. $F_{E}>0$ and $F_{B} >0$. In this region we have:
\begin{eqnarray}
&& 0 < F_{E} < 1/2, \qquad 0<\sigma < 1,
\nonumber\\
&& 1/2 < F_{E} < F_{B} +1,\qquad \sigma <0,
\nonumber\\
&& F_{E} > F_{B} +1,\qquad \sigma > 1.
\label{firstQ}
\end{eqnarray}
For each region of Eq. (\ref{firstQ}) the spectral 
indices have a different functional dependence upon $F_{E}$ and $F_{B}$. In Tab. \ref{Table1} the spectral indices in the first quadrant are 
summarized. Notice that the electric spectral index is always expressible in terms of $n_{B}$ but the relation changes 
from region to region. In Fig. \ref{Figure2} there are peculiar lines dividing the different shaded areas. The straight line 
 $F_{E} = F_{B} +1$ defines a line of singular models separately discussed in Eqs. (\ref{sing1})--(\ref{sing2}) . These are not physical singularities 
 but rather singularities of the parametrization\footnote{ As previously shown in this case the susceptibilities 
 have an exponential dependence on $\eta$ and, consequently, the relation between $\tau$ and $\eta$ is logarithmic. 
 These models are characterized by an electric field than vanishes exactly but besides this they do not have a particularly 
 relevant status from the viewpoint of phenomenology.}. 

Always in Fig. \ref{Figure2} the lines $F_{E} = 1/2$ and $F_{B} = -1/2$ are the 
 zeroes of $\sigma$ and of $1 -\sigma$. Both models are regular, the corresponding spectra have logarithmic 
 corrections either in the electric or in the magnetic case. For the sake of simplicity,  we shall exclude these three lines
 when charting the area of the parameter space in the plane $(F_{B},\, F_{E})$: this is not crucial 
 since we are just excising three lines from a whole plane. Furthermore these cases have been separately discussed
 in the previous sections, they are not more significant than the others and they are also irrelevant for the phenomenological 
 considerations discussed in section \ref{sec5}.
\begin{table}
\begin{center}
\vskip 0.5truecm
\begin{tabular}{| c | l | c | | | c | | |  c | }
\hline
Normalized rates \quad & $n_{B}$ & $n_{E}$   \\ \hline
$0 < F_{E} < 1/2$ &   $n_{B} = (5 + 6 F_{B} - 4 F_{E})/( 1 + F_{B} - F_{E})$ & $n_{E}= 10 - n_{B}$  \\ \hline
$1/2 < F_{E} < F_{B}+1$ &  $n_{B} =(7 + 6 F_{B} - 8 F_{E})/( 1 + F_{B} - F_{E})$&  $n_{E}= n_{B} -2 $ \\ \hline
$F_{E} > F_{B} +1 $ &$n_{B} =  (5 + 6 F_{B} - 4 F_{E})/( 1 + F_{B} - F_{E})$ &  $n_{E} = n_{B} + 2 $ \\ \hline
\hline
\end{tabular}
\caption{First quadrant $F_{E} > 0$ and $F_{B} > 0$.}
\label{Table1}
\end{center}
\end{table}

In Fig. \ref{Figure2} there are finally two particularly simple cases, namely $F_{E} =0$ (with $F_{B}$ free to vary) and $F_{B}=0$ (with 
$F_{E}$ free to vary). These two situations correspond to the situation where one of the gauge couplings is constant.
\begin{table}
\begin{center}
\vskip 0.5truecm
\begin{tabular}{| c | l | c | | | c | | |  c | }
\hline
Normalized rates \quad &  $n_{B}$ & $n_{E}$\\ \hline
$0 < F_{B} < 1/2$ &   $n_{B} =  10 - n_{E}$& $n_{E}= (5 - 6 F_{E} + 4 F_{B})/( 1 + F_{B} - F_{E})$\\ \hline
$1/2 < F_{E} < F_{B}+1$ &$n_{B} =  2+ n_{E} $&  $n_{E}= (7 + 8 F_{B} - 6 F_{E})/( 1 + F_{B} - F_{E})$\\ \hline
$F_{E} > F_{B} +1 $ &$n_{B} = 2-  n_{E}$ &$n_{E} = (5 - 6 F_{E} + 4 F_{B})/( 1 + F_{B} - F_{E})$\\ \hline
\hline
\end{tabular}
\caption{Third quadrant $F_{E} < 0$ and $F_{B}< 0$.}
\label{Table2}
\end{center}
\end{table}

Different portions of Figs. \ref{Figure1} and \ref{Figure2} can be easily related via duality. Equation (\ref{firstQ}) refers to the first quadrant  of Figs. \ref{Figure1} and \ref{Figure2}. Let us now suppose to be interested in the electric and magnetic spectral indices in the third quadrant where $F_{E} <0$ and $F_{B} <0$. In this region both  gauge couplings are decreasing. The three regions of Eq. (\ref{firstQ}) transform under duality into:
\begin{eqnarray}
&& -1/2 < F_{B} < 0, \qquad 0<\sigma < 1,
\nonumber\\
&& -1+ F_{E} < F_{B} < -1/2,\qquad \sigma > 1,
\nonumber\\
&& F_{E} > F_{B} +1,\qquad \sigma <0.
\label{thirdQ}
\end{eqnarray}
The result of Eq. (\ref{thirdQ}) can be easily derived by recalling that, under duality,   $F_{E} \to - F_{B}$ and $F_{B} \to - F_{E}$.
 Notice also that duality relates the first and third quadrant (region by region)  given the transformation properties of $\sigma$. 
The results of the duality transformation are reported in Tab. \ref{Table2}. In the remaining two quadrants the same logic applies so that the spectra, the spectral indices and the other information can be easily deduced by successive duality 
transformations from the first or from the third quadrants\footnote{Notice that from Tab. \ref{Table1} the structure of Tab. \ref{Table2} can be immediately obtained 
by only using duality.}. 

Before discussing in detail the parameter space of the model let us remark that there are, broadly speaking, three distinct 
physical situations. The gauge couplings may be both small at $a_{i}$ and then increase: this is the case of weakly coupled 
initial conditions evolving towards strong coupling at the end of inflation. The second possibility is that the gauge couplings are initially of order $1$ and then decrease towards the end of inflation: this is the case of the strongly coupled initial conditions. The third possibility is 
that initially one of the gauge couplings is strong while the other is weak: this is the case of mixed initial conditions.
In the case of conventional inflationary models  the Universe evolves from strong gravitational coupling to weak gravitational coupling, i.e. the space-time curvature is maximal at the onset of inflation and gets smaller during reheating. It is fair to say that the potential drawbacks of magnetogenesis coincide with the potential drawbacks of conventional models of inflation which are, typically, not geodesically complete in their past history and preceded by a high curvature 
regime.
The considerations reported here can be easily extended to the case of bouncing models evolving from weak gravitational coupling to strong gravitational coupling, i.e. the space-time curvature is small initially and gets larger at the reheating.  

In summary, it would be pedantic to discuss in details all the $6$ regions of the parameter space of Fig. \ref{Figure1} (or Fig. \ref{Figure2}). 
So it is mandatory to complement Figs. \ref{Figure1} and \ref{Figure2} with more physical considerations. 
Fortunately duality comes to our rescue since 
the different portions of the parameter space are all related by duality transformations so that, 
for a complete discussion of the parameter space,  a thorough examination of a single quadrant will suffice. 
Since the possibility of having the simultaneous growth of the gauge couplings during inflation 
is typical of this type of model we shall therefore focus on the first quadrant where  $F_{E} > 0$ and $F_{B}>0$.

\renewcommand{\theequation}{5.\arabic{equation}}
\setcounter{equation}{0}
\section{Charting the parameter space of magnetogenesis}
\label{sec5}
The parameter space of inflationary magnetogenesis will now be charted by discussing in detail 
the first quadrant of Figs. \ref{Figure1} and \ref{Figure2}. Thanks to the duality symmetry 
 the remaining portions of the parameter space will be easy to discuss (see section \ref{sec4} for a specific discussion of this 
 point). Three independent requirements are relevant for the present discussion:  
 the critical density bound (sometimes dubbed backreaction 
constraint), the magnetogenesis requirements and the naturalness of the initial conditions.  
All the mentioned conditions are necessary but none of them is per se sufficient to pin down the allowed region of the 
$(F_{B},\, F_{E})$ plane. However, if they are all jointly verified in a given domain they 
become sufficient.

The logic will be, in short, the following. We shall first identify and classify 
all the domains of the $(F_{B},\, F_{E})$ plane where the backreaction constraints are satisfied. 
Then, in the regions not constrained by backreaction we shall compute precisely the magnetic power spectrum 
by keeping track of all terms and prefactors of the spectra. 
It will then be easier to see in the $(F_{E},\,F_{B})$ plane where 
and how the weakly coupled initial conditions are verified.  

\subsection{Maximal wavenumber of the spectra}
We recall that $k\tau$ can be expressed as:
\begin{equation}
k\tau = \frac{k}{(1 -\epsilon) a H} = \frac{k}{H_{0} } e^{ - N_{\mathrm{max}}} \biggl[1 + \epsilon + {\mathcal O}(\epsilon^2)\biggr],
\label{xx}
\end{equation}
where $H_{0} =2.334\times 10^{-4} \,(h_{0}/0.7) \mathrm{Mpc}^{-1}$ is the present value of the Hubble rate and 
$N_{\mathrm{max}}$ is the maximal number of efolds which are today accessible to our observations  \cite{mgg}(see also \cite{LL}). In practice $N_{\mathrm{max}}$ is determined by fitting the redshifted inflationary event horizon inside the present Hubble radius $H_{0}^{-1}$:
\begin{equation}
e^{N_{\mathrm{max}}} = ( 2 \, \pi  \, \epsilon \, {\mathcal A}_{{\mathcal R}}\, \Omega_{\mathrm{R}0})^{1/4} \,\, \biggl(\frac{M_{\mathrm{P}}}{ H_{0} }\biggr)^{1/2} \biggl(\frac{H_{r}}{H}\biggr)^{\gamma -1/2},
\label{NN1}
\end{equation}
where $\Omega_{\mathrm{R}0}$ is the present critical fraction of radiation (in the concordance model $h_{0}^2 \Omega_{\mathrm{R}0} 
= 4.15 \times 10^{-5}$). The term containing $\gamma$ accounts for the possibility of a delayed reheating ending at a putative scale $H_{r}$ eventually 
much smaller than the inflationary curvature scale $H$. For illustration we shall focus on the simplest case and 
 choose the sudden reheating approximation by setting $\gamma =1 /2$. Different possibilities can be however considered.
A long postinflationary phase dominated by a stiff equation of state has been examined in the context of magnetogenesis 
(see first paper of Ref. \cite{mgg} and references therein). A delayed reheating has the effect of increasing $N_{max}$. The largest 
value of $N_{max}$ in the case of a stiff pos-inflationary phase can be estimated as $N_{max} = 78.3 + (1/3) \ln{\epsilon}$. In the 
sudden reheating approximation we have  $N_{\mathrm{max}} \simeq 63.25 + 0.25 \ln{\epsilon}$ which is 
numerically close to the minimal number of efolds $N_{\mathrm{min}}$ needed to solve the kinematic 
problems of the standard cosmological model (i.e. $N_{\mathrm{min}} \simeq N_{\mathrm{max}}$). Recall, as usual, 
that ${\mathcal A}_{{\mathcal R}}$ denotes the fiducial amplitude of the scalar power 
spectrum at the pivot scale $k_{p}= 0.002\, \mathrm{Mpc}^{-1}$. We shall not deviate from these fiducial values and assume 
the standard concordance lore for all the late time cosmological parameters. 

\begin{figure}[!ht]
\centering
\includegraphics[height=8cm]{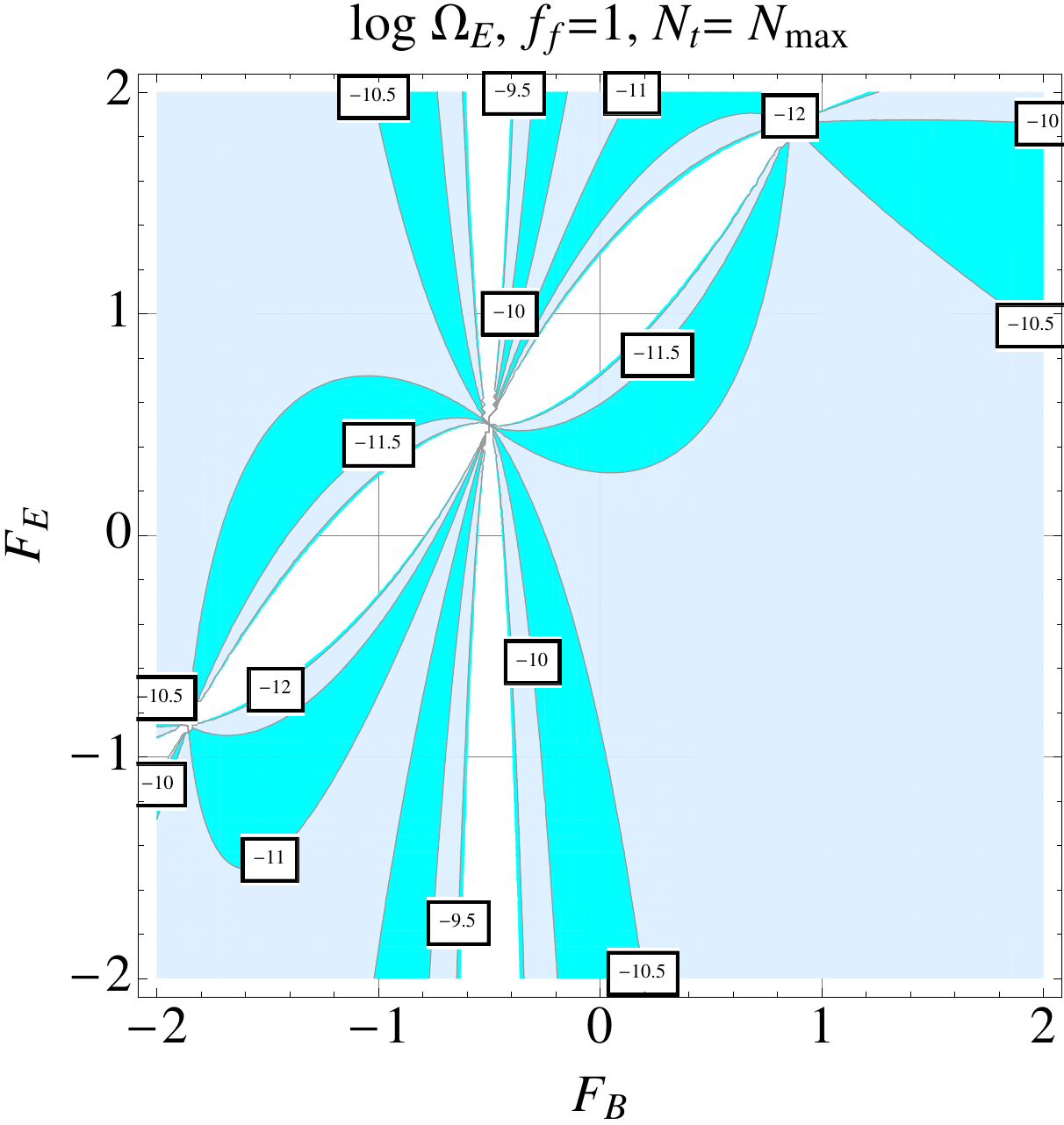}
\includegraphics[height=8cm]{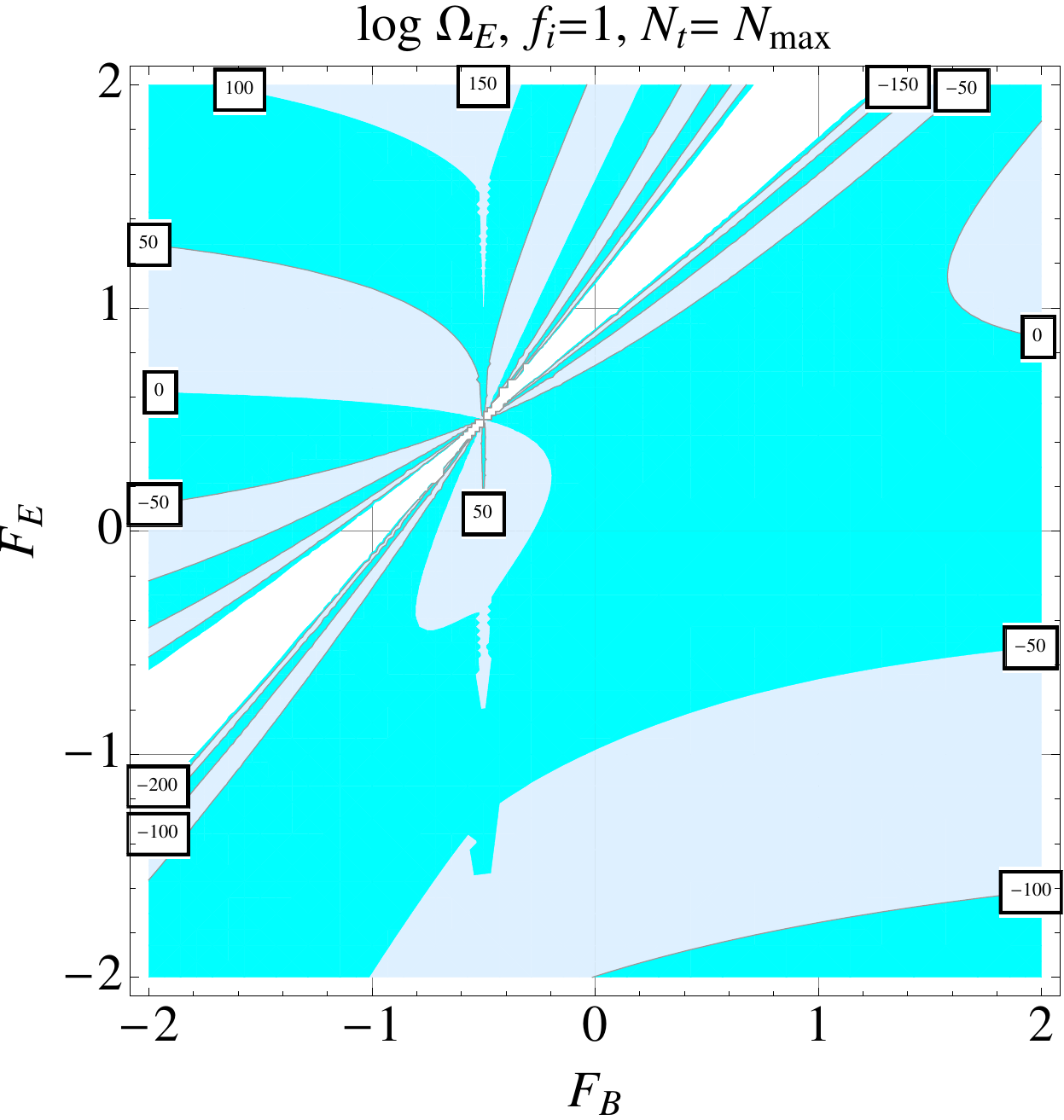}
\includegraphics[height=8cm]{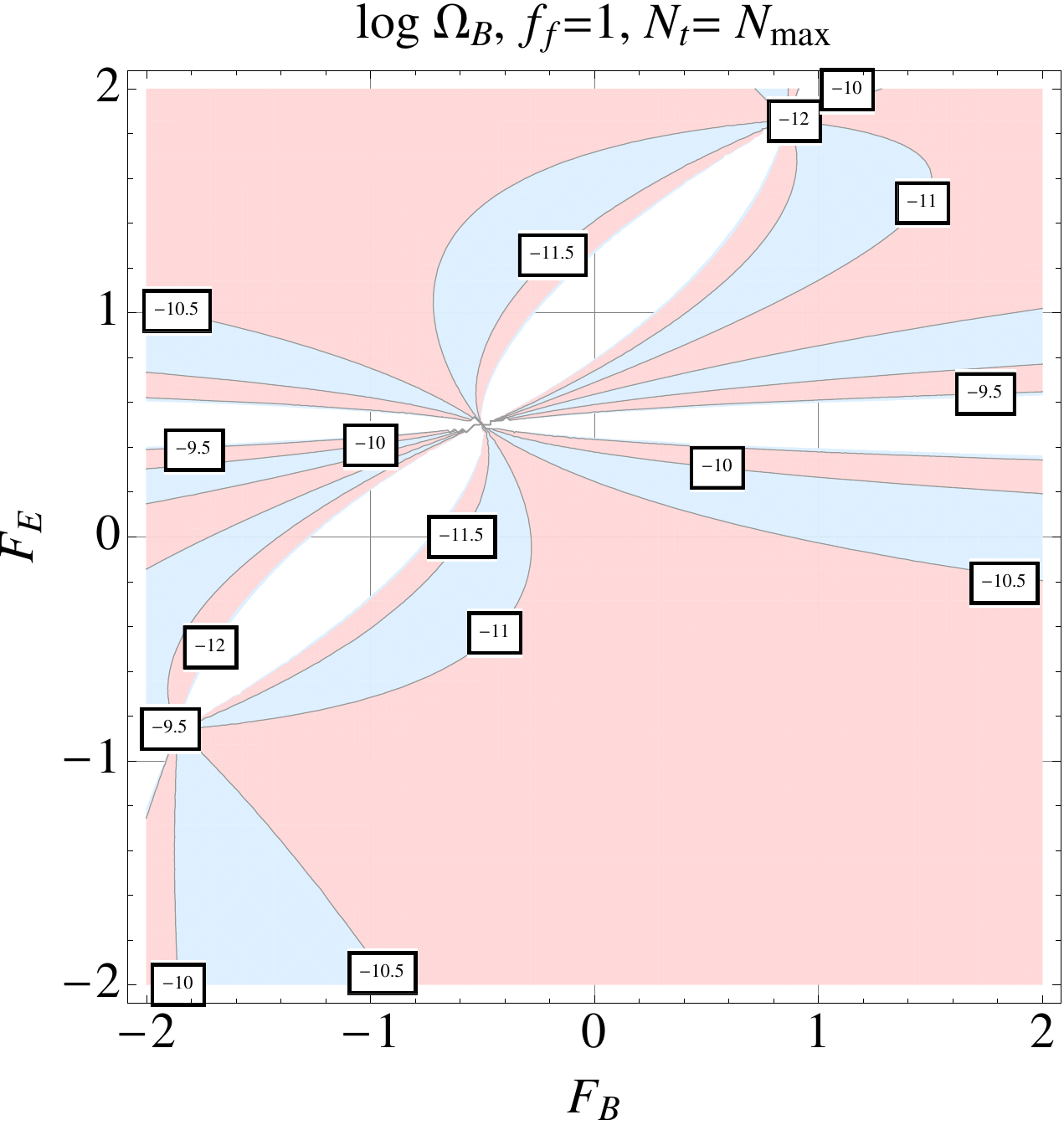}
\includegraphics[height=8cm]{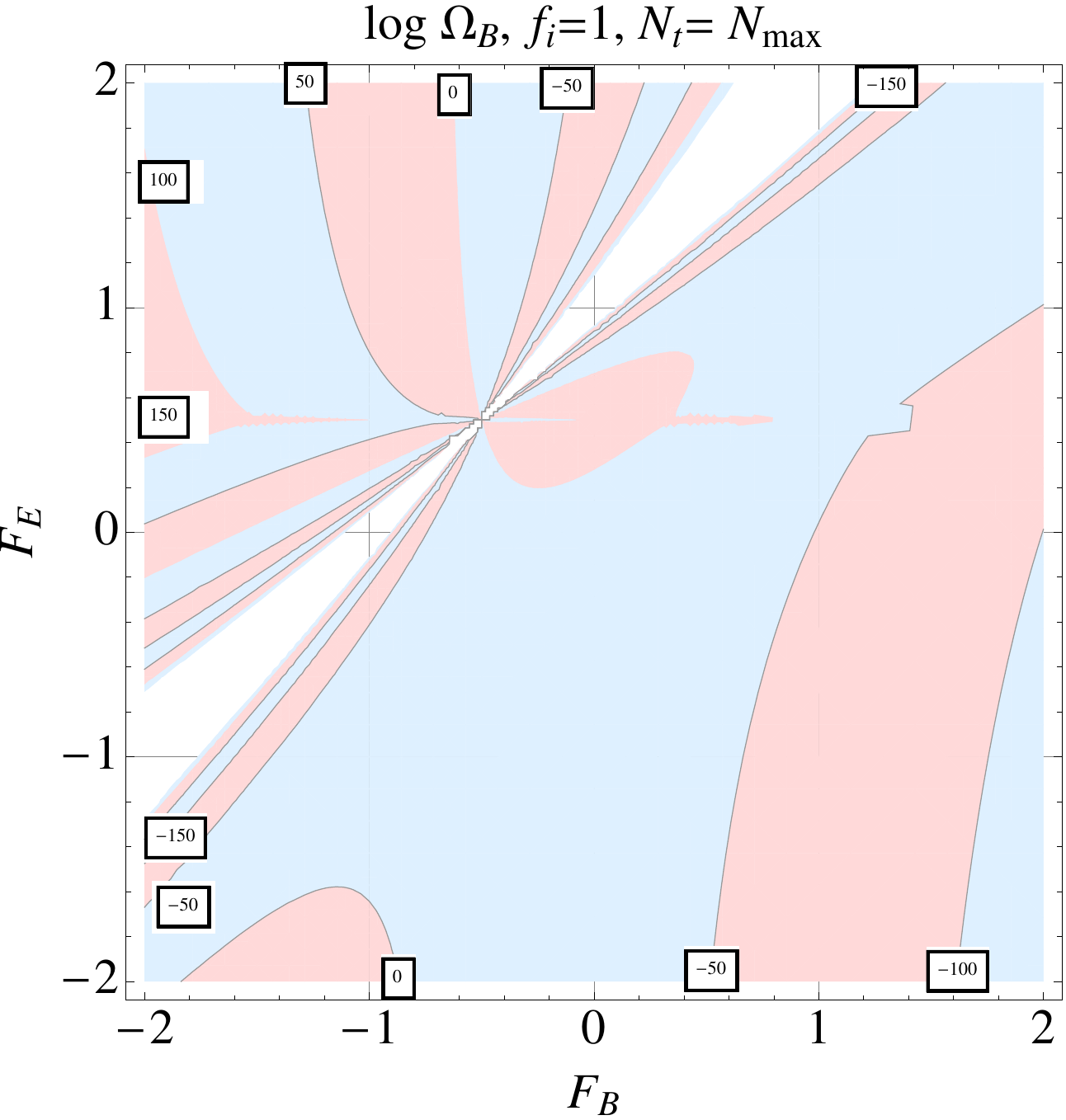}
\caption[a]{The contours for constant $\log{\Omega_{B}}$ and for constant $\log{\Omega_{E}}$. The plots on the left refer to the case  
 $f_{f} =1$ and for $N_{t} = N_{\mathrm{max}}=62.09$; the plots on the right refer to the case  $f_{i} =1$. The various 
labels illustrate the values of the critical fractions on the given curve. On the horizontal axis we illustrate the rate of variation 
of $g_{B}$ while on the vertical axis the rate of variation of $g_{E}$ is illustrated. }
\label{Figure3}      
\end{figure}

In Fig. \ref{Figure3} we present the contours of constant $\log{\Omega_{E}}$ (the common logarithm critical fraction of electric energy density)
and the contours of constant  $\log{\Omega_{B}}$ (the common logarithm of the critical fraction of the magnetic energy density). 
In Fig. \ref{Figure3} the spectra have been evaluated at the maximal wavenumber $k \tau =1$. This is strictly speaking not the highest frequency 
of the electric power spectrum which is exponentially suppressed as soon as the reheating starts\footnote{Imposing the correct boundary conditions between
the inflationary phase and the postinflationary phase entails also an overall suppression of the electric power spectrum in comparison with 
the magnetic power spectrum \cite{mgg}.}. To extend artificially the electric power spectrum beyond 
the frequency fixed by the postinflationary conductivity (for more details see \cite{SUSC1,mgg}) implies that the resulting constraints are 
much stronger. Since we are here trying to constrain the scenario it makes sense to impose here the most demanding requirements at the 
most demanding scale.

In Fig. \ref{Figure3} the  values of the rates encompass all the four quadrants of Figs. \ref{Figure1} and \ref{Figure2}.
This choice has been made also for illustrative reasons since, as explained above, in the remaining figures of the section we shall limit our attention 
to the first quadrant.
The results of Fig. \ref{Figure3} illustrate the domains of the parameter space where the corresponding critical fractions are still  perturbative around the maximal
wavenumber of the spectra. The contours where one of the two critical fractions is larger than $1$ should be ideally excised from the physical 
region of the parameter space. We note that the classes of models with $f_{f} =1$ seem to be comparatively less constrained than the ones with 
$f_{i}=1$. In Fig. \ref{Figure3}  $F_{E} = F_{B}+1$ and $F_{E} =1/2$ have been excised\footnote{As discussed in Eqs. (\ref{sing1}) and (\ref{sing2}) 
this is due  to a singularity of the parametrization (not of the model).} 
In Fig. \ref{Figure3} and in the forthcoming figures we have chosen, as already mentioned, ${\mathcal A}_{{\mathcal R}}= 2.41\times 10^{-9}$ and $\epsilon=0.01$.

The results of Fig. \ref{Figure3} also fix the allowed excursion of $F_{E}$ and $F_{B}$. Indeed at the maximal frequency of the spectrum the amplitude 
depends also on the maximal value of $F_{E}$ and $F_{B}$. In Fig. \ref{Figure3}, for illustrative reasons, we used values $|F_{B}| < 2$ and $|F_{E} |<2$.
Each value of $f_{i}$ (or $f_{f}$) is compatible with a maximal range of  $F_{E}$ and $F_{B}$. For instance, if $f_{f} = {\mathcal O}(1)$ 
we have that, at most $|F_{B}| < 5$ and $|F_{E} |<5$: larger values imply that $\Omega_{E} \simeq {\mathcal O}(1)$ around the maximal 
wavenumber of the spectrum.
\subsection{Backreaction constraints}
If the total energy density of the large-scale modes  does not exceed  the critical energy density the following integral 
\begin{eqnarray}
&& {\mathcal F}(f_{i}, \sigma,\mu) \int_{k_{i}\tau}^{k_{f} \tau} \frac{d x}{x} \biggl[{\mathcal Q}_{B}(\sigma,\mu) x^{5 - 2 |\sigma|}
+{\mathcal Q}_{E}(\sigma,\mu) x^{5 - 2 |\sigma-1|} \biggr],
\nonumber\\
&&{\mathcal F}(f_{i}, \sigma,\mu) = \frac{2\pi}{3} \,\epsilon\, {\mathcal A}_{{\mathcal R}} \,f_{i}^{|\sigma|-1} \, \biggl(\frac{a}{a_{i}}\biggr)^{2 \mu(|\sigma| -1)},
\label{DEF1}
\end{eqnarray}
cannot exceed $1$. In practice it will also have to be much smaller than $1$, i.e. at most $10^{-3}$.
In the most constraining situation the integral goes from $k_{i} = 1/\tau_{i}$ (corresponding to the mode 
leaving the Hubble radius when $a= a_{i}$) up to $k_{f} = 1/\tau_{f}$ 
(corresponding to the mode reentering the Hubble radius when at $a= a_{f}$). Performing explicitly the integrals in the case $ 2 |\sigma| \neq 5$ and  $2 |\sigma-1| \neq 5$, Eq. (\ref{DEF1}) implies\footnote{The cases $ 2 |\sigma| = 5$ and  $2 |\sigma-1| = 5$  must be separately treated since the integral lead to logarithmic divergences. More specifically, if $\sigma = 5/2$ and $f_{f} =1$, then everything is okay besides the logarithmic growth which is not essential for the bound.  Conversely, if $\sigma=5/2$ and $f_{i} =1$ we must anyway demand $\mu <0$.  If $\sigma = -5/2$ and $f_{f} =1$ the spectra are excluded since they get easily overcritical. In the case  $\sigma = -5/2$ and $f_{i} =1$ the spectra are viable provided $3 \mu +2 <0$. 
When $\sigma = 7/2$ and $f_{f} =1$ the spectra are excluded while for  $f_{i} =1$ they are only compatible provided $5 \mu + 2 <0$. 
Finally, if $\sigma = -3/2$ the constraints are under control for $f_{f}=1$ (and for $f_{i} =1$ but provided $\mu <0$).}:
\begin{eqnarray}
{\mathcal F}(f_{i}, \sigma,\mu)\biggl\{\frac{{\mathcal Q}_{B}(\sigma,\mu)x_{i}^{5 - 2 |\sigma|}}{5 - 2 |\sigma|}  \biggl[ \biggl(\frac{a_{i}}{a_{f}} \biggr)^{5 - 2 |\sigma|} 
-1\biggr] + \frac{{\mathcal Q}_{E}(\sigma,\mu)x_{i}^{5 - 2 |\sigma-1|}}{5 - 2 |\sigma-1|}  \biggl[ \biggl(\frac{a_{i}}{a_{f}} \biggr)^{5 - 2 |\sigma-1|} 
-1\biggr] \biggr\}.
\label{DEF2}
\end{eqnarray}
To analyze the requirements imposed by Eq. (\ref{DEF2}) we shall first consider the limit ${\mathcal Q}_{E} \to 1$ and 
${\mathcal Q}_{B} \to 1$. With this approximation it will be easier to pin down the area of the allowed region in the 
$(F_{B},\, F_{E})$ plane. This preliminary step will then be supplemented by the accurate contour plots obtained with the exact form of ${\mathcal Q}_{E}$ and ${\mathcal Q}_{B}$. Indeed the corrections due to ${\mathcal Q}_{E}$ and ${\mathcal Q}_{B}$ can be neglected in the first approximation but they are not irrelevant.

Thus, in the limit ${\mathcal Q}_{E} \to 1$ and 
${\mathcal Q}_{B} \to 1$, Eq. (\ref{DEF2}) implies that the following three inequalities must be separately satisfied in the physical region 
of the parameter space:
\begin{eqnarray}
&& (|\sigma|-1) \biggl[ \frac{\ln{f_{i}}}{N_{t}} + 2 \mu \biggr] \leq 0,
\nonumber\\
&&(|\sigma|-1) \biggl[ \frac{\ln{f_{i}}}{N_{t}} + 2 \mu \biggr] - 5 + 2 |\sigma| \leq 0,
\nonumber\\
&& (|\sigma|-1) \biggl[ \frac{\ln{f_{i}}}{N_{t}} + 2 \mu \biggr] - 5 + 2 |\sigma-1| \leq 0,
\label{DEF3}
\end{eqnarray}
where $N_{t}$ denotes, as usual, the total number of efolds. 
Recalling that $\ln{(f_{f}/f_{i})} =  2(F_{B}-F_{E}) N_{t}$, the two complementary physical situations correspond to $f_{f} =1$ 
and to $f_{i} =1$. Indeed these two cases well represent the effect of the initial conditions on the allowed region of the parameter space. If $f_{i} \neq 1$ (but $f_{i} \simeq {\mathcal O}(1)$) the situation is qualitatively close
to the case $f_{i} = 1$; if $f_{i} \gg 1$ the qualitative features of the spectra are close to the case $f_{f} =1$.
\begin{figure}[!ht]
\centering
\includegraphics[height=8.5cm]{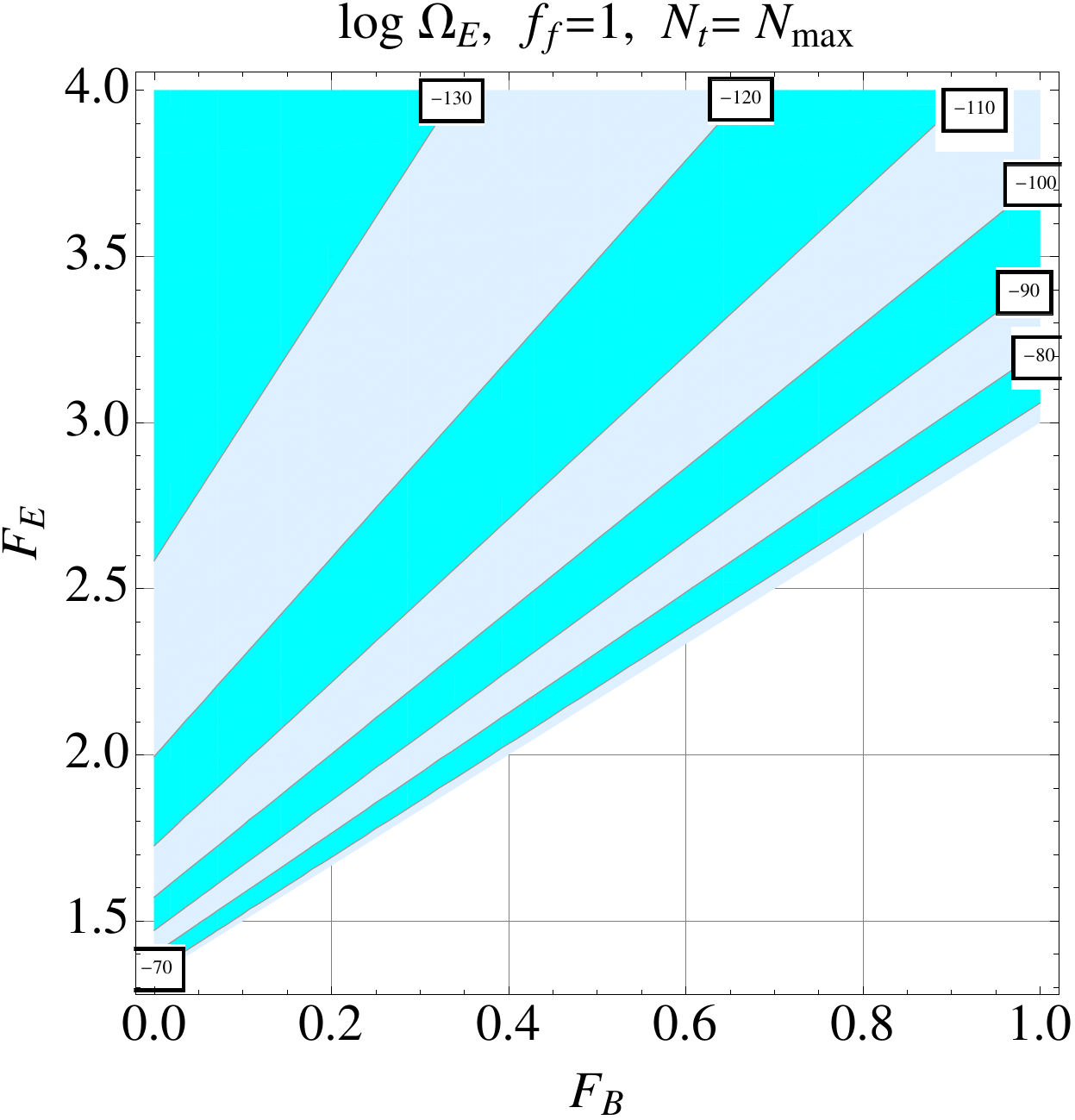}
\includegraphics[height=8.5cm]{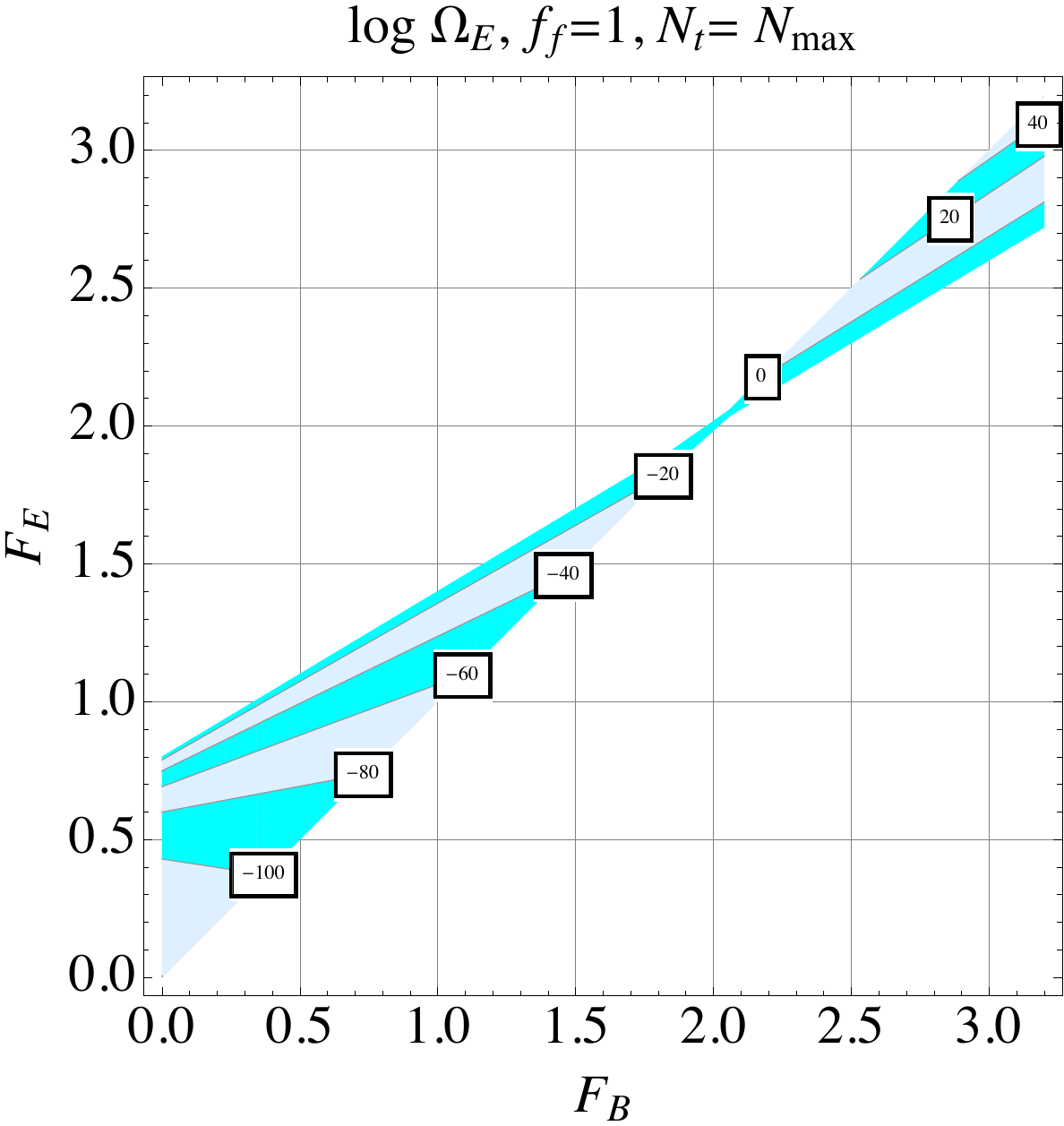}
\caption[a]{The two contour plots illustrate the common logarithm of $\Omega_{E}$ without any simplifying approximation. In the left plot the region of Eq. (\ref{REG1a}) is illustrated. The plot on the right refers instead to the region of Eq. (\ref{REG1b}). In both cases the wavenumber has been chosen to be $10^{-4}\, \mathrm{Mpc}^{-1}$.}
\label{Figure4}      
\end{figure}

\subsubsection{The case $f_{f} =1$}
In this case he backreaction constraints can be evaded provided:
\begin{equation}
5 - 2 |\sigma| \geq 0, \qquad 5 - 2 |\sigma -1| \geq 0.
\label{DEF4}
\end{equation}
When $\sigma >1$ the conditions imposed by Eq. (\ref{DEF4}) imply, respectively, 
$\sigma \leq 5/2$ and  $ \sigma \leq 7/2$.  Thus,  in the domain $\sigma > 1$ the most constraining bound is 
 $\sigma \leq 5/2$. In the region $0 <\sigma <1$ Eq. (\ref{DEF4}) demands that $-3/2 \leq \sigma \leq 5/2$.
But both conditions are always verified in the interval $0 <\sigma <1$. Therefore no supplementary bounds 
arise in this case. Finally, if  $\sigma<0$ Eq. (\ref{DEF4}) demands $\sigma \geq - 5/2$ and  $\sigma \geq - 3/2$.
The most constraining of the two previous conditions is $\sigma \geq -3/2$. In Fig. \ref{Figure2} the three domains 
$\sigma >1$, $0<\sigma <1$ and $ \sigma <0$ have been specifically illustrated in the $(F_{B},\, F_{E})$ plane. 

Collecting together all the relevant requirements, in the case $f_{f} =1$ the bounds of Eqs. (\ref{DEF1}) and (\ref{DEF2}) imply that 
$-3/2 \leq \sigma \leq 5/2$. Recalling Eq. (\ref{defsigma}) we then have the following requirement in the $(F_{B},\, F_{E})$ plane:
\begin{equation}
-3 < \frac{1 - 2 F_{E}}{1 + F_{B} - F_{E}} < 5.
\label{DEF5}
\end{equation}
The inequalities of Eq. (\ref{DEF5}) are verified in the following two non-overlapping regions of the 
first quadrant (i.e. $F_{E} \geq 0$ and $F_{B} \geq 0$) of the $(F_{B},\,F_{E})$ plane:
\begin{eqnarray}
&& F_{E} > F_{B} +1,\qquad  F_{E} \geq \frac{5}{3} F_{B} + \frac{4}{3}, 
\label{REG1a}\\
&& F_{E} < 1 + F_{B},  \qquad F_{E} \leq \frac{3}{5} F_{B} + \frac{4}{5}.
\label{REG1b}
\end{eqnarray}
Notice that in the region of Eq. (\ref{REG1a}) the electric coupling grows always faster than the 
magnetic coupling (i.e. $F_{E} > F_{B}$) and consequently $f$ decreases (i.e. $F<0$).
Conversely in the region determined by Eq. (\ref{REG1b}) we have that $F_{E} \geq F_{B}$ when
$0< F_{B} \leq 2$ and $F_{E} < F_{B}$ whenever $ F_{B} > 2$. 
The straight line $F_{E} = 5 F_{B}/3 + 4/3$ corresponds to the case of a flat magnetic power spectrum (i.e. $n_{B} \to 1$). This is an isospectral line in the 
sense that along this line the magnetic and the electric spectral indices do not change. Conversely the amplitude of the spectrum still has some dependence 
on $F_{B}$. 
\begin{figure}[!ht]
\centering
\includegraphics[height=8.5cm]{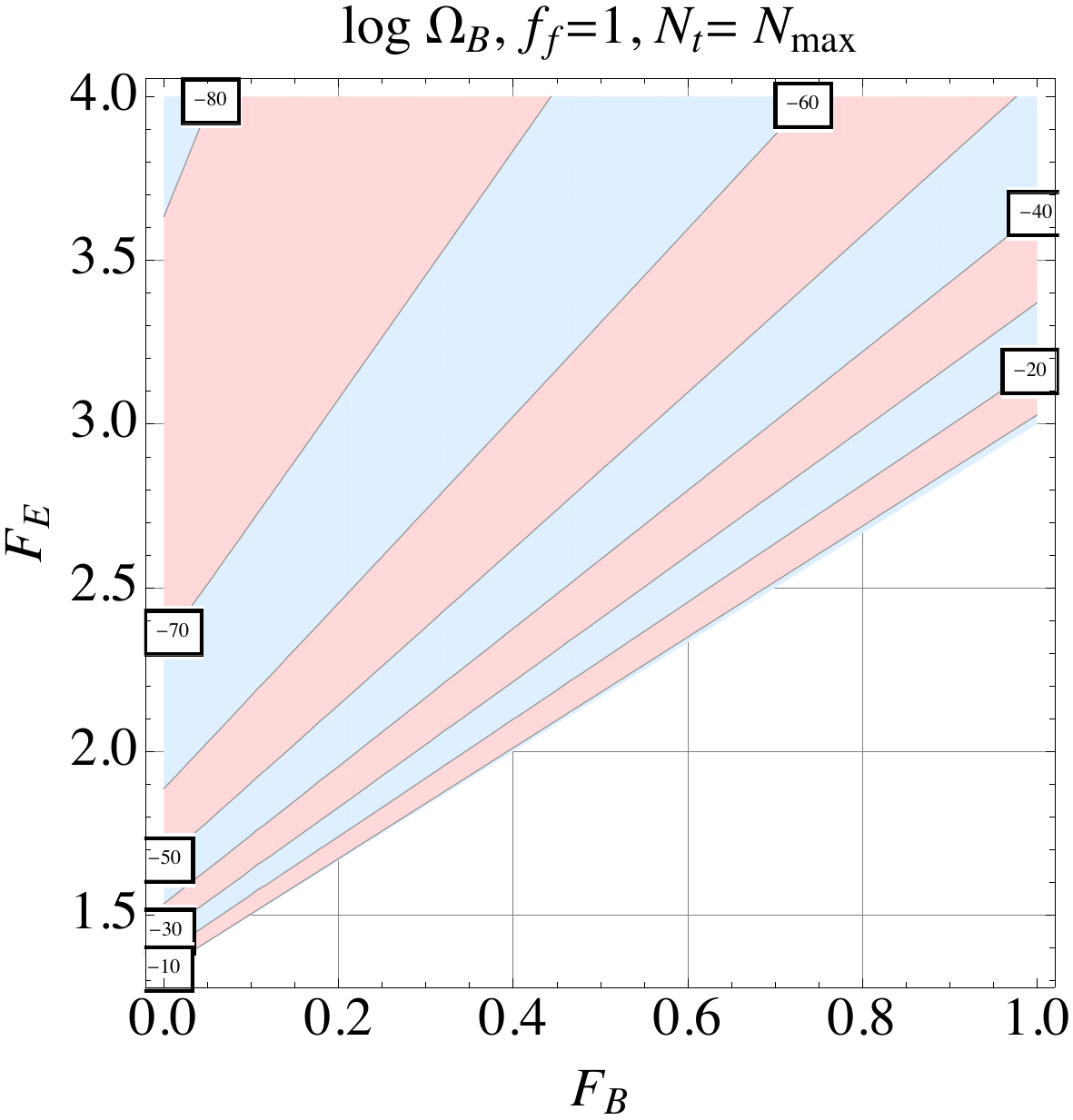}
\includegraphics[height=8.5cm]{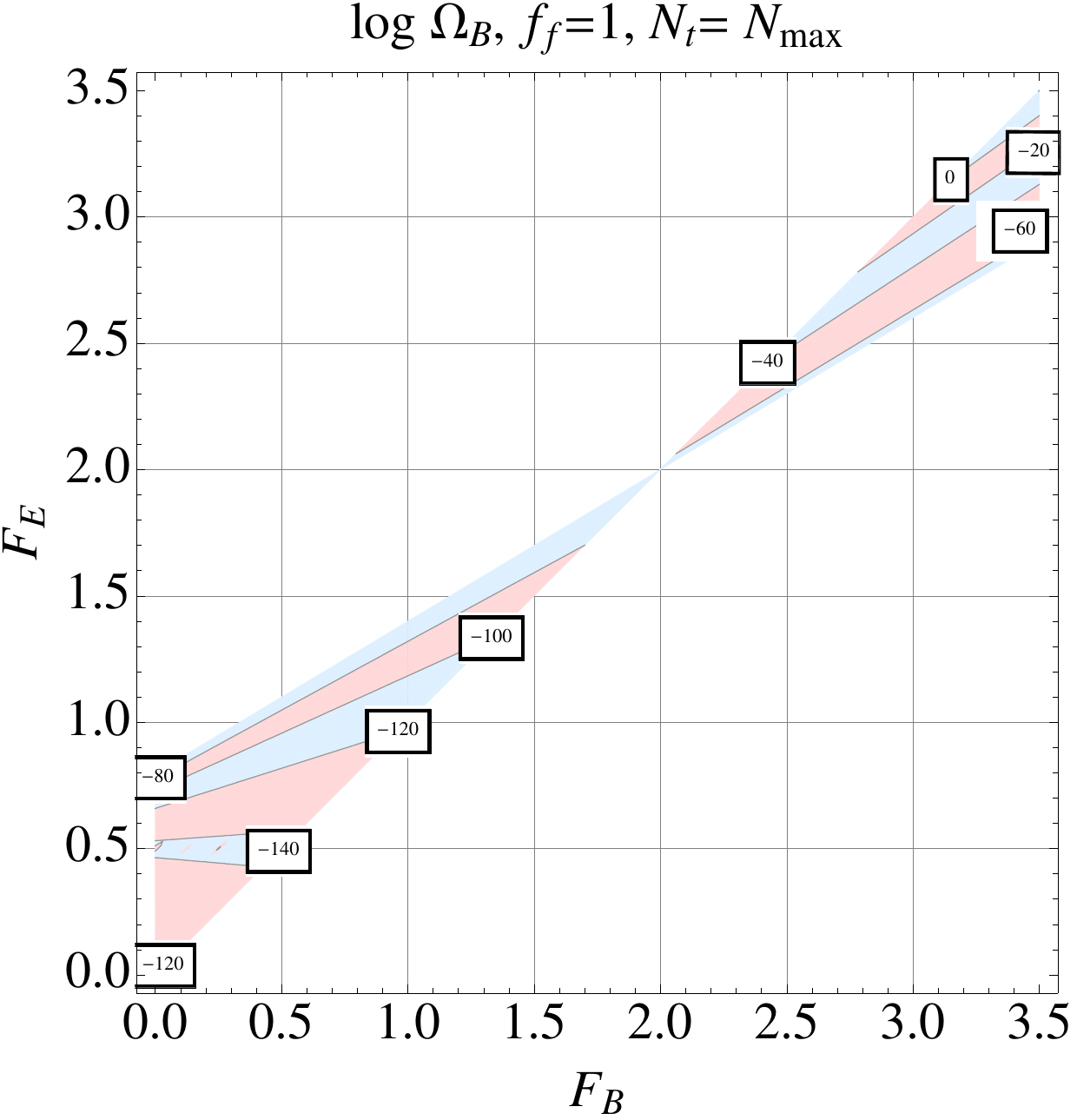}
\caption[a]{The contours of constant $\log{\Omega_{B}}$ in the case of Eq. (\ref{REG1a}) (plot at the right) and in the case of Eq. (\ref{REG1b}) (plot at the left); the typical scale has been chosen, as in Fig. \ref{Figure4}, to coincide with $k = 1\, \mathrm{Mpc}^{-1}$.}
\label{Figure5}      
\end{figure}

The results of Eqs. (\ref{REG1a}) and (\ref{REG1b}) are confirmed by the accurate analysis of the relevant contour plots. 
In the left and right plots of Fig. \ref{Figure4} we illustrate the common logarithm of $\Omega_{E}$ in the case of the regions 
discussed, respectively, in Eqs. (\ref{REG1a}) and (\ref{REG1b}). The power spectra are evaluated for wavenumbers 
comparable with the infrared branch of the spectrum, i.e. typically of the order of $10^{-4}\, \mathrm{Mpc}^{-1}$. Larger 
wavenumbers relevant for the magnetogenesis problem (i.e. $k = {\mathcal O}(\mathrm{Mpc}^{-1})$) lead to slightly larger figures 
which are however always of the same order. This is due to the quasi-flatness of the spectra in this domain. At the highest frequency
the plots of Fig. \ref{Figure3} apply and, even there, the energy densities are much smaller than the critical one.
In both plots  $N_{t} = N_{\mathrm{max}}=62.09$; the various 
labels illustrate the values of the critical fraction on the given curve. The shaded areas illustrates the regions where, according to Eqs. (\ref{REG1a}) 
and (\ref{REG1b}) the backreaction constraints are satisfied. 

In Fig. \ref{Figure5} we illustrate the same cases discussed in Fig. \ref{Figure4} but in terms of the common logarithm of $\Omega_{B}$. The 
fiducial scale is always $k = 10^{-4}\, \mathrm{Mpc}^{-1}$ and the other parameters are also unchanged in comparison with Fig. \ref{Figure4}.
The combination of Figs. \ref{Figure3}, \ref{Figure4} and \ref{Figure5} shows that 
the backreaction constraints are satisfied both for the electric and for the magnetic fields in the physical range of wavenumbers. In the last 
part of this section the safe regions pinned down by the backreaction constraints will be confronted with the magnetogenesis 
requirements. 

\subsubsection{The case $f_{i} =1$}
We are now going to address the complementary case where $f_{i} =1$.  According to Eq. (\ref{DEF3}) in the case $f_{i} =1$ the following inequalities must be satisfied:
 \begin{eqnarray}
 && - 2 \mu ( |\sigma| -1) + 5 - 2 |\sigma| \geq 0,
 \nonumber\\
 && - 2 \mu  ( |\sigma| -1) + 5 - 2 |\sigma -1| \geq 0,
 \nonumber\\
 && - 2 \mu (|\sigma| -1) \geq 0.
 \label{DEF6}
 \end{eqnarray}
 If $\sigma >1$, Eq. (\ref{DEF6}) implies that the two most constraining inequalities are
 $2\mu (\sigma -1) \leq 0$  and $5 - 2 \sigma \geq 2 \mu (\sigma -1)$. In the plane $(F_{B}, F_{E})$
 this requirement translates into the following triplet of inequalities: 
 \begin{equation}
 F_{E} > F_{B}+1,\qquad F_{E} < F_{B}, \qquad F_{B} \geq -2;
 \label{DEF7}
 \end{equation}
notice that $F_{B} \geq -2$ is automatically 
 satisfied since we work in the region $F_{B}\geq 0$ and $F_{E} \geq 0$.
 
 If $\sigma < 0$ the conditions of Eq. (\ref{DEF6}) become 
 \begin{equation}
 2 \mu (\sigma +1) \geq 0, \qquad 5 + 2 \sigma \geq - 2 \mu (\sigma +1), \qquad 5 + 2 \sigma \geq 2 - 2 \mu (\sigma +1).
 \label{DEF8}
 \end{equation}
 The last inequality of Eq. (\ref{DEF8}) is the most constraining so that $-3/2 - \mu (\sigma +1) \leq \sigma < 0$.
 In terms of $F_{E}$ and $F_{B}$ this range of $\sigma$ translates into:
 \begin{equation}
 \frac{1 - 2 F_{E}}{2 ( 1 + F_{B} - F_{E})} \leq 0, \qquad \frac{(F_{B} - F_{E}) ( 3 + 2 F_{B} - 4 F_{E})}{1 + F_{B} - F_{E}} \geq 0, \qquad F_{E} \leq 1 + \frac{F_{B}}{2}.
\label{DEF9}
 \end{equation}
\begin{figure}[!ht]
\centering
\includegraphics[height=8.4cm]{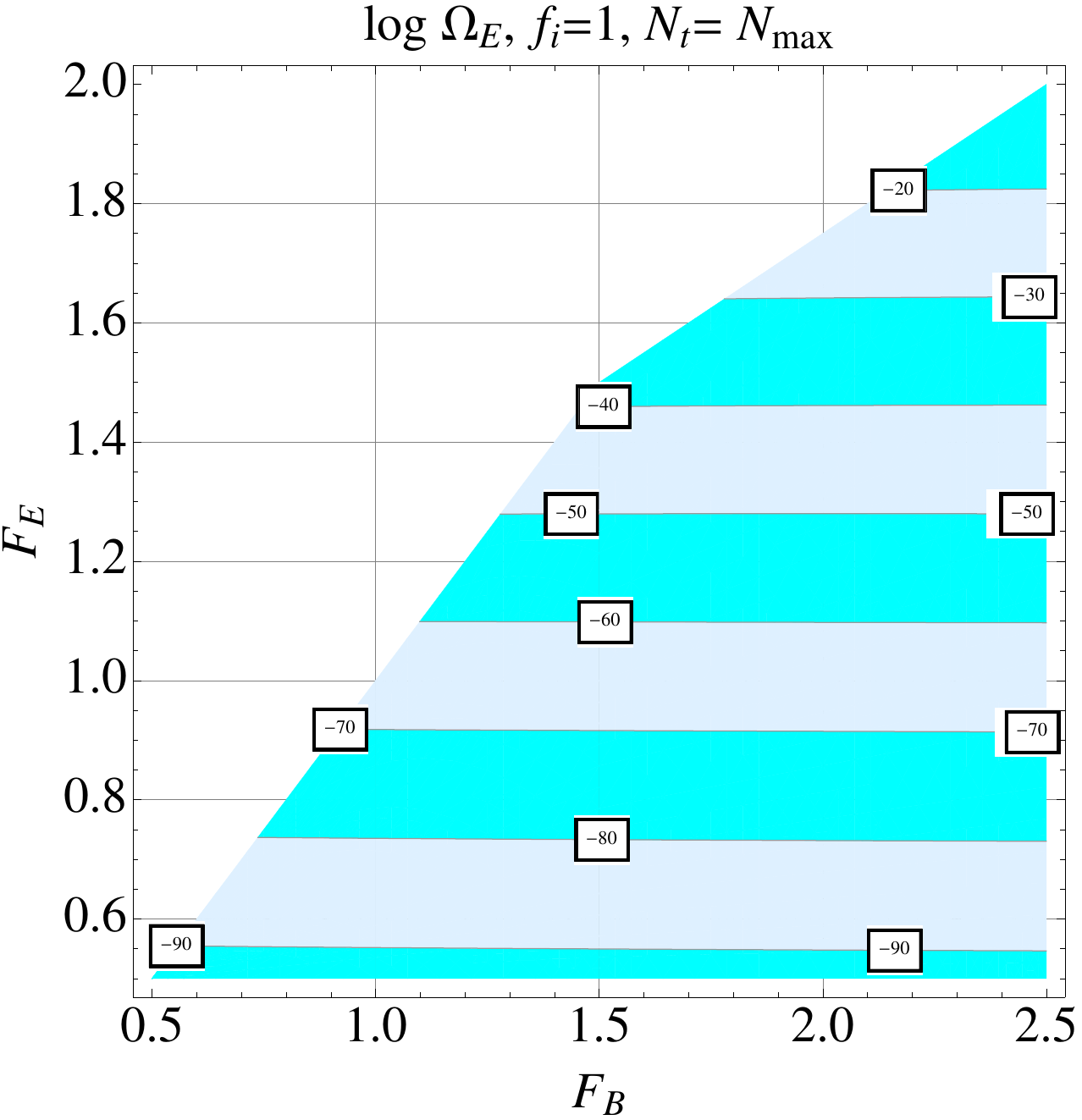}
\includegraphics[height=8.4cm]{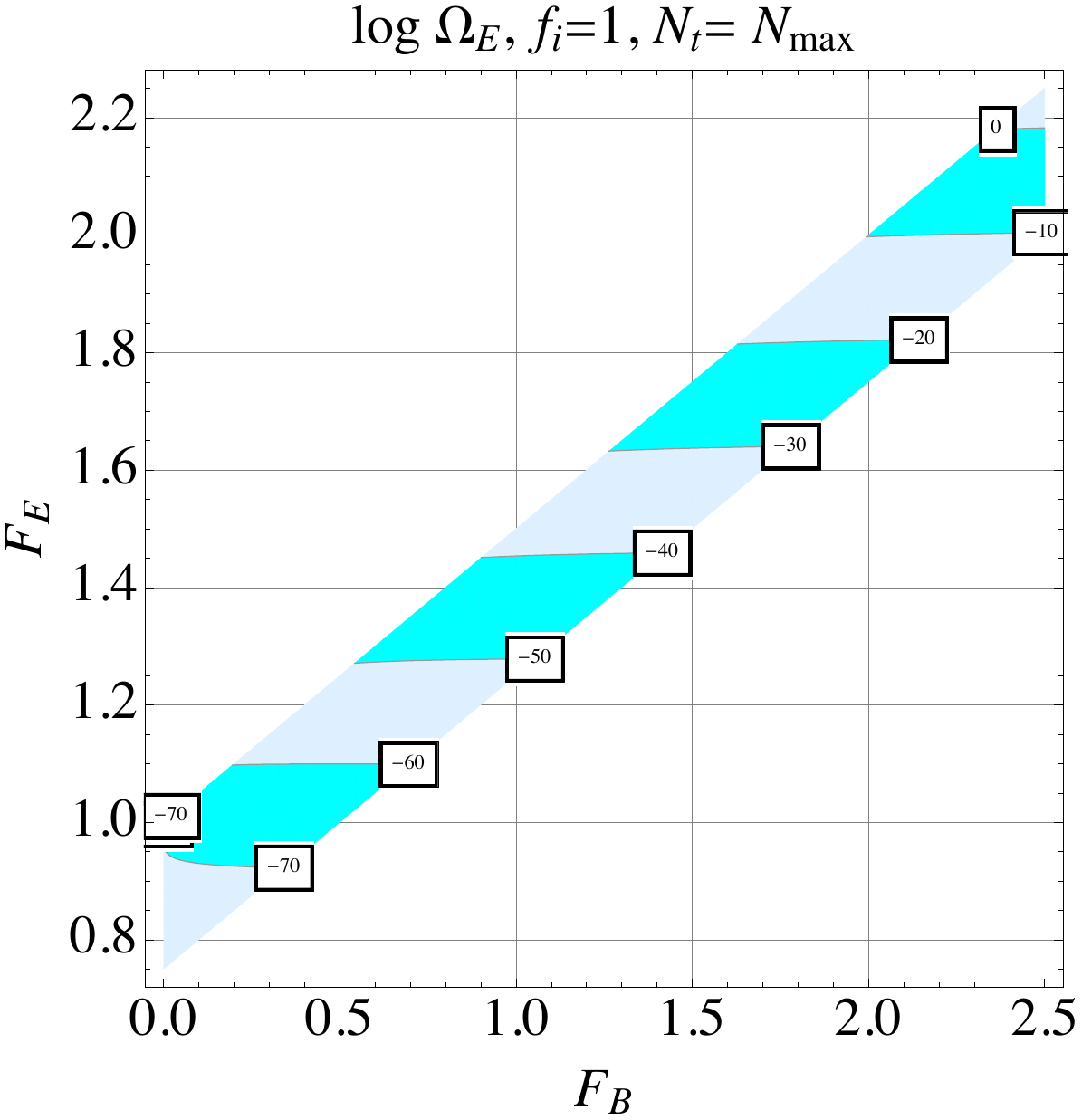}
\caption[a]{ The common logarithm of $\Omega_{E}$ in the region (\ref{REG2a}) (plot on the left) and in the region (\ref{REG2b}) (plot on the right). The same 
parameters and the same scales of Figs. \ref{Figure4} and \ref{Figure5} have been employed.}
\label{Figure6}      
\end{figure}
If $1 + F_{B} - F_{E} >0$, then $F_{E} > 1/2$ and the inequalities defining the allowed regions are\footnote{If $1 - F_{E} + F_{B} <0$, Eq. (\ref{DEF9}) demands that  $F_{E} <1/2$ which is not compatible with $F_{E} > F_{B} +1$. So this possibility must be discarded.}:
 \begin{equation}
 (F_{B} - F_{E}) ( 3 + 2 F_{B} -4 F_{E}) \geq 0, \qquad F_{E} < 1 + F_{B}/2.
 \end{equation}
For $F_{B} > F_{E}$  the allowed region is:
 \begin{eqnarray}
&& 1/2 < F_{B} \leq 3/2, \qquad 1/2< F_{E} < F_{B},
 \nonumber\\
&& F_{B} \geq 3/2, \qquad 3/2 \leq F_{E} \leq 3/4 + F_{B}/2.
\label{REG2a}
 \end{eqnarray}
 For $F_{B} < F_{E}$  the allowed region is:
 \begin{equation}
 0 \leq F_{B} \leq 3/2,\qquad 3/4 + F_{B}/2\leq F_{E} \leq 1 + F_{B}/2
 \label{REG2b}
 \end{equation}
 Finally in the region $0 < \sigma < 1$ the backreaction constraints are satisfied in the region $F_{E} < 1/2$.

\begin{figure}[!ht]
\centering
\includegraphics[height=8.5cm]{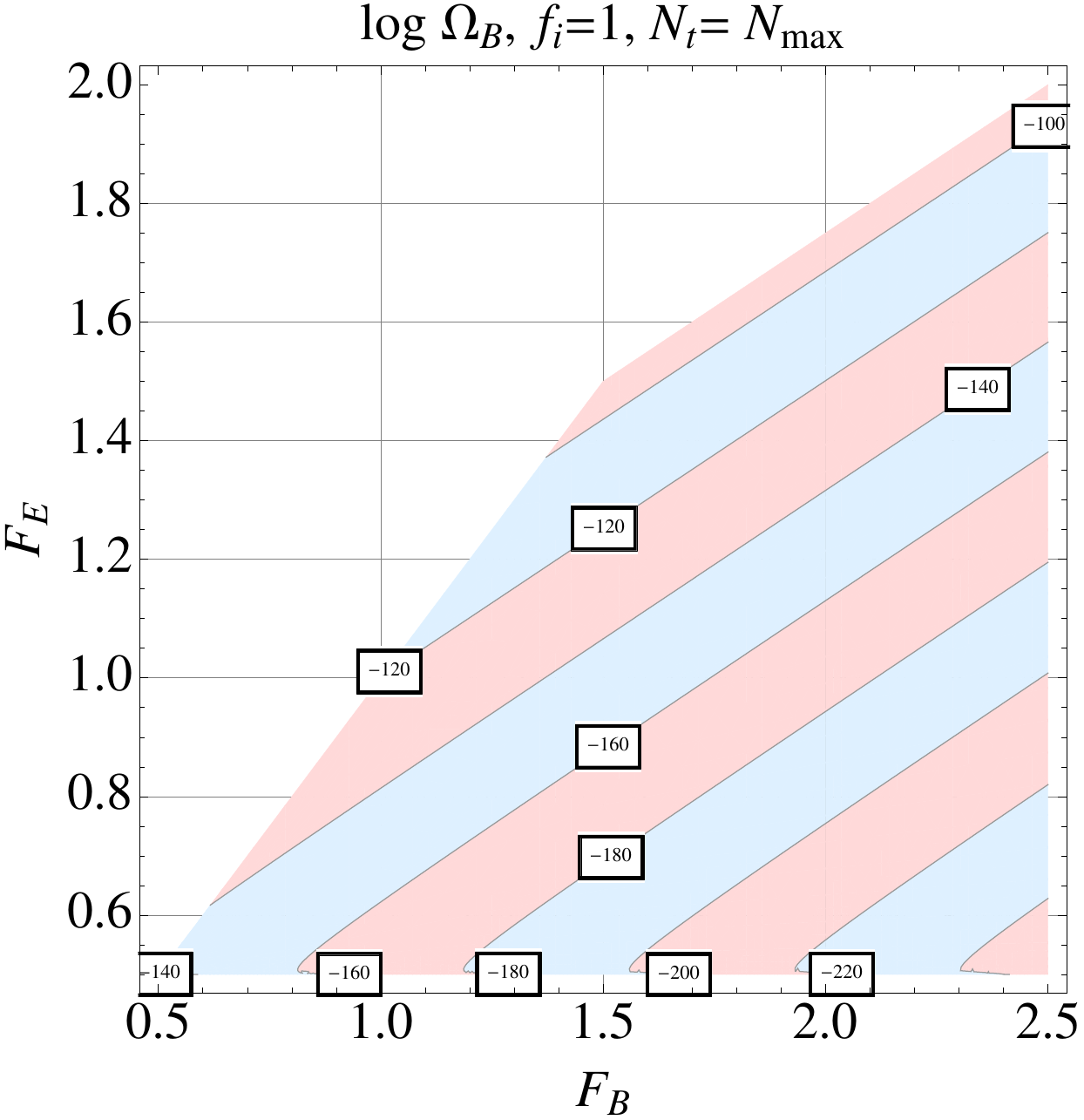}
\includegraphics[height=8.5cm]{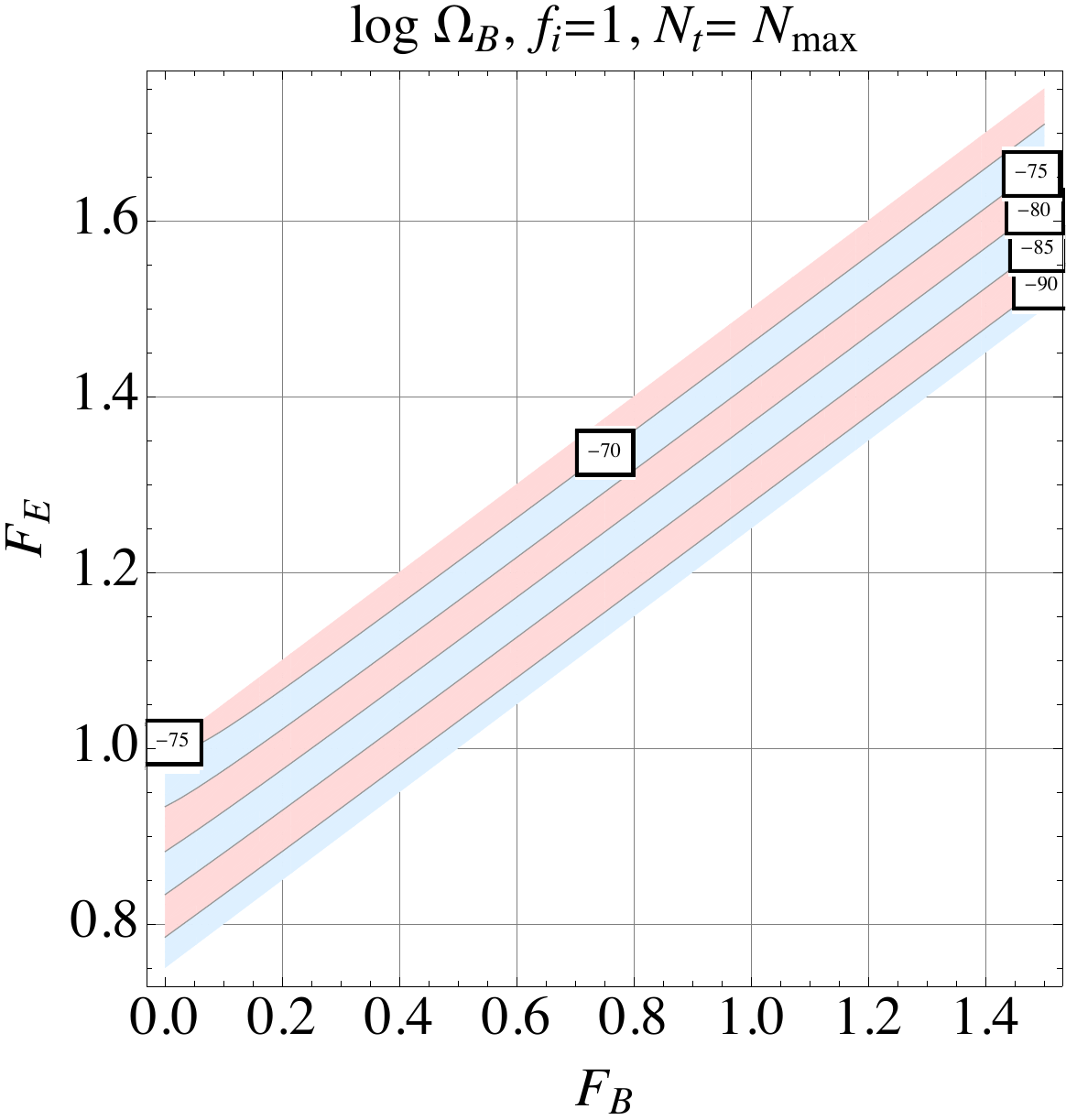}
\caption[a]{  The common logarithm of $\Omega_{B}$ in the region (\ref{REG2a}) (plot on the left) and in the region (\ref{REG2b}) (plot on the right). The same 
parameters and the same scales of Figs. \ref{Figure4} and \ref{Figure5} have been employed.}
\label{Figure7}      
\end{figure}

In Figs. \ref{Figure6} and \ref{Figure7} we illustrate, respectively, the common logarithm of $\Omega_{E}$ and of $\Omega_{B}$.
The contour plots on the left refer to the region of Eq. (\ref{REG2a}) while the contour plots on the right illustrate the parameter range of Eq. (\ref{REG2b}). 
The fiducial set of cosmological parameters is the same as in the previous figures.

\subsection{Dualizing the contour plots}

The region of the parameter space explored so far coincides with the first quadrant 
of the $(F_{B}, F_{E})$ plane. Using duality the contours plots of the other portions 
of the parameter space can be easily obtained and studied, as specifically discussed in sec. \ref{sec4}. Let us just give an example 
of this strategy. Let us suppose, for 
instance, to be interested in the case where both rates are negatives, namely $F_{B} <0$ and $F_{E} <0$, as it happens 
in the third quadrant of Figs. \ref{Figure1} and \ref{Figure2}. 

Recalling therefore Eqs. (\ref{firstQ}) and (\ref{thirdQ}) (and also Tabs. \ref{Table1} and \ref{Table2}) we can easily dualize
all the contour plots obtained so far in the regions where the backreaction constraints are safely satisfied. 
\begin{figure}[!ht]
\centering
\includegraphics[height=8cm]{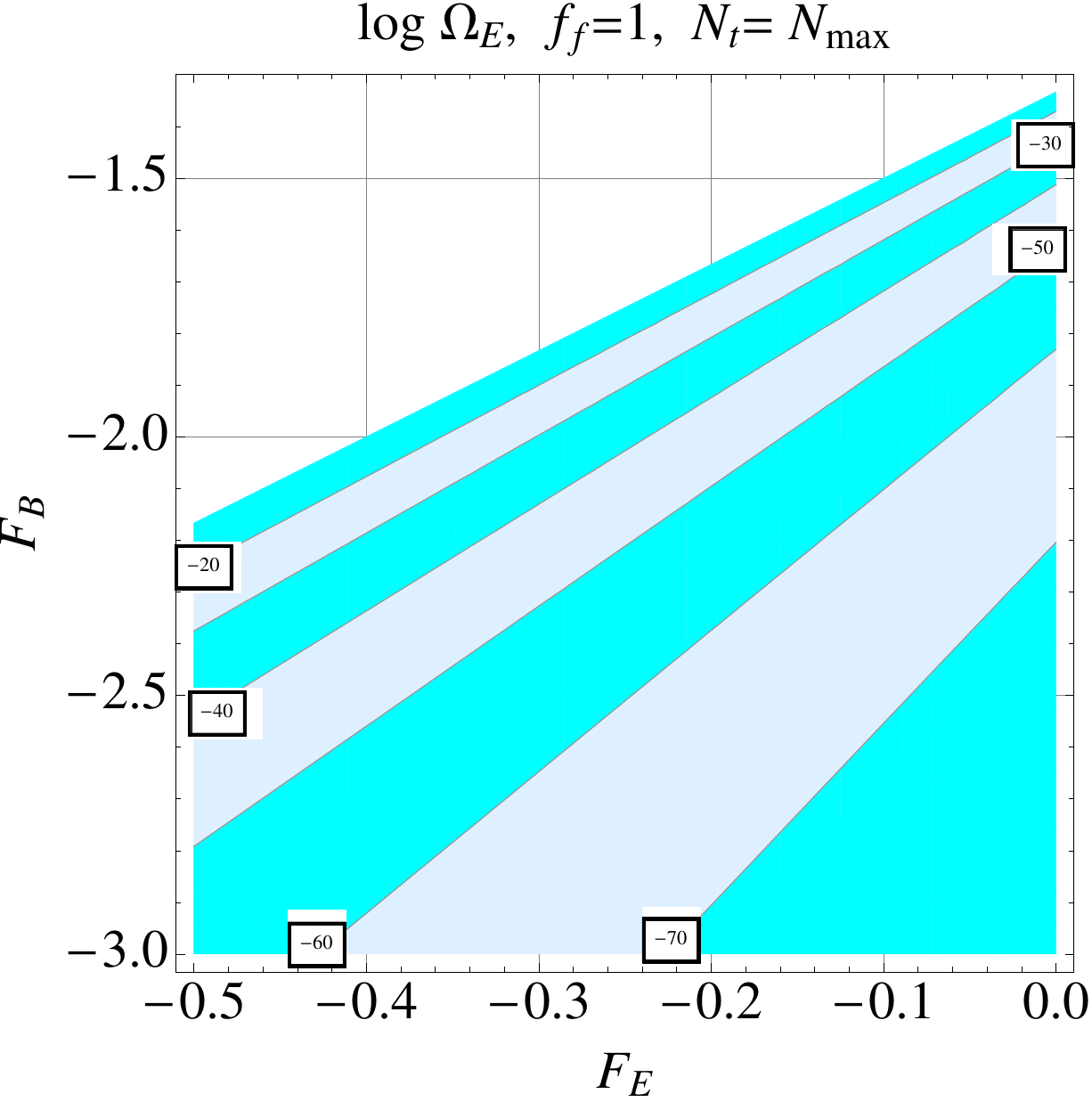}
\includegraphics[height=8cm]{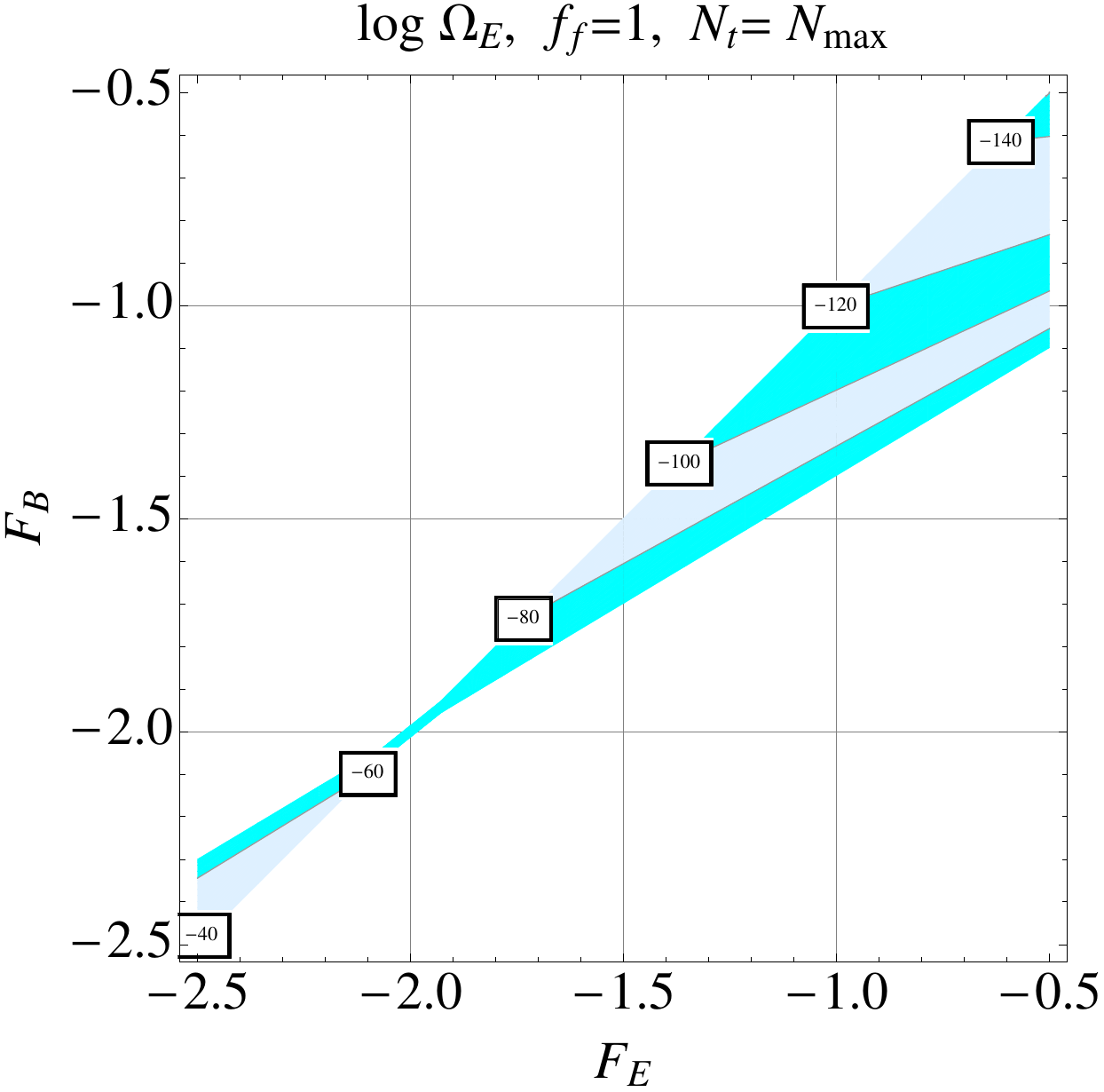}
\caption[a]{We illustrate the contour plots obtained by duality from Fig. \ref{Figure4}. Note that the axes 
are interchanged thanks to duality. The notations and the parameters are the same as in Fig. \ref{Figure4}.}
\label{Figure8}      
\end{figure}
In Figs. \ref{Figure8} abd \ref{Figure9} we illustrate the contours plots dual to the ones reported in Figs. \ref{Figure4} and \ref{Figure5}.
\begin{figure}[!ht]
\centering
\includegraphics[height=8cm]{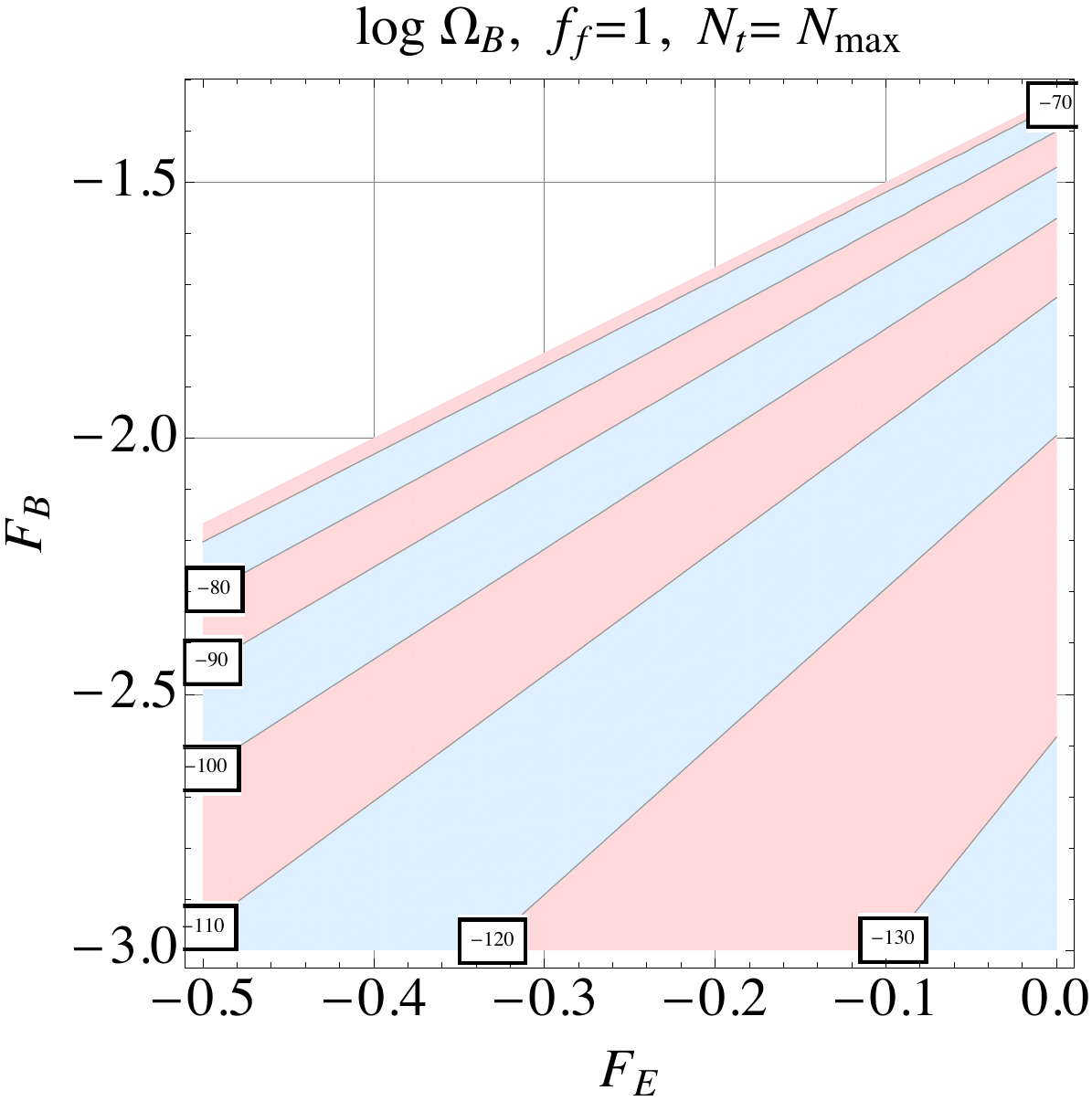}
\includegraphics[height=8cm]{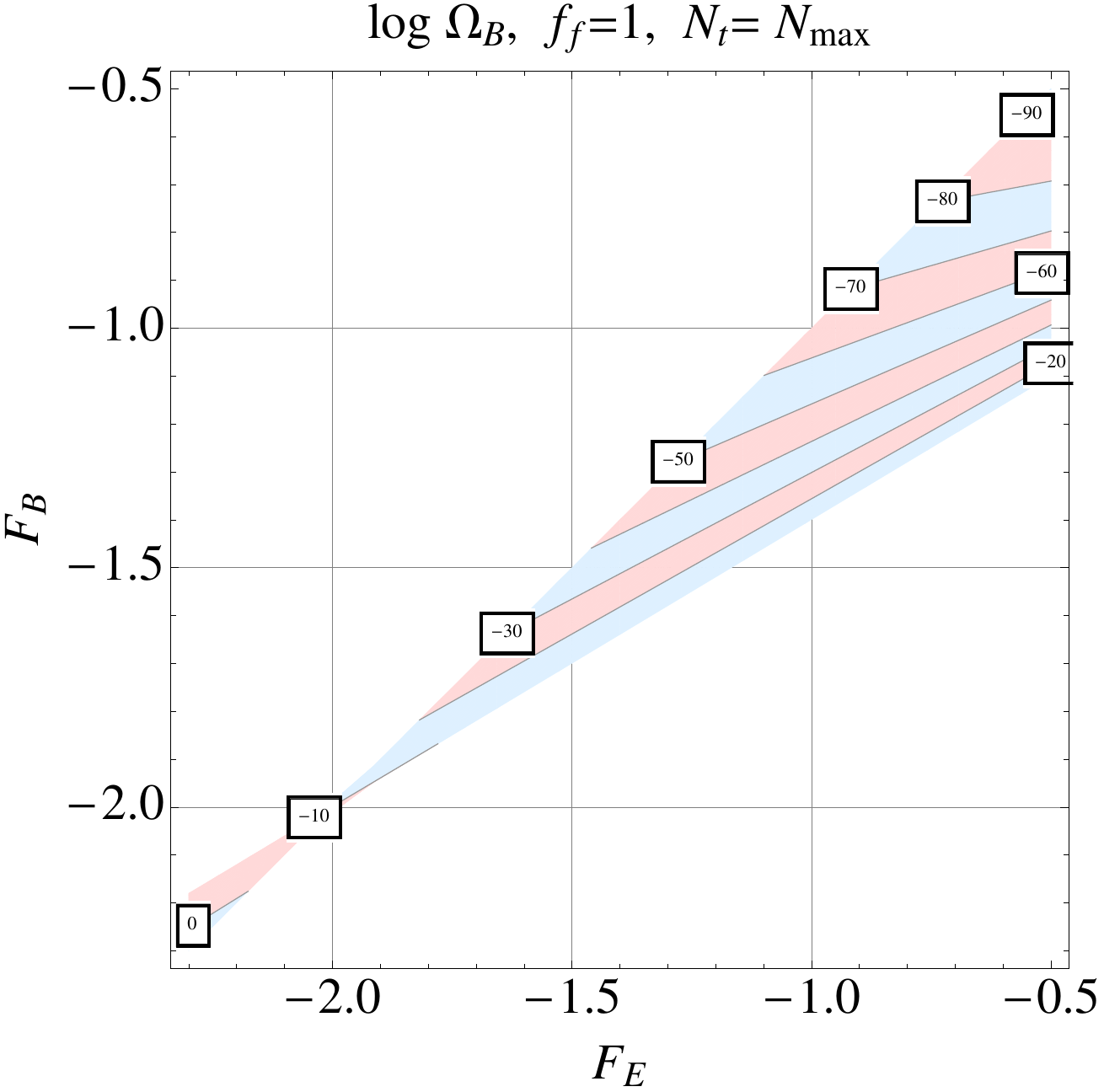}
\caption[a]{ We illustrate the contour plots obtained by duality from Fig. \ref{Figure5}. Note that the axes 
are interchanged thanks to duality.The notations and the parameters are the same as in Fig. \ref{Figure5}.}
\label{Figure9}      
\end{figure}
Since, under duality, $F_{B} \to - F_{E}$ and $F_{E} \to - F_{B}$ the axes of Figs. \ref{Figure8} and \ref{Figure9} are interchanged in comparison 
with the axes of Figs. \ref{Figure4} and \ref{Figure5}.  The same exercise discussed in the case $f_{f}=1$ can be also carried on 
in all the other relevant cases of phenomenological interest and also in the remaining quadrants of the parameter space. 

The plots reported in Figs. \ref{Figure8} and \ref{Figure9} can be related, by symmetry, to the plots of Figs. \ref{Figure4} and \ref{Figure5}.
Of course also the domains of the spectrum must be dualized as specifically discussed in section \ref{sec4}. It is however appropriate 
to stress that  Figs. \ref{Figure8} and \ref{Figure9} have been obtained by plotting explicitly the relevant critical fractions in the dual regions 
of Figs. \ref{Figure4} and \ref{Figure5}. The obtained results show indeed that, a posteriori, duality can be directly applied also to the exclusion 
plots without plotting explicitly the relevant spectra. 

\subsection{Magnetogenesis requirements}

Having determined the regions of the parameter space where the backreaction constraints are enforced we can address the problem 
of the magnetogenesis constraints. Let us therefore focus our attention on the first quadrant, as in the previous part of this section. 
In units of $\mathrm{nG}^2$  the magnetic power spectrum can be written as follows:
\begin{eqnarray}
\frac{P_{B}}{\mathrm{nG}^2} &=&  10^{-3.05} \, \biggl(\frac{h_{0}^2 \Omega_{R0}}{4.15 \times 10^{-5}}\biggr) \, \biggl(\frac{{\mathcal A}_{\mathcal R}}{2.41\times 10^{-9}}\biggr) \,
\biggl(\frac{\epsilon}{0.01}\biggr)  {\mathcal Q}_{B}(n_{B},\mu) 
\nonumber\\
&\times& \biggl(\frac{k}{H_{0}}\biggr)^{n_{B}-1} \,e^{- N_{max}(n_{B}-1)} \, f_{i}^{(4 - n_{B})/2} \, e^{ \mu N_{t} (4 - n_{B})}.
\label{mg1}
\end{eqnarray}
The magnetogenesis requirements 
 roughly demand that the magnetic fields at the time of the gravitational collapse of the protogalaxy should be approximately larger than a (minimal) power spectrum which can be estimated between $10^{-32}$ nG and $10^{-22}$ nG. The most optimistic estimate is derived by assuming 
that every rotation of the galaxy would increase the magnetic field of one efold. The number of galactic rotations since the 
collapse of the protogalaxy can be estimated between $30$ and $35$, 
leading approximately to  a purported growth of $13$ orders of magnitude. During collapse of the protogalaxy compressional amplification will 
increase the field of about $5$ orders of magnitude. Thus the required seed field at the onset of the gravitational collapse must be, at least, as large as $10^{-15}$ nG or, more realistically, larger than $10^{-11}$ nG. 

The simplest way to implement the magnetogenesis requirements is therefore to plot the magnetic power spectrum 
in units of $\mathrm{nG}^2$ for all the regions where the backreaction constraints are satisfied. The areas of the parameter 
space where 
\begin{equation}
\log{\biggl(\frac{P_{B}}{\mathrm{nG}^2} \biggr)} \geq - \lambda, \qquad 22 < \lambda < 32,
\label{mg3}
\end{equation}
will therefore offer viable models of magnetogenesis. Depending on the sign of the rates and on their relative magnitude 
it will therefore be possible to see whether the selected models will evolve either from a strongly coupled or from a weakly coupled regime.

Let us therefore start with the case where $f_{f} =1$. The relevant regions of the parameter space 
have been already discussed in Figs. \ref{Figure4} and \ref{Figure5}. For the same regions we shall now 
plot the common logarithm of $P_{B}/\mathrm{nG}^2$ and check if and where the condition of Eq. (\ref{mg3}) is 
satisfied. 
\begin{figure}[!ht]
\centering
\includegraphics[height=8.5cm]{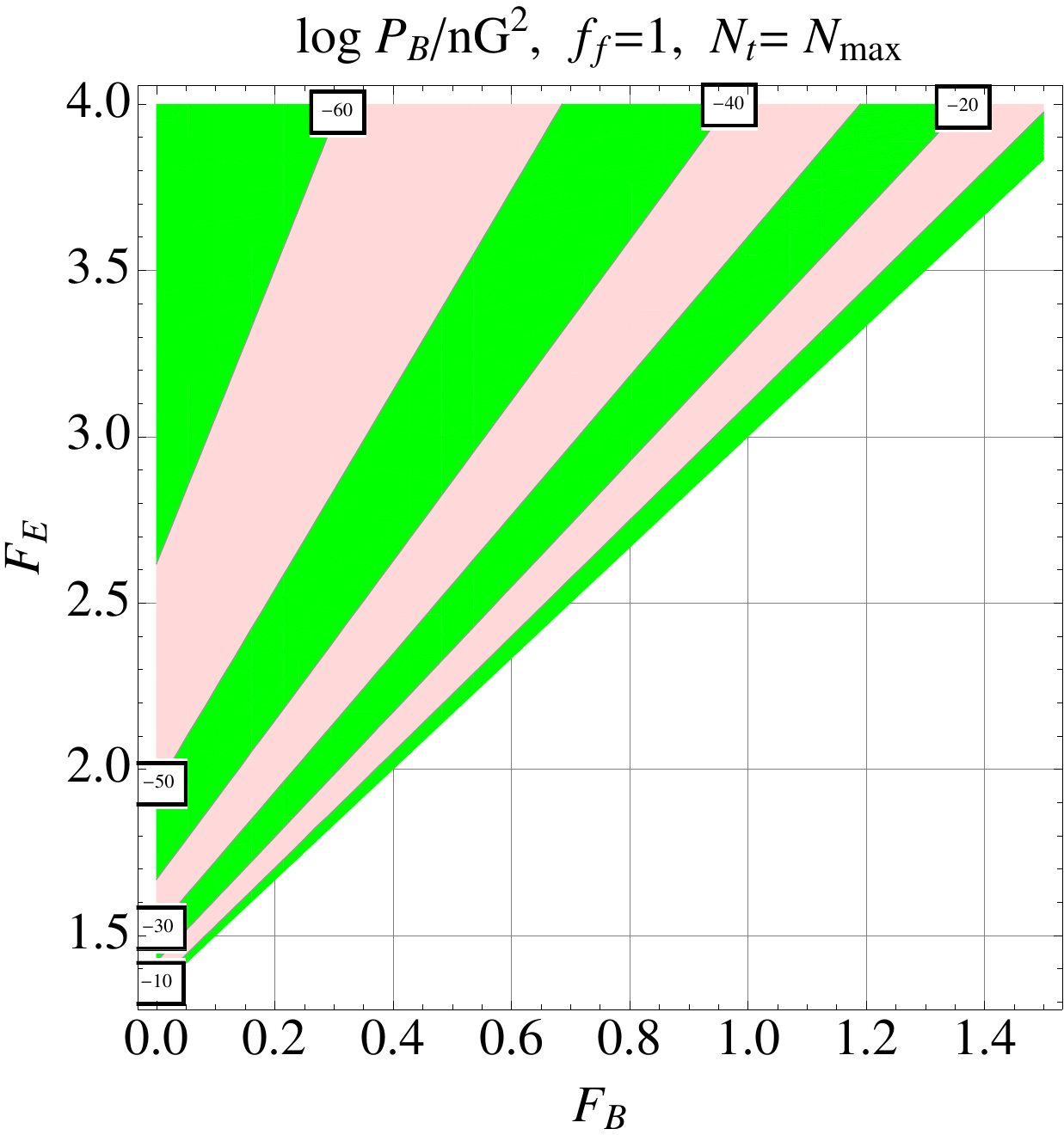}
\includegraphics[height=8.5cm]{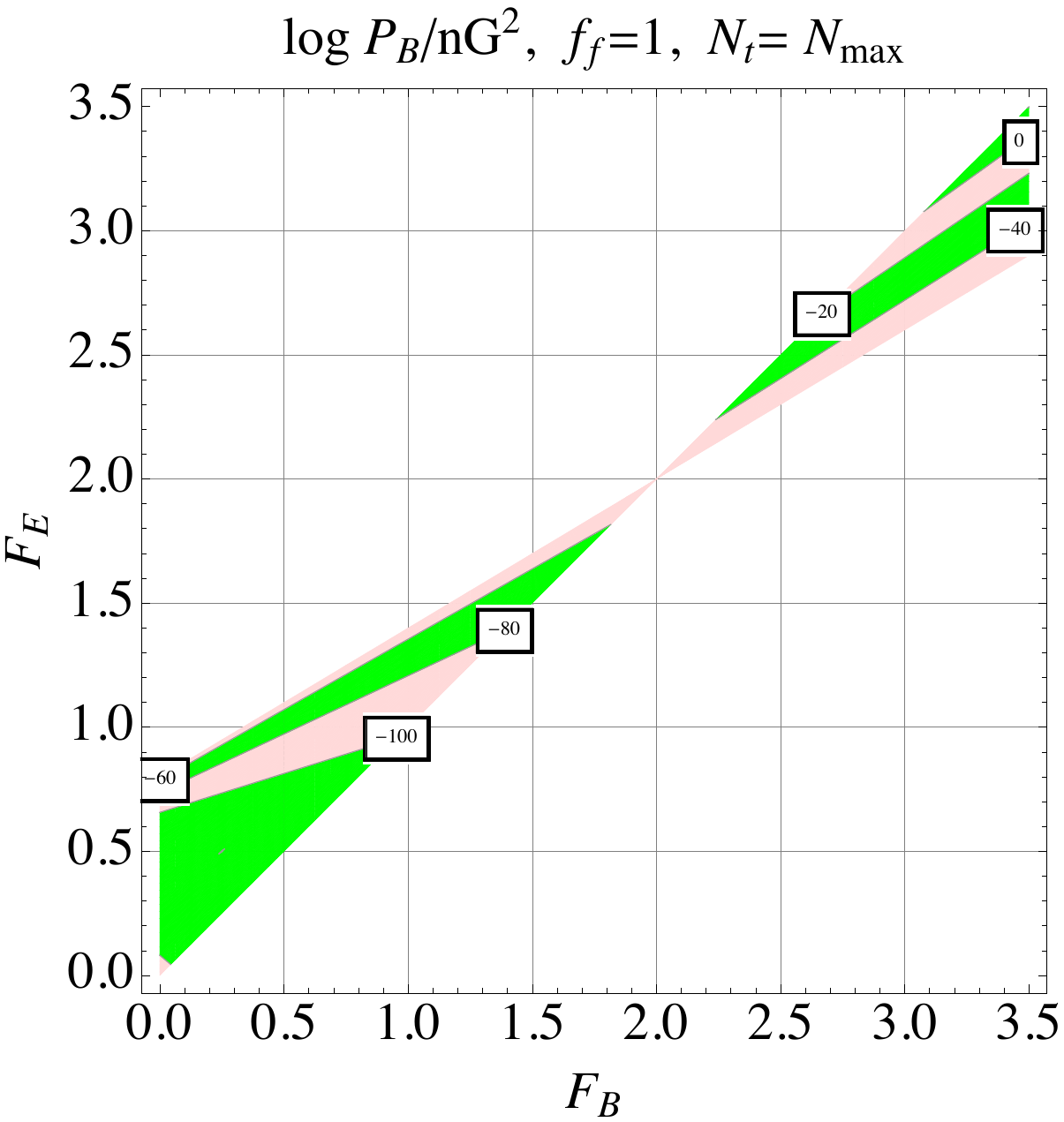}
\caption[a]{We illustrate the magnetic power spectrum in units of $\mathrm{nG}^2$ for the regions discussed in Figs. \ref{Figure4} and \ref{Figure5} 
where the backreaction constraints are all satisfied. }
\label{Figure10}      
\end{figure}
The result of this procedure is reported in Fig. \ref{Figure10} where the labels in the contour plot  refer 
to the values of the power spectrum in units of $\mathrm{nG}^2$.  The plot on the left in Fig. \ref{Figure10} refers to the 
same region discussed in the plots on the left in Figs. \ref{Figure4} and \ref{Figure5}; similarly the plots on the right 
correspond to the plots on the right in Figs. \ref{Figure4} and \ref{Figure5}. We clearly see 
that the requirement of Eq. (\ref{mg3}) is satisfied in two distinct regions: a stripe of the left plot and a smaller stripe in the plot 
on the right. According to Fig. \ref{Figure4}, however, the  region in the plot on the right must be excluded since 
in that part the electric fields are overcritical. We are therefore left with a region which can be analytically 
expressed as
\begin{eqnarray}
&& 5 F_{B}/3 + 4/3 \leq F_{E} \leq \frac{{\mathcal K} -6}{{\mathcal K}-4} F_{B} +  \frac{{\mathcal K} -5}{{\mathcal K}-4},
\label{slice}\\
&& {\mathcal K}(\lambda, k, N_{max}, f_{f}) = \frac{3.05 - \lambda + \log{(k/H_{0})} - N_{max} \log{e} - 2 \log{f_{f}}}{\log{(k/H_{0})} - N_{max} \log{e} - 0.5 \log{f_{f}}}.
\end{eqnarray}
In the case of the fiducial set of parameters corresponding to the previous figures we have that the above slice becomes
\begin{eqnarray}
&& 5 F_{B}/3 + 4/3 \leq F_{E} \leq 1.46 +1.91 F_{B}, \qquad \lambda= 22,
\label{slice2}\\
&& 5 F_{B}/3 + 4/3 \leq F_{E} \leq 1.56 +2.13 F_{B}, \qquad \lambda= 32.
\label{slice3}
\end{eqnarray}

The lower bound in Eq. (\ref{slice}) corresponds to the slightly distorted straight line appearing both in Figs. \ref{Figure4} and \ref{Figure10}.
The upper bound has been obtained from the approximate shape of the slice defined by the corresponding contour plot. 
In this region both gauge couplings are increasing (no strong coupling problem) and furthermore $f$ decreases from an initially 
large value to $f_{f} =1$.
 
The same analysis discussed in the case $f_{f} =1$ can now be extended to the case $f_{i}=1$ already discussed 
in Fig. \ref{Figure6} and \ref{Figure7}.
\begin{figure}[!ht]
\centering
\includegraphics[height=8.5cm]{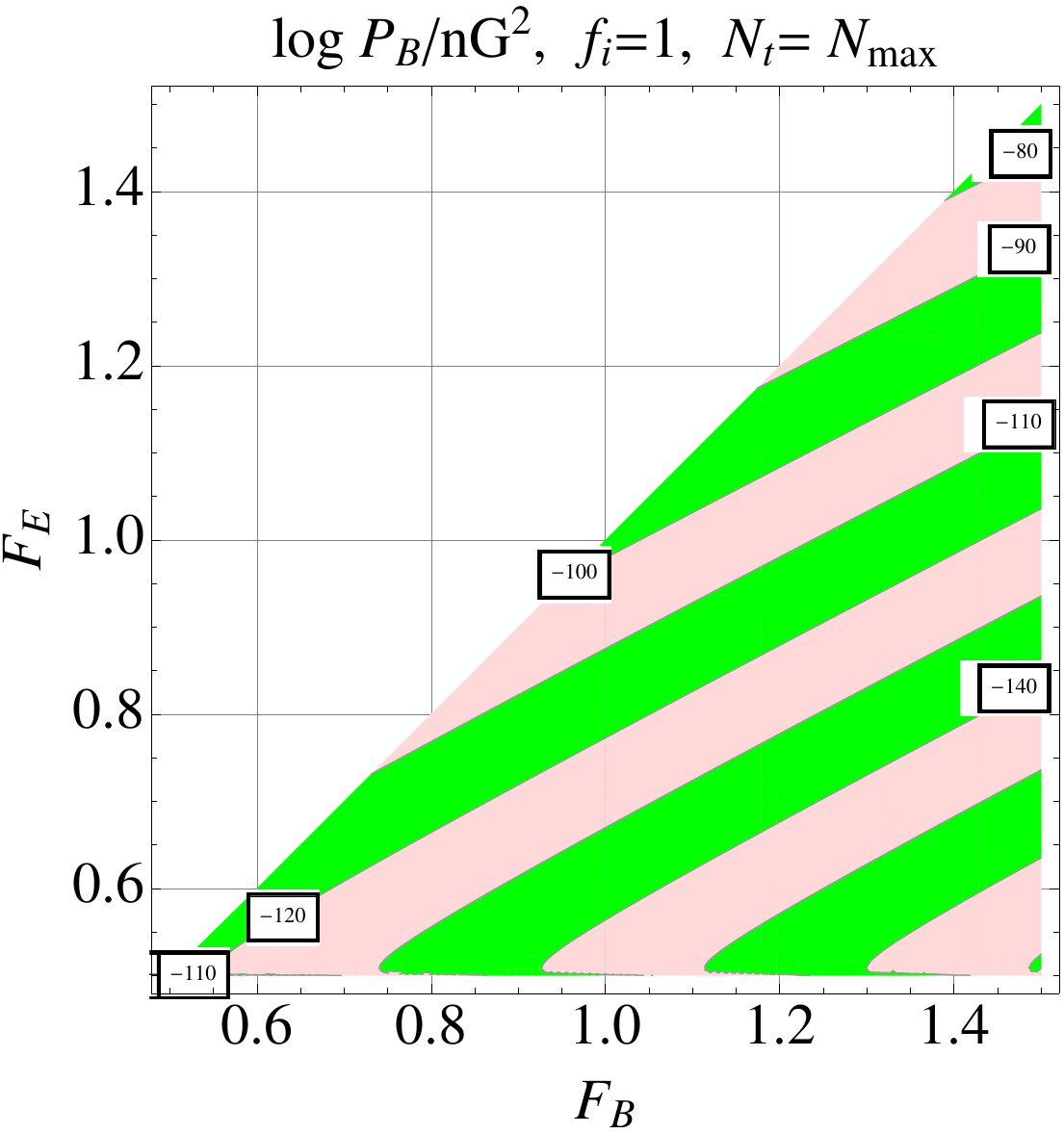}
\includegraphics[height=8.5cm]{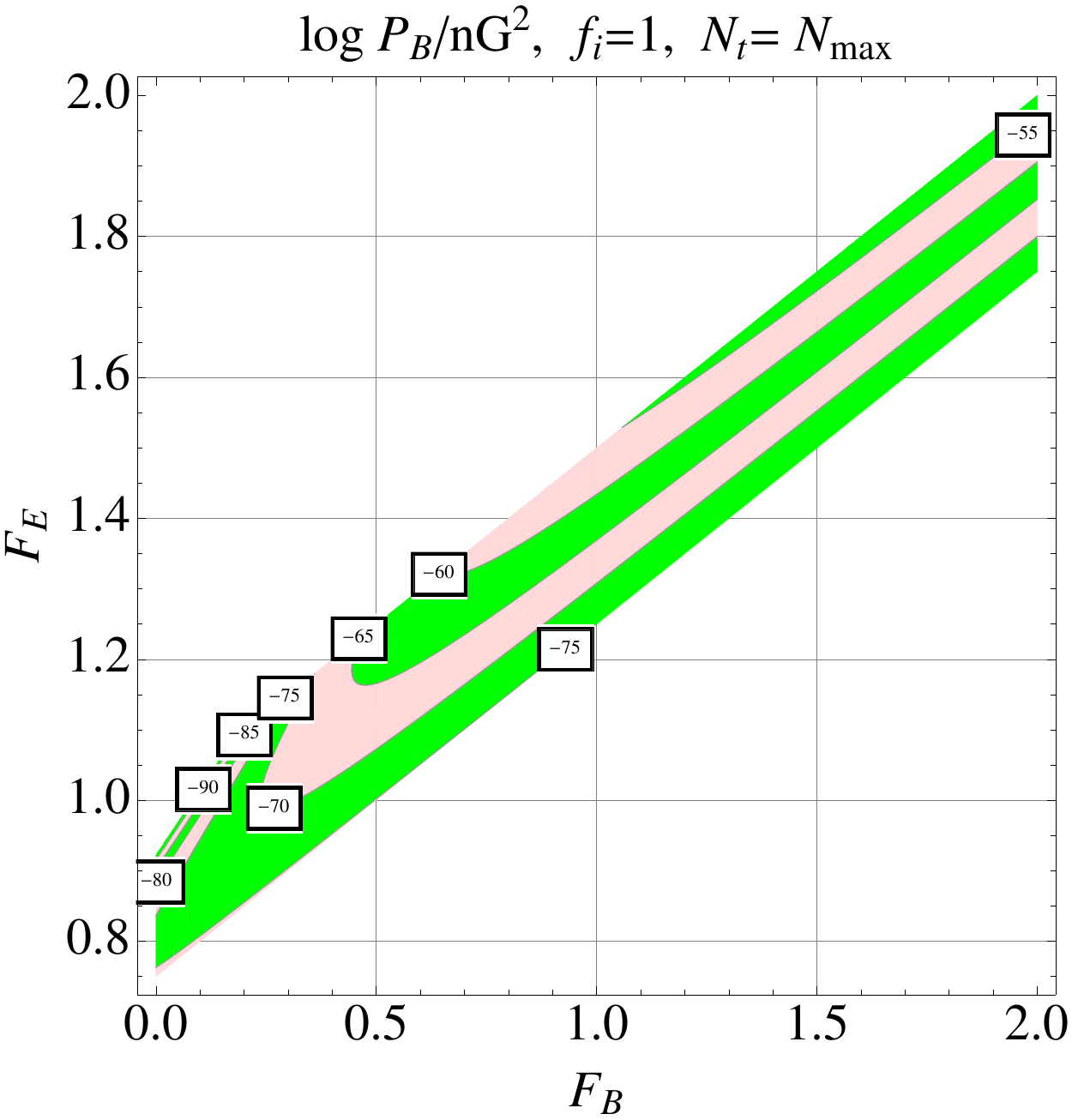}
\caption[a]{We illustrate the magnetic power spectrum in units of $\mathrm{nG}^2$ for the regions discussed in Figs. \ref{Figure6} and \ref{Figure7} 
where the backreaction constraints are all satisfied. }
\label{Figure11}      
\end{figure}
The result of this analysis is reported in Fig. \ref{Figure11} where we clearly see that the magnetogenesis constraints are not satisfied. 

As already mentioned in the case of the backreaction constraints, the various regions of the parameter space can be 
easily dualized with the purpose of discussing a given region of the parameter space once the results on the first 
quadrant are known.
\begin{figure}[!ht]
\centering
\includegraphics[height=8cm]{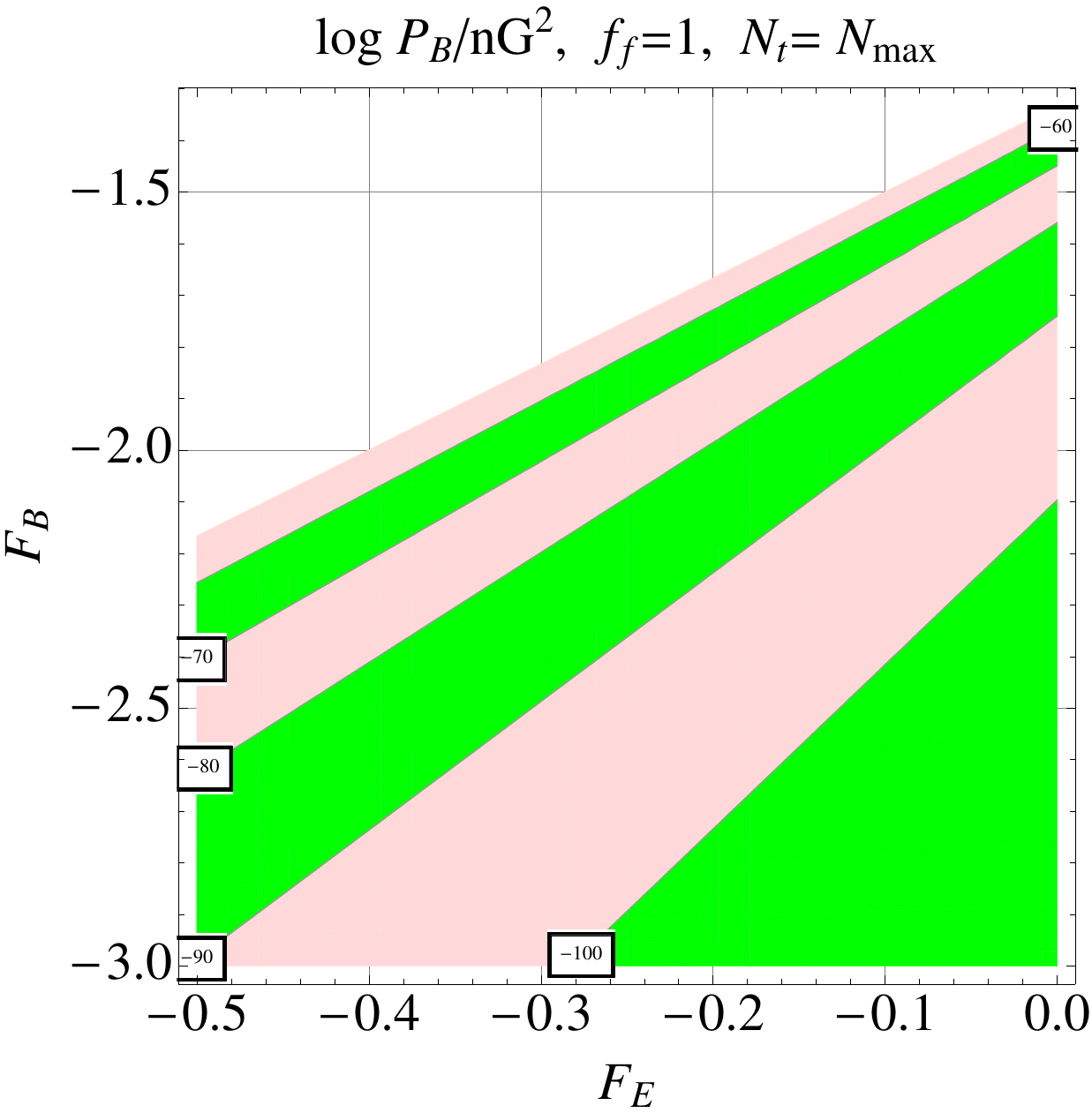}
\includegraphics[height=8cm]{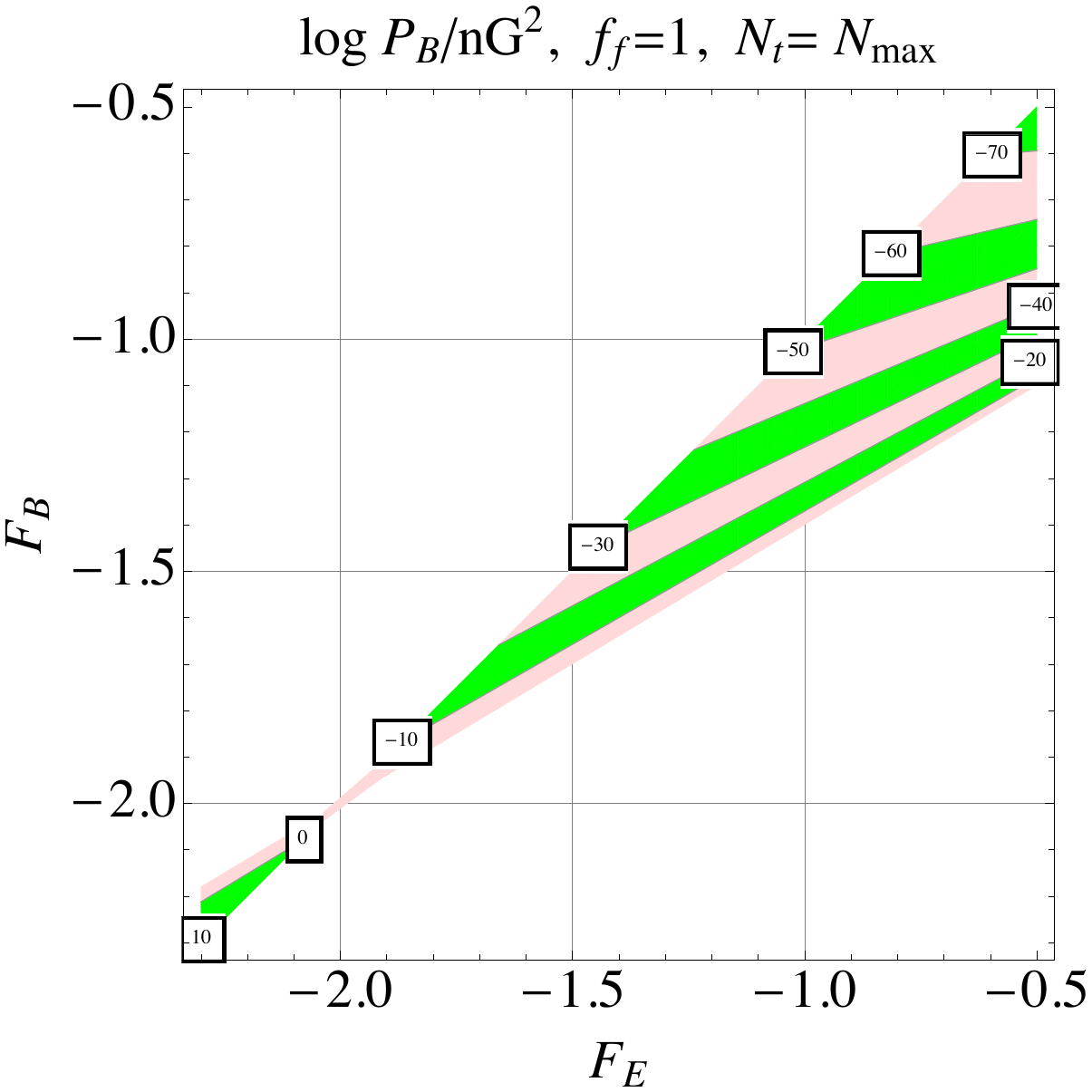}
\caption[a]{We illustrate the magnetic power spectrum in units of $\mathrm{nG}^2$ for the regions discussed in Figs. \ref{Figure8} and \ref{Figure9} 
where the backreaction constraints are all satisfied. }
\label{Figure12}      
\end{figure}
An example along this line is obtained in Fig. \ref{Figure12} where the magnetogenesis constraints are illustrated in the case of the dual 
regions already studied in Figs. \ref{Figure8} and \ref{Figure9}.  We can therefore see that also the region of Fig. \ref{Figure12}
leads to a viable class of magnetogenesis models where, however, the rates are negative and the gauge couplings decrease rather
than increasing. There are some who say that these initial data are unnatural. Some other authors claim instead that the gauge coupling should follow the gravitational coupling.
In conventional inflationary models the strong coupling is at the beginning (since the singularity is at the beginning) while in bouncing models the singularity 
is at the end of the inflationary phase. So the most natural choice would be to ask for decreasing gauge couplings in conventional inflationary models and increasing gauge couplings in bouncing models (see, e.g. third paper in Ref. \cite{mgg}). In this scenario both options can be realized in non-overlapping domains of the parameter 
space.
 
The cases $f_{f}=1$ and $f_{i}=1$ represent in practice the two extremes of the whole range of physical models where $f_{i} \geq 1$. The cases 
where $f_{i} = {\mathcal O}(1)$ are very close to the patterns discussed in the $f_{i}=1$ case. Conversely when $f_{i} \gg 1$ the situation 
is similar to the case $f_{f} = {\mathcal O}(1)$. It is clear that by appropriately scanning the parameter space we could easily 
pass from an allowed region in the plane $(F_{B},\, F_{E})$ to a tuning volume in the space $(F_{B},\, F_{E},\, f_{i})$. In that space the range 
of variation will be given exactly by the excursion discussed in the present paper, i.e. $ 1 \leq f_{i} \leq f^{(max)}_{i}$ where 
$f^{(max)}_{i}$ coincides with $\exp{[(F_{E} - F_{B}) N_{t}]}$ and it corresponds to $f_{f}=1$.

Before closing this section let us also comment on the possible excursion of $N_{max}$. As already mentioned if the reheating 
is delayed $N_{max}$ can increase and get larger by even 15 efolds \cite{mgg,LL}. In this case the effect on the allowed region is minimal. For instance 
if $N_{max} =78$  the area of Eq. (\ref{slice2}) becomes  $F_{B}/3 + 4/3 \leq F_{E} \leq 1.42 +1.84 F_{B}$ which is not different, for 
all practical purposes, from the previous expression.

\renewcommand{\theequation}{6.\arabic{equation}}
\setcounter{equation}{0}
\section{Concluding remarks}
\label{sec6}

Magnetogenesis models based on derivative couplings have been comprehensively analyzed. 
A similar kind of framework arises in the relativistic generalization of Van der Waals interactions. 
The key aspect of this class of models is that the electric and the magnetic susceptibilities 
are not bound to coincide all the time, i.e. $\chi_{E} \neq \chi_{B}$. This implies that also the corresponding gauge couplings $g_{E}$ and $g_{B}$ may evolve in time at different rates. After  presenting a general decomposition of the various coupling functions, 
the quantization of the system has been approached in terms of a newly defined time coordinate 
reducing to conformal time in the case when the electric and the magnetic susceptibilities coincide. 

It turns out that the power spectra depend on three different quantities:  the normalized rates of variation of the electric and magnetic gauge 
couplings (i.e.  $F_{E}$ and $F_{B}$) and also $f_{i}$, i.e the amplitude of $g_{B}^2/g_{E}^2$ at the onset of the dynamical evolution that has been taken to coincide with the beginning of the inflationary phase.  A fourth parameter measures the strength of one of the two gauge couplings at the beginning of inflation. 
The parameter space of the model has been scanned by using duality in order to relate the  
different quadrants of the $(F_{B},\, F_{E})$ plane. The physical content of the whole parameter space can therefore be reduced, via duality,  to the 
careful analysis of the region where $F_{E}$ and $F_{B}$ are both positive semidefinite. 

While the arbitrary variation of $F_{B}$ and $F_{E}$ is probably the most relevant aspect, 
a full account of the arbitrary variation of $f_{i}$ is not particularly significant at this stage of the analysis of the model. To explain the general  trend it is sufficient to examine the case where $f_{i} = {\mathcal O}(1)$ and the case where $f_{f} = {\mathcal O}(1)$, recalling that $f_{f}$ denotes the value of $f$ at the end of inflation.
The parameter space of this magnetogenesis scenario has then be accurately charted in the $(F_{B},\, F_{E})$ plane. As a consequence of this analysis 
the models where $f_{i} = {\mathcal O}(1)$ are favoured when at least one of the gauge couplings is initially strong while the models 
$f_{f} = {\mathcal O}(1)$ are favoured when both gauge couplings are small at the beginning of inflation. 

There exist wide regions in the parameter space where backreaction effects are negligible, gauge couplings are initially small and $f_{f} ={\mathcal O}(1)$. This area is defined by the corresponding excursion of the normalized rates\footnote{The curves defining the allowed region depend on the 
total number of efolds which will be taken to coincide with $N_{\mathrm{max}} \simeq 63.25 + 0.25 \ln{\epsilon}$ , i.e. the maximal number of efolds today accessible by observations} in the $(F_{B},\, F_{E})$ plane: 
\begin{equation}
5 F_{B}/3 + 4/3 \leq F_{E} \leq 1.46 + 1.91 F_{B}, \qquad 0 < F_{B} < 5.
\label{area}
\end{equation}
The functional dependence of the magnetic spectral index upon $F_{B}$ and $F_{E}$ changes from quadrant to quadrant (and also within the same quadrant).
In the area defined by Eq. (\ref{area}) it is given by:
\begin{equation}
n_{B} = \frac{5 + 6 F_{B} - 4 F_{E}}{1 + F_{B} - F_{E}}, \qquad 1 \leq n_{B} \leq 1.9.
\label{area2}
\end{equation}
The second relation appearing in Eq. (\ref{area2}) follows from Eq. (\ref{area}) and from the explicit expression 
of $n_{B}$ in the allowed region. In the range of Eqs. (\ref{area}) and (\ref{area2}) 
the magnetic power spectrum at the scale of the protogalactic collapse is always larger 
than $10^{-22}\, \mathrm{nG}^2$. As in the case of curvature perturbations, the scale-invariant 
limit of the magnetic power spectrum holds for $n_{B}\to 1$ where its amplitude is:
\begin{equation}
\lim_{n_{B} \to 1}\frac{P_{B}}{\mathrm{nG}^2} =  10^{-2.805} \, \biggl(\frac{h_{0}^2 \Omega_{R0}}{4.15 \times 10^{-5}}\biggr) \, \biggl(\frac{{\mathcal A}_{\mathcal R}}{2.41\times 10^{-9}}\biggr) \,
\biggl(\frac{\epsilon}{0.01}\biggr)  \biggl(\frac{1 + 2 F_{B}}{5}\biggr)^4.
\label{area3}
\end{equation}
The scale-invariant amplitude still depends on $F_{B}$. If the magnetogenesis constraints are relaxed by requiring that the magnetic power spectrum at the  protogalactic collapse exceeds only 
$10^{-32}\, \mathrm{nG}^2$ the range of allowed values of $n_{B}$ gets slightly wider, namely $1 \leq n_{B} \leq 2.2$.
\begin{table}
\begin{center}
\vskip 0.5truecm
\begin{tabular}{| c | | | c | l | c | | | c | | |  c | }
\hline
 $F_{B}$ \quad & $F_{E} = 5 F_{B}/3 + 4/3$ & $F_{E}= 1.46 + 1.91 F_{B}$& $F_{E} =1.56 + 2.13 F_{B}$   \\ \hline
$0$ & $\sqrt{P_{B}} = 10^{-2.80} \,\mathrm{nG}$ & $\sqrt{P_{B}} = 10^{-12.25} \,\mathrm{nG}$   &  $\sqrt{P_{B}} = 10^{-16.69} \,\mathrm{nG}$ \\ \hline
$2$ & $\sqrt{P_{B}} = 10^{-1.40} \,\mathrm{nG} $&  $\sqrt{P_{B}} = 10^{-10.91} \,\mathrm{nG} $ &  $\sqrt{P_{B}} = 10^{-15.86} \,\mathrm{nG} $\\ \hline
$4 $ & $\sqrt{P_{B}} = 10^{-0.89} \,\mathrm{nG} $ &   $\sqrt{P_{B}} = 10^{-10.48} \,\mathrm{nG}$ &  $\sqrt{P_{B}} = 10^{-15.52} \,\mathrm{nG}$\\ \hline
\hline
\end{tabular}
\caption{Typical values of the magnetic power spectra in the allowed region and at the fiducial scale $k = 1\, \mathrm{Mpc}^{-1}$.}
\label{Table3}
\end{center}
\end{table}
\begin{table}
\begin{center}
\vskip 0.5truecm
\begin{tabular}{| c | | | c | l | c | | | c | | |  c | }
\hline
 $F_{B}$ \quad & $F_{E} = 5 F_{B}/3 + 4/3$ & $F_{E}= 1.46 + 1.91 F_{B}$ & $F_{E} =1.56 + 2.13 F_{B}$   \\ \hline
$0$ &  $n_{B} =1$ &       $n_{B} =1.826$                 &  $n_{B} =2.214$ \\ \hline
$2$ &  $n_{B} =1$ &       $n_{B} =1.807$                 &  $n_{B} =2.226$\\ \hline
$4$ &  $n_{B} =1$ &       $n_{B} =1.804$                 &  $n_{B} =2.228$\\ \hline
\hline
\end{tabular}
\caption{Typical values of the magnetic spectral index in the allowed region.}
\label{Table4}
\end{center}
\end{table}
In Tab. \ref{Table3} we illustrate the values of the magnetic power spectra in the allowed region defined by Eqs. (\ref{area}) and (\ref{area2}).
In Tab. \ref{Table4}, for the same portion of the parameter space, we report the magnetic spectral indices. Notice that the boundary lines 
of the allowed region are isospectral in the sense that, along these lines, the magnetic spectral index does not change 
or it changes very little. It is interesting to stress that, in this model, $f_{f} \to 1$ implies the equality of the gauge couplings at the end of inflation but not 
necessarily the coincidence of the rates. On the contrary, if the rates are coincident during inflation (i.e. $F_{E} \to F_{B}$) 
the conventional situation is recovered and the two gauge couplings evolve at the same rate. The curve $F_{E} =F_{B}$ is not isospectral and it is not included 
in the area defined by Eqs. (\ref{area}) and (\ref{area2}). This occurrence shows, once more, that when $F_{E} \neq F_{B}$ the parameter space 
gets indeed wider in comparison with the conventional situation.

All in all if the magnetic and the electric susceptibilities do not coincide the allowed regions in the parameter space of inflationary magnetogenesis gets wider in comparison with the conventional class of models. According to some, the natural initial conditions for inflationary magnetogenesis strictly demand minute gauge couplings at the beginning of inflation and larger gauge 
couplings at reheating. According to a different way of thinking gauge coupling should follow the gravitational coupling:  as the production of a flat spectrum of curvature perturbations demands a strong gravitational coupling in the past, similarly a quasiflat magnetic field spectrum is realized in the case of a decreasing gauge coupling which gets progressively smaller during inflation. In this scenario both options are plausible in different regions of the parameter space.

\end{document}